\documentclass[fleqn,11pt]{report}
\usepackage{pictex}
\usepackage{makeidx}
\usepackage{amsfonts}
\usepackage{latexsym}
\begin{document}
%
% packages needed for math fonts
% these require in the tex-file: \usepackage{amsfonts}
\newcommand{\bb}[1]{\mathbb{#1}}
\newcommand{\R}{\bb{R}}
\newcommand{\C}{\bb{C}}
\newcommand{\Z}{\bb{Z}}
\newcommand{\bbI}{\bb{I}} 

\newcommand{\QFT}{Quantum Field Theory}
\newcommand{\be}{\begin{equation}}
\newcommand{\ee}{\end{equation}}
\newcommand{\bea}{\begin{eqnarray}}
\newcommand{\eea}{\end{eqnarray}}
\newcommand{\bean}{\begin{eqnarray*}}
\newcommand{\eean}{\end{eqnarray*}}
\newcommand{\bdm}{\begin{displaymath}}
\newcommand{\edm}{\end{displaymath}}
\newcommand{\noi}{\noindent}
\newcommand{\nn}{\nonumber}
\newcommand{\ul}{\underline}
\newcommand{\ulin}[1]{\ul{#1}\index{#1}}
\newcommand{\emphin}[1]{\emph{#1}\index{#1}}
\renewcommand{\bold}[1]{{\bf {#1}}}
\newcommand{\boldin}[1]{{\bf {#1}}\index{#1}}
\newcommand{\I}{\;\,}
\newcommand{\conj}[1]{\overline{#1}}
\newcommand{\thdots}{.\!\!\:^{\displaystyle \cdot}\!\!\:.}
\newcommand{\psic}{\overline{\psi}}
\newcommand{\epsuvst}{\epsilon_{\mu\nu\sigma\tau}}
\newcommand{\Lie}{\pounds}
\newcommand{\mcH}{\mathcal{H}}
\newcommand{\mcM}{\mathcal{M}}
\newcommand{\mcD}{\mathcal{D}}
\newcommand{\subtext}[1]{\mbox{\scriptsize {#1}}}
\newcommand{\smfrac}[2]{\frac{\scriptstyle {#1}}{\scriptstyle {#2}}}
\newcommand{\scrfrac}[2]{ {\displaystyle \frac{#1}{#2} } }
\newcommand{\bgfrac}[2]{\displaystyle \frac{\displaystyle {#1}}{\displaystyle {#2}}}
%approximately less than and approx greater than
\newcommand{\lsim}{\stackrel{\scriptstyle <}{\scriptstyle \sim}}
\newcommand{\gsim}{\stackrel{\scriptstyle >}{\scriptstyle \sim}}
% boxed equations
\newcommand{\bebox}[1]{\be
	\fbox{ ${\displaystyle {#1}}$}
	\ee}
%% derivatives
\newcommand{\dd}[1]{\frac{d}{d{#1}}}
\newcommand{\pdoL}[1]{\frac{\partial\mathcal{L}}{\partial{#1}}}
\newcommand{\pd}[2]{\frac{\partial{#1}}{\partial{#2}}}
\newcommand{\fltpd}[2]{\partial{#1}/\partial{#2}}
\newcommand{\bgpd}[2]{\bgfrac{\partial{#1}}{\partial{#2}}}
\newcommand{\pdd}[2]{\frac{\partial^2{#1}}{\partial{#2}^2}}
\newcommand{\pdel}[2]{\frac{\delta{#1}}{\delta{#2}}}
\newcommand{\half}{\frac{1}{2}}
%% integrals
\newcommand{\dx}[2]{d^{#1}\!{#2}\,}
%% special expressions
\newcommand{\dtp}{\frac{d^3\vec{p}}{(2\pi)^32E_p}}
\newcommand{\dtpp}{\frac{d^3\vec{p}'}{(2\pi)^32E_{p'}}}
\newcommand{\dtpdtpp}{\frac{d^3\vec{p}'d^3\vec{p}}{(2\pi)^32E_{p'}(2\pi)^32E_p}}
\newcommand{\phip}{\phi^{(+)}(x)}
\newcommand{\phim}{\phi^{(-)}(x)}
\newcommand{\ofx}{(\vec{x})}
\newcommand{\ofxpt}{(\vec{x}\,',t)}
\newcommand{\feyn}{\Delta_F(x-y)}
\newcommand{\FTfeyn}{\overline{\Delta}_F(p)}
\newcommand{\Lagd}{\mathcal{L}}
\newcommand{\nota}{\not{\!a}}
\newcommand{\Tr}{\mathrm{Tr}}
\newcommand{\tr}{\mathrm{tr}}
\newcommand{\transpose}[1]{{#1}^{\mathsf{T}}}
%% add dirac notation
\newcommand{\ket}[1]{\left|{#1}\right>}
\newcommand{\bra}[1]{\left<{#1}\right|}
\newcommand{\expt}[1]{\left<{#1}\right>}
\newcommand{\avg}{\expt}
\newcommand{\braket}[3]{\left<{#1}\left|{#2}\right|{#3}\right>}
%% matrices
\newcommand{\matrixtwo}[4]{\left(\begin{array}{cccc}
	{#1} & {#2} \\ {#3} & {#4} \end{array}\right) }
%% Black Holes specific
\newcommand{\Schr}{1-\frac{2M}{r}}
\newcommand{\SchR}{1-\frac{2M}{R}}
\newcommand{\gamsym}[3]{\left\{ { {#1} \atop {#2}\;{#3} } \right\}}
\newcommand{\mcN}{\mathcal{N}}
\newcommand{\lrpd}[1]{\stackrel{\leftrightarrow}{\partial}^{#1}}
\newcommand{\scri}{\Im}
\newcommand{\vac}{\ket{\mbox{vac}}}
%% Cosmology specific
\newcommand{\Dim}[1]{{^{^{({#1})}}}\!}
\newcommand{\rhobar}{\overline{\rho}}
\newcommand{\gbar}{\overline{\gbar}}
%% From Applications of Diff. Geom.
\newcommand{\Hstar}[1]{{^{\displaystyle \ast}}\!{#1}}
\newcommand{\Cinf}{\mathcal{C}^{\infty}}
%% CFT specific
\newcommand{\zbar}{\overline{z}}
\newcommand{\zetab}{\overline{\zeta}}
\newcommand{\PB}[2]{\left\{{#1},{#2}\right\}_{\mbox{\tiny PB}}}
\newcommand{\gtlt}{{\scriptscriptstyle >} \atop {\scriptscriptstyle <}}
\newcommand{\ltgt}{{\scriptscriptstyle <} \atop {\scriptscriptstyle >}}
\newcommand{\psiT}{\transpose{\psi}}

\thispagestyle{empty}
\begin{center}
\Huge{Black Holes}

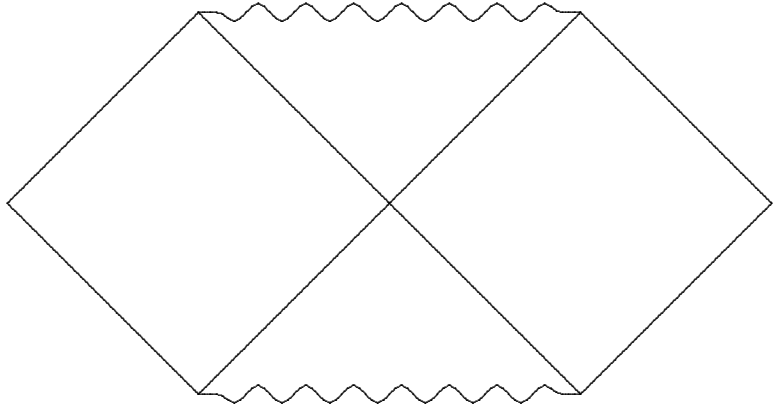
\vspace{1in}
\begin{center}\input{cover-BH.pictex}\end{center}
\vspace{1in}

\Large{Lecture notes \\
 by \\
\vspace{0.25in}
Dr. P.K. Townsend \\
\vspace{0.25in} 
DAMTP, University of Cambridge, \\ Silver St., Cambridge, U.K. }

\end{center}

\clearpage

\section*{Acknowledgements}

These notes were written to accompany a course taught in
Part III of the Cambridge University Mathematical Tripos. 
There are occasional references to questions on four 'example sheets', which can be found in the Appendix.
The writing of these course notes has greatly benefitted from 
discussions with Gary Gibbons and Stephen Hawking. The organisation of the 
course was based on unpublished notes of Gary Gibbons and owes much to the
1972 Les Houches and 1986 Carg{\' e}se lecture notes of Brandon Carter, and to
the 1972 lecture notes of Stephen Hawking. Finally, I am very grateful to 
Tim Perkins for typing the notes in \LaTeX, producing the diagrams, and putting
it all together.

\clearpage

\tableofcontents
%
% Black Hole notes by P. Townsend
% typed by Tim Perkins

\chapter{Gravitational Collapse}

\section{The Chandrasekhar Limit}

A Star is a self-gravitating ball of hydrogen atoms supported by thermal
pressure $P \sim nkT$ where $n$ is the number density of atoms. In equilibrium,
\be
E=E_{\subtext{grav}}+E_{\subtext{kin}}
\ee
is a minimum.  For a star of mass $M$ and radius $R$
\bea
E_{\subtext{grav}} & \sim & -\frac{GM^2}{R} \\
E_{\subtext{kin}} & \sim & nR^3\left<E\right>
\eea
where $\left<E\right>$ is average kinetic energy of atoms.  Eventually, fusion 
at the core must stop, after which the star cools and contracts.  Consider the
possible final state of a star at $T=0$. The pressure $P$ does not go to zero as
$T\to 0$ because of \emph{degeneracy pressure}\index{degenerate pressure}. 
Since $m_e\ll m_p$ the electrons become degenerate first, at a number density 
of one electron in a cube of side $\sim$ Compton wavelength.
\be
n_e^{-1/3} \sim \frac{\hbar}{\left<p_e\right>}, \quad 
\left<p\right>=\mbox{average electron momentum}
\ee

\subsection*{Can electron degeneracy pressure support a star from collapse at 
$T=0$?}

Assume that electrons are \emph{non-relativistic}.  Then
\be
\left<E\right> \sim \frac{\left<p_e\right>^2}{m_e}.
\ee

So, since $n=n_e$,
\be
E_{\subtext{kin}}\sim \frac{\hbar^2 R^2 r_e^{2/3}}{m_e}.
\ee
Since $m_e \ll m_p$, $M\approx n_eR^3m_e$, so 
\fbox{$n_e\sim\bgfrac{M}{m_pR^3}$} and 
\be
E_{\subtext{kin}} \sim \underbrace{ \frac{\hbar^2}{m_e}
\left( \frac{M}{m_p}\right)^{5/3} }_{\mbox{constant for}\atop\mbox{fixed $M$}}
\frac{1}{R^2}.
\ee
Thus 
\be
E\sim -\frac{\alpha}{R}-\frac{\beta}{R^2},\quad \alpha,
\beta\mbox{independent of $R$}.
\ee
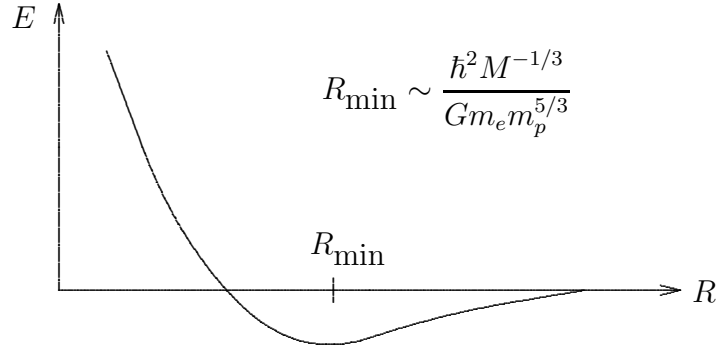
\begin{center}\input{fig1-1.pictex}\end{center}
The collapse of the star is therefore prevented. It becomes a 
\emph{White Dwarf}\index{white dwarf} or a cold, dead star supported by electron
degeneracy pressure. \\

At equilibrium
\be
n_e \sim \frac{M}{m_p R^3_{\subtext{min}}} \
\left( \frac{m_e G}{\hbar^2}\left(Mm_p^2\right)^{2/3}\right)^3.
\ee
But the validity of non-relativistic approximation requires that
$\left<p_e\right> \ll m_e c$, i.e.
%end page 2
\bea
\frac{\left<p_e\right>}{m_e} = \frac{\hbar n_e^{1/3}}{m_e}\ll c \\
\mbox{or} \quad n_e \ll \left( \frac{m_e c}{\hbar} \right)^2.
\eea

For a White Dwarf this implies
\bea
\frac{m_e G}{\hbar^2}\left(M m_p^2\right)^{2/3} \ll \frac{m_e c}{\hbar} \\
\mbox{or} \quad M \ll \frac{1}{m_p^2}\left(\frac{\hbar c}{G}\right)^{3/2}.
\eea

For sufficiently large $M$ the electrons would have to be relativistic, in which
case we must use
\bea
\lefteqn{\avg{E}  =  \avg{p_e}c=\hbar c n_e^{1/3} }\\
\Rightarrow \quad E_{\subtext{kin}} & \sim & 
n_eR^3\avg{E}\sim \hbar cR^3n_e^{4/3} \\
 & \sim & \hbar c R^3\left(\frac{M}{m_pR^3}\right)^{4/3}\sim \hbar c 
\left(\frac{M}{m_p}\right)^{4/3}\frac{1}{R}
\eea
So now,
\be
E \sim -\frac{\alpha}{R}+\frac{\gamma}{R}.
\ee
Equilibrium is possible only for
\be
\gamma=\alpha \quad \Rightarrow \quad M \sim \frac{1}{m_p^2}
\left(\frac{\hbar c}{G}\right)^{3/2}.
\ee

For smaller $M$, $R$ must increase until electrons become non-relativistic, in
which case the star is supported by electron degeneracy pressure, as we just
saw.  For larger $M$, $R$ must continue to decrease, so electron degeneracy
pressure cannot support the star. There is therefore a critical mass
$M_C$
\be
M_C \sim \frac{1}{m_p^2}\left(\frac{\hbar c}{G}\right)^{3/2} 
\quad \Rightarrow \quad R_C \sim \frac{1}{m_em_p}\left(
\frac{\hbar^3}{Gc}\right)^{1/2}
\ee
above which a star cannot end as a White Dwarf.  This is the 
\emph{Chandrasekhar limit}\index{Chandrasekhar limit}.  Detailed calculation
gives $M_C\simeq 1.4 M_{\odot}$.

\section{Neutron Stars}

The electron energies available in a White Dwarf are of the order of the 
Fermi energy.  Necessarily $E_F\stackrel{\scriptstyle <}{\scriptstyle \sim}
m_ec^2$ since the electrons are otherwise relativistic and cannot support the
star.  A White Dwarf is therefore stable against inverse $\beta$-decay
\be
e^-+p^+\to n+\nu_e
\ee
since the reaction needs energy of at least $(\Delta m_n)c^2$ where $\Delta m_n$ is the neutron-proton mass difference.  Clearly $\Delta m > m_e$ ($\beta$-decay would otherwise be impossible) and in fact $\Delta m\sim 3m_e$.  So we need energies of order of $3m_ec^2$ for inverse $\beta$-decay.  This is not available in White Dwarf stars but for $M>M_C$ the star must continue to contract until $E_F\sim (\Delta m_n)c^2$.  At this point inverse $\beta$-decay can occur.  The reaction cannot come to equilibrium with the reverse reaction
\be
n+\nu_e\to e^-+p^+
\ee 
because the neutrinos escape from the star, and $\beta$-decay, 
\be
n \to e^-+p^+\bar{\nu}_e
\ee
cannot occur because all electron energy levels below $E<(\Delta m_n)c^2$ are 
filled when $E>(\Delta m_n)c^2$.  Since inverse $\beta$-decay removes the
electron degeneracy pressure the star will undergo a catastrophic collapse to
nuclear matter density, at which point we must take \emph{neutron-degeneracy
pressure} into account.

\subsubsection{Can neutron-degeneracy pressure support the star against 
collapse?}

The ideal gas approximation would give same result as before but with 
$m_e\to m_p$.  The critical mass $M_C$ is \emph{independent} of $m_e$ and so is
unaffected, but the critical radius is now
\be
\left(\frac{m_e}{m_p}\right)R_C \sim \frac{1}{m_p^2}
\left(\frac{\hbar^3}{Gc}\right)^{1/2}\sim \frac{GM_C}{c^2}
\ee
which is the Schwarzschild radius, so the neglect of GR effects was
not justified.  Also, at nuclear matter densities the ideal gas approximation is
not justified.  A perfect fluid approximation is reasonable (since viscosity
can't help).  Assume that $P(\rho)$ ($\rho=$ density of fluid) satisfies
\bea
& \mbox{i)} & P\ge 0 \quad \mbox{(local stability).} \\
& \mbox{ii)} & P' < c^2 \quad \mbox{(causality).} 
\eea
Then the \emph{known behaviour} of $P(\rho)$ at low nuclear densities gives
\be
M_{\subtext{max}} \sim 3M_{\odot}.
\ee
More massive stars must continue to collapse either to an unknown new  
ultra-high density state of matter or to a black hole. The latter is more
likely. In any case, there must be {\it some} mass at which gravitational
collapse to a black hole is unavoidable because the density at the
Schwarzschild radius decreases as the total mass increases. In the limit of
very large mass the collapse is well-approximated by assuming the collapsing
material to be a pressure-free ball of fluid. We shall consider this
case shortly.
% end of p.5

\chapter{Schwarzschild Black Hole}

\section{Test particles: geodesics and affine parameterization}

Let $\mathcal{C}$ be a timelike curve with endpoints $A$ and $B$.  The action 
for a particle of mass $m$ moving on $\mathcal{C}$ is
\be
I=-mc^2\int^B_A d\tau
\ee 
where $\tau$ is proper time on $\mathcal{C}$.  Since
\be
d\tau = \sqrt{-ds^2}=\sqrt{-dx^{\mu}dx^{\nu}g_{\mu\nu}}=
\sqrt{-\dot{x}^{\mu}\dot{x}^{\nu}g_{\mu\nu}} d\lambda
\ee
where $\lambda$ is an arbitrary parameter on $\mathcal{C}$ and 
$\dot{x}^{\mu}=\frac{dx^{\mu}}{d\lambda}$, we have
\be
I\left[x\right] = -m\int^{\lambda_B}_{\lambda_A} 
d\lambda\sqrt{ -\dot{x}^{\mu}\dot{x}^{\nu}g_{\mu\nu}} \quad (c=1)
\ee
The particle worldline, $\mathcal{C}$, will be such that 
$\delta I/\delta x(\lambda)=0$.  By definition, this is a \emphin{geodesic}. 
For the purpose of finding geodesics, an equivalent action is
\be
I\left[x,e\right]= \half \int^{\lambda_B}_{\lambda_A}d\lambda 
\left[ e^{-1}(\lambda)\dot{x}^{\mu}\dot{x}^{\nu}g_{\mu\nu}-m^2e(\lambda)\right]
\ee
where $e(\lambda)$ (the `einbein'\index{einbein}) is a new independent function.
\paragraph{Proof of equivalence} (for $m\neq 0$)
\be
\frac{\delta I}{\delta e} =0 \quad \Rightarrow \quad e = \frac{1}{m}
\sqrt{ -\dot{x}^{\mu}\dot{x}^{\nu}g_{\mu\nu}} =
\frac{1}{m}\frac{d\tau}{d\lambda} 
\label{eq:einbein_one}
\ee
and (exercise)
\be
\frac{\delta I}{\delta x^{\mu}} = 0 \quad \Rightarrow 
\quad D_{(\lambda)}\dot{x}^{\mu}=(e^{-1}\dot{e})\dot{x}^{\mu}
\label{eq:einbein_two}
\ee
where
\be
D_{(\lambda)}V^{\mu}(\lambda)\equiv \frac{d}{d\lambda}V^{\mu}+
\dot{x}^{\nu}\gamsym{\mu}{\rho}{\nu}V^{\rho}
\ee
If (\ref{eq:einbein_one}) is substituted into (\ref{eq:einbein_two}) we get 
the EL equation $\delta I/\delta x^{\mu}=0$ of the original action $I[x]$
(exercise), hence equivalence. \\

The freedom in the choice of parameter $\lambda$ is equivalent to the freedom 
in the choice of function $e$.  Thus any curve $x^{\mu}(\lambda)$ for which
$t^{\mu}=\dot{x}^{\mu}(\lambda)$ satisfies
\be
D_{(\lambda)}t^{\mu}V^{\mu}=f(x)t^{\mu} \quad \mbox{(arbitrary $f$)}
\ee is a geodesic.  Note that for any vector field on $\mathcal{C}$, 
$V^{\mu}(x(\lambda))$,
\bea
t^{\nu}D_{\nu}V^{\mu} & \equiv & t^{\nu}\partial_{\nu}V^{\mu}+
t^{\nu}\gamsym{\mu}{\nu}{\rho}V^{\rho} \\
 & = & \frac{d}{d\lambda}V^{\mu} +\dot{x}^{\nu}
\gamsym{\mu}{\nu}{\rho}V^{\rho} \\
 & = & D_{(\lambda)}V^{\mu}
\eea
Since t is \emph{tangent} to the curve $\mathcal{C}$, a vector field $V$ on 
$\mathcal{C}$ for which
\be
D_{(\lambda)}=f(\lambda)V^{\mu} \quad \mbox{(arbitrary $f$)}
\ee 
is said to be \emph{parallely transported}\index{parallel transport} along the 
curve.  A geodesic is therefore \emph{a curve whose tangent is parallely
transported along it} (w.r.t. the affine connection). \\

A natural choice of parameterization is one for which
\be
D_{(\lambda)}t^{\mu}=0 \quad (t^{\mu}=\dot{x}^{\mu})
\ee
This is called \emph{affine parameterization}\index{affine parameter}.  For a 
timelike geodesic it corresponds to $e(\lambda)=$ constant, or
\be
\lambda \propto \tau +\mbox{constant}
\ee

The einbein form of the particle action has the advantage that we can take the 
$m\to 0$ limit to get the action for a massless particle.  In this case
\be
\frac{\delta I}{\delta e} = 0 \quad \Rightarrow \quad ds^2=0 \quad (m=0)
\ee
while (\ref{eq:einbein_two}) is unchanged.  We still have the freedom to 
choose $e(\lambda)$ and the choice $e=$ constant is again called affine
parameterization. \\

\noi\fbox{\parbox{\textwidth}{
\subsubsection{Summary}

Let $t^{\mu}=\bgfrac{dx^{\mu}(\lambda)}{d\lambda}$ and ${\displaystyle 
\sigma = \left\{ \begin{array}{cc} 1 & m\neq 0 \\ 0 & m=0 
\end{array}\right\} }$.

Then
\bebox{
\begin{array}{rcl} t\cdot Dt^{\mu} & \equiv & D_{(\lambda)}t^{\mu} = 0 \\
ds^2 & = & -\sigma d\lambda^2 \end{array}}
are the equations of affinely-parameterized timelike or null geodesics. \\
}}
% end of page 8
\section{Symmetries and Killing Vectors}

Consider the transformation
\be
x^{\mu}\to x^{\mu}-\alpha k^{\mu}(x), \quad (e\to e)
\ee
Then (Exercise)
\be
I\left[x,e\right] \to I\left[x,e\right]-\frac{\alpha}{2}
\int^{\lambda_B}_{\lambda_A}d\lambda\,
e^{-1}\dot{x}^{\mu}\dot{x}^{\nu}
\left(\Lie_{k}g\right)_{\mu\nu}+\mathcal{O}\left(\alpha^2\right)
\ee
where
\bea
\left(\Lie_{k}g\right)_{\mu\nu} & = & k^{\lambda}g_{\mu\nu,\lambda}+
k^{\lambda}_{\I,\mu}g_{\lambda\nu}+k^{\lambda}_{\I,\nu}g_{\lambda\mu} \\
 & = & 2D_{(\mu}k_{\nu)} \quad \mbox{(Exercise)}
\eea
Thus the action is invariant to first order if 
\be
\Lie_k g=0
\ee
A vector field $k^{\mu}(x)$ with this property is a 
\emph{Killing vector}\index{Killing!vector} field.  $k$ is associated with a
symmetry of the particle action and hence with a conserved charge. This charge
is (Exercise)
\be
Q=k^{\mu}p_{\mu}
\ee
where $p_{\mu}$ is the particle's 4-momentum.
\bea
p_{\mu} & = & \pdoL{\dot{x}^{\mu}}=e^{-1}\dot{x}^{\nu}g_{\mu\nu} \\
 & = & m\frac{dx^{\nu}}{d\tau}g_{\mu\nu} \quad \mbox{when } m\neq 0
\eea
\paragraph{Exercise} Check that the Euler-Lagrange equations imply
\bdm
\frac{dQ}{d\lambda}=0
\edm

Quantize, $p_{\mu}\to -i\partial/\partial x^{\mu}\equiv -i\partial_{\mu}$.  
Then
\be
Q \to -ik^{\mu}\partial_{\mu}
\ee
Thus the components of $k$ can be viewed as the components of a 
\emph{differential operator} in the basis $\left\{\partial_{\mu}\right\}$.
\be
k\equiv k^{\mu}\partial_{\mu}
\ee
It is convenient to identify this operator with the vector field.  Similarly 
for all other vector fields, e.g. the tangent vector to a curve
$x^{\mu}(\lambda)$ with affine parameter $\lambda$.
\be
t=t^{\mu}\partial_{\mu}=\frac{dx^{\mu}}{d\lambda}\partial_{\mu}=
\frac{d}{d\lambda}
\ee
For any vector field, $k$, local coordinates can be found such that 
\be
k=\partial/\partial\xi
\ee 
where $\xi$ is one of the coordinates. In such a coordinate system
\be
\Lie_kg_{\mu\nu}=\pd{}{\xi}g_{\mu\nu}
\ee
So $k$ is Killing if $g_{\mu\nu}$ is independent of $\xi$.

e.g. for Schwarzschild $\partial_t g_{\mu\nu}=0$, so $\partial/\partial t$ 
is a Killing vector field.  The conserved quantity is
\be
mk^{\mu}\frac{dx^{\nu}}{d\tau}g_{\mu\nu} = mg_{00}\frac{dt}{d\tau}
=-m\varepsilon \quad (\varepsilon=\mbox{ energy/unit mass})
\ee

% p11 of notes
\section{Spherically-Symmetric Pressure Free Collapse}

While it is impossible to say with complete confidence that a real star of 
mass $M\gg 3M_{\odot}$ will collapse to a BH, it is easy to invent idealized,
but physically possible, stars that definitely do collapse to black holes. One
such `star' is a spherically-symmetric ball of `dust' (i.e. zero pressure
fluid).  \emphin{Birkhoff's theorem} implies that the metric outside the star is
the \emphin{Schwarzschild metric}.  Choose units for which 
\be
G=1, \quad c=1.
\ee
Then
\be
ds^2= -\left(\Schr\right)dt^2+\left(\Schr\right)^{-1}dr^2+r^2d\Omega^2 
\ee
where 
\be
d\Omega^2=d\theta^2+\sin^2\theta d\varphi^2 \quad \mbox{(metric on a unit 
2-sphere)} 
\ee
This is valid outside the star but also, by continuity of the metric, at the 
surface.  If $r=R(t)$ on the surface we have 
\be
ds^2=-\left[ \left(\SchR\right)-\left(\SchR\right)^{-1}
\dot{R}^2\right]dt^2+R^2d\Omega^2, \quad \left(\dot{R}=\frac{d}{dt}R\right) 
\ee
On the surface zero pressure and spherical symmetry implies that a point on 
the surface follows a \emph{radial timelike geodesic}, so $d\Omega^2=0$ and
$ds^2=-d\tau^2$, so
\be
1=\left[\left(\SchR\right)-\left(\SchR\right)^{-1}\dot{R}^2\right]
\left(\frac{dt}{d\tau}\right)^2 
\label{eq:collapse_star}
\ee
But also, since $\partial/\partial t$ is a Killing vector we have 
\emph{conservation of energy}: 
\be
\varepsilon=-g_{00}\frac{dt}{d\tau}=\left(\SchR\right)\frac{dt}{d\tau} 
\quad \mbox{(energy/unit mass)}
\ee
$\varepsilon$ is \emph{constant on the geodesics}.  Using this in 
(\ref{eq:collapse_star}) gives
\be
1=\left[\left(\SchR\right)-\left(\SchR\right)^{-1}\dot{R}^2\right]
\left(\SchR\right)^{-2}\varepsilon^2
\ee
or
\bebox{
\dot{R}^2=\frac{1}{\varepsilon^2}\left(\SchR\right)^2\left(\frac{2M}{R}-1+
\varepsilon^2\right)
\label{eq:collapse_dagger}}
($\varepsilon < 1$ for gravitationally bound particles).
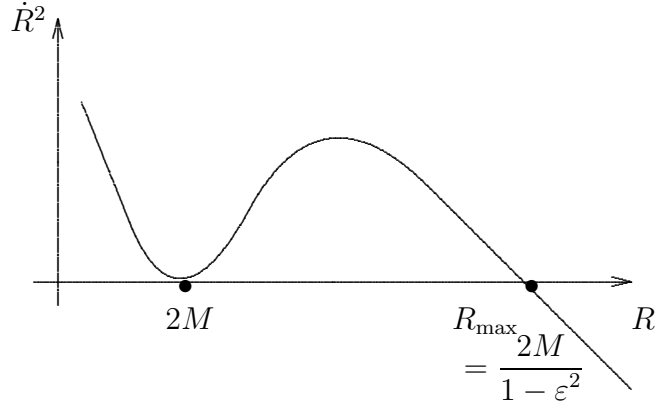
\begin{center}\input{p12-1.pictex}\end{center}
$\dot{R}=0$ at $R=R_{\subtext{max}}$ so we consider collapse to begin with 
zero velocity at this radius.  $R$ then decreases and approaches $R=2M$
asymptotically as $t\to\infty$.  So an observer `sees' the star contract at most
to $R=2M$ but no further. \\

However from the point of view of an observer on the surface of the star, the 
relevant time variable is proper time\index{proper time} along a radial
geodesic, so use
\be
\frac{d}{dt}=\left(\frac{dt}{d\tau}\right)^{-1}\frac{d}{d\tau}=
\frac{1}{\varepsilon}\left(\SchR\right)\frac{d}{d\tau}
\ee
to rewrite (\ref{eq:collapse_dagger}) as
\bebox{
\left(\frac{dR}{d\tau}\right)^2=\left(\frac{2M}{R}-1+\varepsilon^2\right)=
(1-\varepsilon^2)\left(\frac{R_{\subtext{max}}}{R}-1\right)}
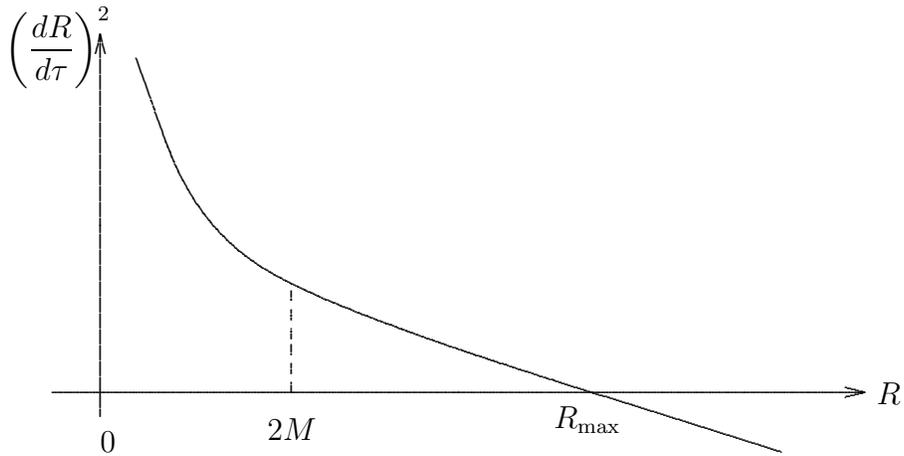
\begin{center}\input{p13-1.pictex}\end{center}
Surface of the star falls from $R=R_{\subtext{max}}$ through $R=2M$ in 
\emph{finite proper time}.  In fact, it falls to $R=0$ in proper time
\be
\tau=\frac{\pi M}{(1-\varepsilon)^{3/2}} \quad\mbox{(Exercise)}
\ee
Nothing special happens at $R=2M$ which suggests that we investigate the 
spacetime near $R=2M$ in coordinates adapted to infalling observers.  It is
convenient to choose \emph{massless} particles. \\

On radial null geodesics in Schwarzschild spacetime
\be
dt^2=\frac{1}{\left(\Schr\right)^2}dr^2\equiv\left(dr^*\right)^2
\ee
where
\be
r^*=r+2M\ln\left| \frac{r-2M}{2M}\right| 
\ee
is the \emphin{Regge-Wheeler radial coordinate}.  As $r$ ranges from $2M$ 
to $\infty$, $r^*$ ranges from $-\infty$ to $\infty$.  Thus 
\be
d(t\pm r^*)=0 \quad  \mbox{on radial null geodesics}
\ee
Define the ingoing radial null coordinate $v$ by 
\be
v=t+r^*,\quad  -\infty<v<\infty
\ee
and rewrite the Schwarzschild metric in \emph{ingoing Eddington-Finkelstein 
coordinates}\index{Eddington-Finkelstein coordinates!ingoing}
($v,r,\theta,\phi$).
\bea
ds^2 & = & \left(\Schr\right)\left(-dt^2+d{r^*}^2\right)+r^2d\Omega^2 \\
 & = & -\left(\Schr\right)dv^2+2dr\,dv+r^2d\Omega^2 
\eea
This metric is \emph{initially} defined for $r>2M$ since the relation 
$v=t+r^*(r)$ between $v$ and $r$ is only defined for $r>2M$, but it can now be
\emph{analytically continued} to all $r>0$.  Because of the $dr\,dv$ cross-term
the metric in EF coordinates is \emph{non-singular at $r=2M$}, so the
singularity in Schwarzschild coordinates was really a coordinate singularity. 
There is nothing at $r=2M$ to prevent the star collapsing through $r=2M$.  This
is illustrated by a \emphin{Finkelstein diagram}, which is a plot of $t^*=v-r$
against $r$:
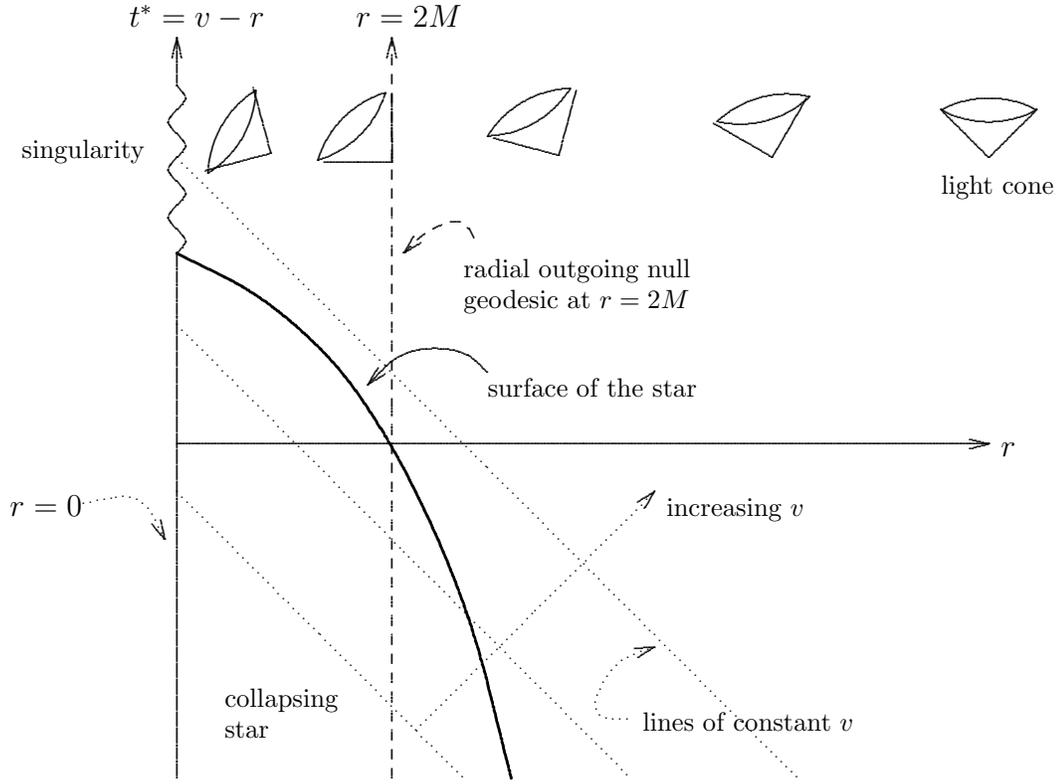
\begin{center}\input{p15-1.pictex}\end{center}
The light cones distort as $r\to 2M$ from $r>2M$, so that no future-directed 
timelike or null worldline can reach $r>2M$ from $r\le 2M$.

\paragraph{Proof}  When $r\le 2M$,
\bea
2dr\,dv & = & -\left[ -ds^2 + \left(\frac{2M}{r}-1\right)dv^2+ r^2d\Omega^2 
\right] \\
 & \le & 0 \quad\mbox{when $ds^2\le 0$}
\eea
for all timelike or null worldlines $dr\,dv\le 0$.  $dv>0$ for future-directed 
worldlines, so $dr\le 0$ with equality when $r=2M$, $d\Omega=0$ (i.e. ingoing
radial null geodesics at $r=2M$).

\subsection{Black Holes and White Holes}

No signal from the star's surface can escape to infinity once the surface has 
passed through $r=2M$.  The star has collapsed to a \emphin{black hole}.  For
the external observer, the surface never actually reaches $r=2M$, but as $r\to
2M$ the redshift of light leaving the surface increases \emph{exponentially}
fast and the star effectively disappears from view within a time $\sim MG/c^3$. 
The late time appearance is dominated by photons escaping from the unstable
photon orbit at $r=3M$. \\

The hypersurface $r=2M$ acts like a one-way membrane.  This may seem 
paradoxical in view of the time-reversibility of Einstein's equations.  Define
the \emph{outgoing} radial null coordinate $u$ by
\be
u=t-r^*,\quad -\infty<u<\infty 
\ee
and rewrite the Schwarzschild metric in \emph{outgoing Eddington-Finkelstein 
coordinates}\index{Eddington-Finkelstein coordinates!outgoing}
($u,r,\theta,\phi$). 
\be
ds^2=-\left(\Schr\right)du^2-2dr\,du+r^2d\Omega^2
\ee
This metric is initially defined only for $r>2M$ but it can be analytically 
continued to all $r>0$.  However the $r<2M$ region in outgoing EF coordinates is
\ul{not} the same as the $r<2M$ region in ingoing EF coordinates.  To see this,
note that for $r\le 2M$
\bea
2dr\,du & = & -ds^2+\left(\frac{2M}{r}-1\right)du^2+r^2d\Omega^2 \\ 
 & \ge & 0 \quad \mbox{when } ds^2\le 0
\eea
i.e. $dr\,du\ge 0$ on timelike or null worldlines.  But $du>0$ for 
future-directed worldlines so $dr\ge 0$, with equality when $r=2M$, $d\Omega=0$,
and $ds^2=0$.  In this case, a star with a surface at $r<2M$ must \emph{expand}
and explode through $r=2M$, as illustrated in the following Finkelstein
diagram\index{Finkelstein diagram}.
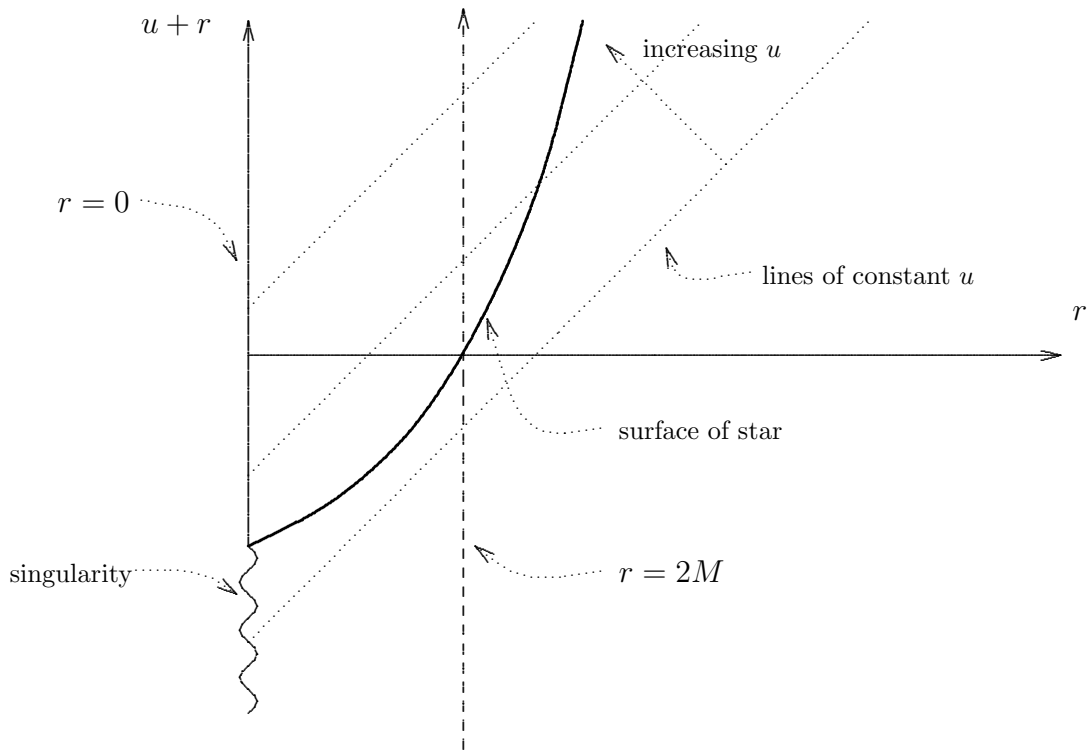
\begin{center}\input{p17-1.pictex}\end{center}
This is a \emphin{white hole}, the time reverse of a black hole.  Both black 
and white holes are allowed by G.R. because of the time reversibility of
Einstein's equations, but white holes require very special initial conditions
near the singularity, whereas black holes do not, so only black holes can occur
in practice (cf. irreversibility in thermodynamics).

\subsection{Kruskal-Szekeres Coordinates}\index{Kruskal-Szekeres coordinates}

The exterior region $r>2M$ is covered by both ingoing \emph{and} outgoing 
Eddington-Finkelstein coordinates, and we may write the Schwarzschild metric in
terms of $(u,v,\theta,\phi)$
\be
ds^2=-\left(\Schr\right)du\,dv +r^2d\Omega^2 
\ee
We now introduce the new coordinates $(U,V)$ defined (for $r>2M$) by
\be
U=-e^{-u/4M},\quad V=e^{v/4M}
\ee
in terms of which the metric is now 
\bebox{
ds^2=\frac{-32M^3}{r}e^{-r/2M}dU\,dV+r^2d\Omega^2
}
where $r(U,V)$ is given implicitly by $UV=-e^{r^*/2M}$ or
\bebox{
UV=-\left( \frac{r-2M}{2M}\right) e^{r/2M}
}
We now have the Schwarzschild metric in KS coordinates $(U,V,\theta,\phi)$.  
Initially the metric is defined for $U<0$ and $V>0$ but it can be extended by
analytic continuation to $U>0$ and $V<0$.  Note that $r=2M$  corresponds to
$UV=0$, i.e. \ul{either} $U=0$ \ul{or} $V=0$.  The singularity at $r=0$
corresponds to $UV=1$. \\

It is convenient to plot lines of constant $U$ and $V$ (outgoing or ingoing
radial null geodesics) at $45^{^0}$, so the spacetime diagram now looks like 
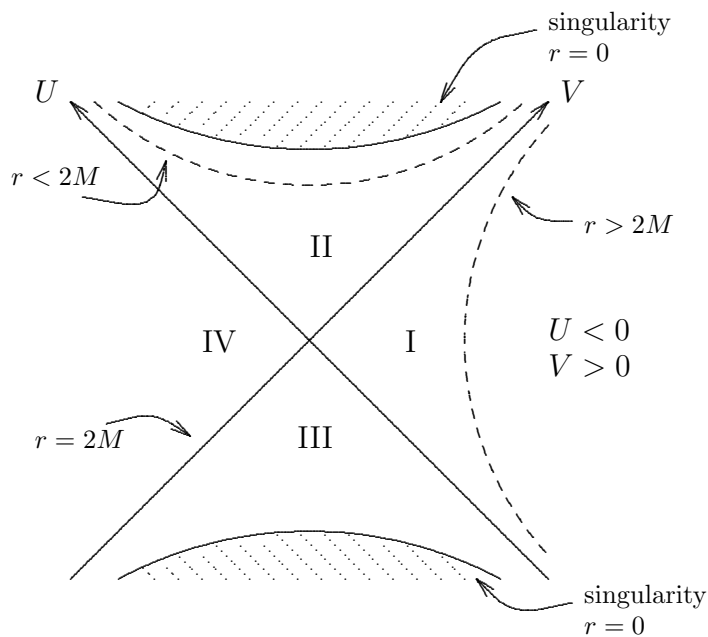
\begin{center}\input{p19-1.pictex}\end{center}
There are four regions of Kruskal spacetime, depending on the signs of $U$ and 
$V$.  Regions I and II are also covered by the ingoing Eddington-Finkelstein
coordinates.  These are the only regions relevant to gravitational collapse
because the other regions are then replaced by the star's interior, e.g. for
collapse of homogeneous ball of pressure-free fluid: 
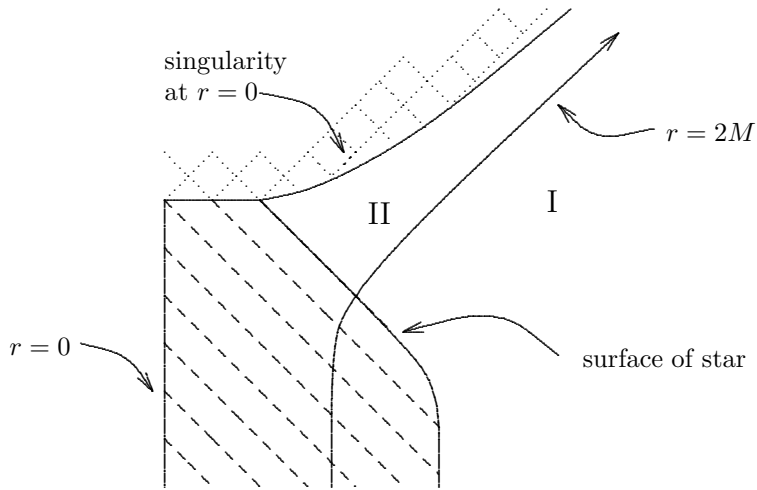
\begin{center}\input{p19-2.pictex}\end{center}
Similarly, regions I and III are those relevant to a white hole.

\subsubsection{Singularities and Geodesic Completeness}

A singularity of the metric is a point at which the determinant of either it or
its inverse vanishes. However, a singularity of the metric may be simply due
to a failure of the coordinate system. A simple two-dimensional example is the
origin in plane polar coordiates, and we have seen that the singularity of
the Schwarzschild metric at the Schwarzschild radius is of this type. Such
singularities are removable. If no coordinate system exists for which the
singularity is removable then it is irremovable, i.e. a genuine singularity of
the spacetime. Any singularity for which some scalar constructed from the
curvature tensor blows up as it is approached is irremovable. Such singularities
are called `curvature singularities'. The singularity at $r=0$ in the
Schwarzschild metric is an example. Not all irremovable singularities are
`curvature singularities', however. Consider the singularity at the tip of a
cone formed by rolling up a sheet of paper. All curvature invariants remain
finite as the singularity is approached; in fact, in this two-dimensional
example the curvature tensor is everywhere zero. If we could assign a curvature
to the singular point at the tip of the cone it would have to be infinite but,
strictly speaking, we cannot include this point as part of the manifold since
there is no coordinate chart that covers it. 

We might try to make a virtue of this necessity: by excising the regions
containing irremovable singularities we apparently no longer have to worry about
them. However, this just leaves us with the essentially equivalent problem of
what to do with curves that reach the boundary of the excised region. There is
no problem if this boundary is at infinity, i.e. at infinite affine parameter
along all curves that reach it from some specified point in the interior, but
otherwise the inability to continue all curves to all values of their affine
parameters may be taken as the defining feature of a `spacetime singularity'.
Note that the concept of affine parameter is not restricted to geodesics, e.g.
the affine parameter on a timelike curves is the proper time on the curve
regardless of whether the curve is a geodesic. This is just as well, since there
is no good physical reason why we should consider only geodesics. Nevertheless,
it is virtually always true that the existence of a singularity as just
defined can be detected by the incompleteness of some geodesic, i.e. there is
some geodesic that cannot be continued to all values of its affine parameter.
For this reason, and because it is simpler, we shall follow the common practice
of defining a spacetime singularity in terms of `geodesic incompleteness'.
Thus, {\it a spacetime is non-singular if and only if all geodesics can be
extended to all values of their affine parameters}, changing coordinates if
necessary. 

In the case of the Schwarzschild vacuum solution, a particle on an ingoing radial
geodesics will reach the coordinate singularity at $r=2M$ at finite affine
parameter but, as we have seen, this geodesic can be continued into region II
by an appropriate change of coordinates. Its continuation will then approach the
curvature singularity at $r=0$, coming arbitrarily close for finite affine
parameter. The excision of any region containing $r=0$ will therefore lead to a
incompleteness of the geodesic. The vacuum Schwarzschild solution is therefore
singular. The singularity theorems of Penrose and Hawking show that geodesic
incompleteness is a \emph{generic feature} of gravitational collapse, and not
just a special feature of spherically symmetric collapse.

\subsubsection{Maximal Analytic Extensions}

Whenever we encounter a singularity at finite affine parameter along some
geodesic (timelike, null, or spacelike) our first task is to identify it as
removable or irremovable. In the former case we can continue through it by a
change of coordinates.  By considering all geodesics we can construct in
this way the \emphin{maximal analytic extension} of a given spacetime in which
\emph{any geodesic that does not terminate on an irremovable singularity can be
extended to arbitrary values of its affine parameter}. The Kruskal manifold
is the maximal analytic extension of the Schwarzschild solution, so no more
regions can be found by analytic continuation.

\subsection{Eternal Black Holes}

A black hole formed by gravitational collapse is not time-symmetric because it 
will continue to exist into the indefinite future but did not always exist in
the past, and vice-versa for white holes.  However, one can imagine a
time-symmetric eternal black hole that has always existed (it could equally well
be called an eternal white hole, but isn't).  In this case there is no matter
covering up part of the Kruskal spacetime and all four regions are relevant.  
In region I
\be
\frac{U}{V}=e^{-t/2M}
\ee
so hypersurfaces of constant Schwarzschild time $t$ are straight lines through 
the origin in the Kruskal spacetime.
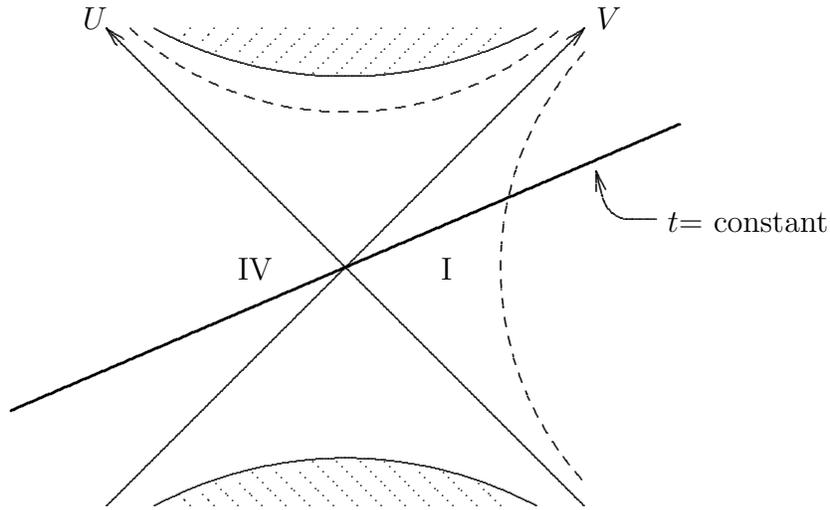
\begin{center}\input{p21-1.pictex}\end{center}
These hypersurfaces have a part in region I and a part in region IV.  Note that 
$(U,V)\to (-U,-V)$ is an isometry of the metric so that region IV is isometric
to region I. \\

To understand the geometry of these $t=$ constant hypersurfaces it is 
convenient to rewrite the Schwarzschild metric in \emphin{isotropic coordinates}
$(t,\rho,\theta,\phi)$, where $\rho$ is the new radial coordinate
\be
r=\left(1+\frac{M}{2\rho}\right)^2\rho
\ee
Then (\bold{Exercise})
\be
ds^2 = -\left( \frac{ 1-\frac{M}{2\rho} }{ 1+\frac{M}{2\rho}} 
\right)^2 dt^2 + \left(1+\frac{M}{2\rho}\right)^4 \underbrace{ \left[
d\rho^2+\rho^2d\Omega^2 \right] }_{\mbox{flat 3-space metric}} 
\ee
In isotropic coordinates, the $t=$ constant hypersurfaces are 
\emph{conformally flat}, but to each value of $r$ there corresponds \emph{two}
values of $\rho$
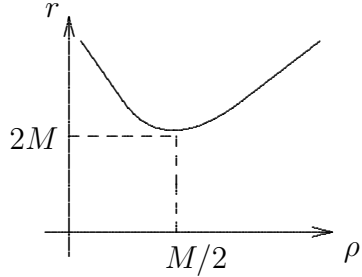
\begin{center}\input{p22-1.pictex}\end{center}
The two values of $\rho$ are exchanged by the isometry, $\rho\to M^2/4\rho$ 
which has $\rho=M/2$ as its fixed `point', actually a fixed 2-sphere of radius
$2M$.  This isometry corresponds to the $(U,V)\to (-U,-V)$ isometry of the
Kruskal spacetime.  The isotropic coordinates cover only regions I and IV since
$\rho$ is complex for $r<2M$.
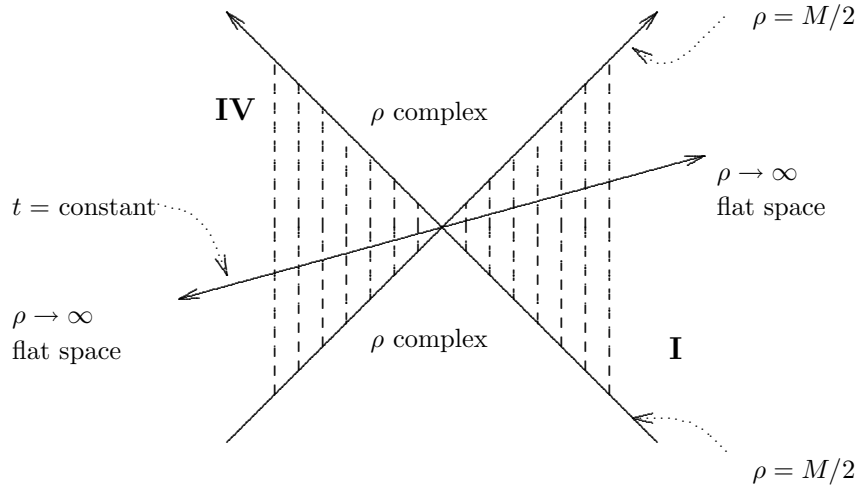
\begin{center}\input{p22-2.pictex}\end{center}
As $\rho\to M/2$ from either side the radius of a 2-sphere of constant 
$\rho$ on a $t=$ constant hypersurface decreases to minimum of $2M$ at
$\rho=M/2$, so $\rho=M/2$ is a \emph{minimal 2-sphere}.  It is the midpoint of
an \emphin{Einstein-Rosen bridge} connecting spatial sections of regions I and
IV.  
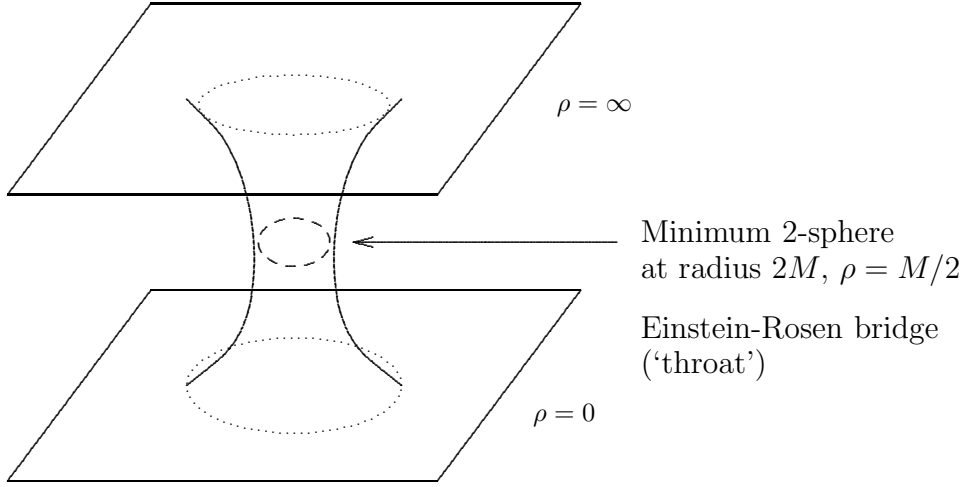
\begin{center}\input{p23-1.pictex}\end{center}

\subsection{Time translation in the Kruskal Manifold}

The time translation $t\to t+c$, which is an isometry of the Schwarzschild 
metric becomes
\be
U\to e^{-c/4M}U,\quad V\to e^{c/4M}V 
\ee
in Kruskal coordinates and extends to an isometry of the entire Kruskal 
manifold.  The infinitesimal version
\be
\delta U = -\frac{c}{4M}U, \quad \delta V=\frac{c}{4M}V
\ee
is generated by the Killing vector field
\be
k=\frac{1}{4M}\left(V\pd{}{V}-U\pd{}{U}\right)
\ee
which equals $\fltpd{}{t}$ in region I.  It has the following properties
\newcounter{kproperties}
\begin{list}{(\roman{kproperties})}
{\usecounter{kproperties}}
\item $k^2=-\left(1-\frac{2M}{r}\right)\quad \Rightarrow \left\{\mbox{
\begin{tabular}{lcl}
timelike & in & I \& IV \\
spacelike & in & II \& III \\
null & on & $r=2M$, i.e. $\{U=0\}\cup\{V=0\}$
\end{tabular}}\right.$

\item $\{U=0\}$ and $\{V=0\}$ are \emphin{fixed sets} on $k$. \\

On $\left\{\mbox{\begin{tabular}{cc}
$\left\{U=0\right\}$ & $k=\fltpd{}{v}$ \\ $\left\{V=0\right\}$ & 
$k=\fltpd{}{u}$ \end{tabular}}\right\}$ where $v,u$ are EF null coordinates.

$\thdots$ $v$ is the natural group parameter on $\left\{U=0\right\}$.  
Orbits of $k$ correspond to $-\infty<v<\infty$, (where $v$ is well-defined).

\item Each point on the \emphin{Boyer-Kruskal axis}, $\{U=V=0\}$ (a 2-sphere) 
is a \emphin{fixed point} of $k$.

\end{list}
The orbits of $k$ are shown below
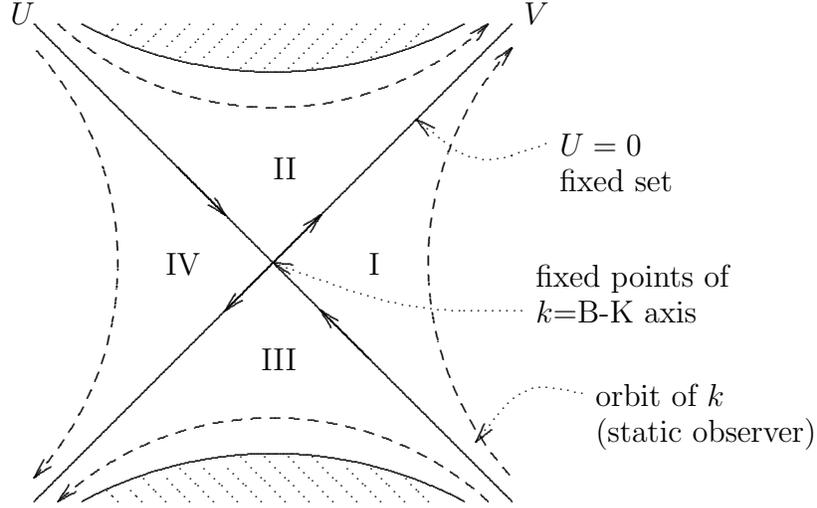
\begin{center}\input{p24-1.pictex}\end{center}

\subsection{Null Hypersurfaces}

Let $S(x)$ be a smooth function of the spacetime coordinates $x^{\mu}$ and 
consider a family of hypersurfaces $S= $ constant.  The vector fields normal to
the hypersurface are
\be
l=\tilde{f}(x)\left(g^{\mu\nu}\partial_{\nu}S\right)\pd{}{x^{\mu}}
\ee
where $\tilde{f}$ is an arbitrary non-zero function.  If $l^2=0$ for a 
particular hypersurface, $\mcN$, in the family, then $\mcN$ is said to be a
\emphin{null hypersurface}.

\paragraph{Example} Schwarzschild in ingoing Eddington-Finkelstein coordinates 
$(r,v,\theta,\phi)$ and the surface $S=r-2M$.
\bea
l & = & \tilde{f}(r)\left[\left(1-\frac{2M}{r}\right)\pd{S}{r}\pd{}{r}+
\pd{S}{r}\pd{}{v}+\pd{S}{v}\pd{}{r}\right] \\
 & = & \tilde{f}(r)\left[\left(1-\frac{2M}{r}\right)\pd{}{r}+\pd{}{v}\right]
\eea
while
\bea
l^2 & = & g^{\mu\nu}\partial_{\mu}S\partial_{\nu}S\tilde{f}^2 \\
 & = & g^{rr}\tilde{f}^2 = \left(1-\frac{2M}{r}\right)\tilde{f}^2
\eea
so \emph{$r=2M$ is a null hypersurface}, and 
\be
\left.l\right|_{r=2M} = \tilde{f}\pd{}{v}
\ee

\subsubsection{Properties of Null Hypersurfaces}

Let $\mathcal{N}$ be a null hypersurface with normal $l$.  A vector $t$, 
tangent to $\mathcal{N}$, is one for which $t\cdot l=0$.  But, since
$\mathcal{N}$ is null, $l\cdot l=0$, so \emph{$l$ is itself a tangent vector},
i.e. 
\be
l^{\mu}=\frac{dx^{\mu}}{d\lambda}
\ee
for some null curve $x^{\mu}(\lambda)$ in $\mcN$.

\paragraph{Proposition}  The curves $x^{\mu}(\lambda)$ are \emph{geodesics}.

\paragraph{Proof}  Let $\mathcal{N}$ be the member $S=0$ of the family of 
(not necessarily null) hypersurfaces $S=$ constant.  Then
$l^{\mu}=\tilde{f}g^{\mu\nu}\partial_{\nu}S$ and hence
\bea
l\cdot Dl^{\mu} & = & \left(l^{\rho}\partial_{\rho}\tilde{f}\right)
g^{\mu\nu}\partial_{\nu}S +\tilde{f}g^{\mu\nu}l^{\rho}D_{\rho}
\partial_{\nu}S \\
 & = & \left(l\cdot \partial\ln\tilde{f}\right)l^{\mu} +
\tilde{f}g^{\mu\nu}l^{\rho}D_{\nu}\partial_{\rho}S \quad 
\mbox{(by symmetry of
$\Gamma$)} \\
 & = & \left(\frac{d}{d\lambda}\ln\tilde{f}\right)l^{\mu}+
l^{\rho}\tilde{f}D^{\mu}\left(\tilde{f}^{-1}l_{\rho}\right) \\
 & = & \left(\frac{d}{d\lambda}\ln\tilde{f}\right)l^{\mu}+ 
l^{\rho}D^{\mu}l_{\rho}  -\left(\partial^{\mu}\ln\tilde{f}\right) l^2 \\
 & = & \left(\frac{d}{d\lambda}\ln\tilde{f}\right)l^{\mu}+
\half l^{2,\mu} -\left(\partial^{\mu}\ln\tilde{f}\right) l^2
\eea
Although $\left.l^2\right|_{\mcN}=0$ it doesn't follow that 
$\left.l^{2,\mu}\right|_{\mcN}=0$ unless the whole family of hypersurfaces $S=$
constant is null.  However since $l^2$ is constant on $\mcN$,
$t^{\mu}\partial_{\mu}l^2=0$ for any vector $t$ tangent to $\mcN$.  Thus
\be
\left.\partial_{\mu}l^2\right|_N \propto l_{\mu}
\ee
and therefore
\be
\left.l\cdot Dl^{\mu}\right|_N\propto l^{\mu}
\ee
i.e. $x^{\mu}(\lambda)$ is a geodesic (with tangent $l$).  The function 
$\tilde{f}$ can be chosen such that $l\cdot Dl=0$, i.e. so that $\lambda$ is an
affine parameter.

\paragraph{Definition}  The null geodesics $x^{\mu}(\lambda)$ with affine 
parameter $\lambda$, for which the tangent vectors $dx^{\mu}/d\lambda$ are
normal to a null hypersurface $\mathcal{N}$, are the \emph{generators of
$\mathcal{N}$}.

\paragraph{Example} $\mcN$ is $U=0$ hypersurface of Kruskal spacetime.  
Normal to $U=$ constant is
\bea
l & = & -\frac{\tilde{f}r}{32M^3}e^{r/2M}\pd{}{V} \\
\left.l\right|_N & = & -\frac{ \tilde{f}e}{16 M^2}\pd{}{V} \quad 
\mbox{since $r=2M$ on $\mathcal{N}$}
\eea
Note that $l^2\equiv 0$, so $l^2$ and $l^{2,\mu}$ both vanish on 
$\mathcal{N}$; this is because $U=$ constant is null for \emph{any} constant,
not just zero.  thus $l\cdot Dl=0$ if $\tilde{f}$ is \emph{constant}.  Choose
$\tilde{f}=-16M^2e^{-1}$.  Then 
\be
l=\pd{}{V}
\ee
is normal to $U=0$ and \emph{V is an affine parameter for the generator of 
this null hypersurface}.

\subsection{Killing Horizons}

\paragraph{Definition}  A null hypersurface $\mathcal{N}$ is a Killing 
horizon\index{Killing!horizon} of a Killing vector field $\xi$ if, on
$\mathcal{N}$, $\xi$ is normal to $\mcN$. \\

Let $l$ be normal to $\mathcal{N}$ such that $l\cdot Dl^{\mu}=0$ (affine 
parameterization).  Then, since, on $\mathcal{N}$, 
\be
\xi=fl
\ee
for some function $f$, it follows that
\bebox{
\xi\cdot D\xi^{\mu}=\kappa\xi^{\mu}, \quad \mbox{on }\mathcal{N}
}
where $\kappa = \xi \cdot \partial\ln\left|f\right|$ is called the 
\emphin{surface gravity}.

\subsubsection{Formula for surface gravity}

Since $\xi$ is normal to $\mcN$, \emphin{Frobenius' theorem} implies that
\bebox{
\left.\xi_{[\mu}D_{\nu}\xi_{\rho]}\right|_{\mathcal{N}}=0
\label{eq:normal_star}
}
where `$[\quad]$' indicates total anti-symmetry in the enclosed indices, 
$\mu,\nu,\rho$.  For a Killing vector field $\xi$, $D_{\mu}\xi_{\nu} =
D_{[\mu}\xi_{\nu]}$ (i.e. symmetric part of $D_{\mu}\xi_{\nu}$ vanishes). In
this case (\ref{eq:normal_star}) can be written as
\be
\left.\xi_{\rho}D_{\mu}\xi_{\nu}\right|_{\mathcal{N}} + 
\left.\left(
\xi_{\mu}D_{\nu}\xi_{\rho}-\xi_{\nu}D_{\mu}\xi_{\rho}\right)
\right|_{\mathcal{N}}
= 0
\ee
Multiply by $D^{\mu}\xi^{\nu}$ to get
\be
\left.\xi_{\rho}\left(D^{\mu}\xi^{\nu}\right)\left(D_{\mu}\xi_{\nu}\right)
\right|_{\mathcal{N}}  = 
-\left.2\left(D^{\mu}\xi^{\nu}\right)\xi_{\mu}\left(D_{\nu}\xi_{\rho}\right)
\right|_{\mathcal{N}}
\qquad \mbox{(since $D^{\mu}\xi^{\nu}=D^{[\mu}\xi^{\nu]}$)} 
\ee
or
\bea
\left.\xi_{\rho}\left(D^{\mu}\xi^{\nu}\right)\left(D_{\mu}\xi_{\nu}\right)
\right|_{\mathcal{N}}  & = & -\left.2\left(\xi\cdot
D\xi^{\nu}\right)D_{\nu}\xi_{\rho}\right|_{\mathcal{N}} \\
 & = & -\left. 2\kappa \xi\cdot D \xi_{\rho}\right|_{\mathcal{N}} 
\qquad \mbox{(for Killing horizon)} \\
 & = & -\left.2\kappa^2\xi_{\rho}\right|_{\mathcal{N}} 
\eea
Hence, except at points for which $\xi=0$,
\be
\fbox{ ${\displaystyle \kappa^2=-\left.\half\left(D^{\mu}\xi^{\nu}\right)
\left(D_{\mu}\xi_{\nu}\right)\right|_{\mathcal{N}} }$}
\label{eq:normal_dagger}
\ee
It will turn out that all points at which $\xi=0$ are limit points of orbits 
of $\xi$ for which $\xi\neq 0$, so continuity implies that this formula is valid
even when $\xi=0$ (Note that $\xi=0 \not\Rightarrow D_{\mu}\xi_{\nu}=0$). \\

\fbox{\parbox{6in}{
\paragraph{Killing Vector Lemma}

For a Killing vector field $\xi$
\be
\fbox{ ${ \displaystyle D_{\rho}D_{\mu}\xi^{\nu}=
R^{\nu}_{\I \mu\rho\sigma}\xi^{\sigma} }$}
\ee
where $R^{\nu}_{\I \mu\rho\sigma}$ is the Riemann tensor.

\bold{Proof: Exercise} (Question II.1)
}}

\paragraph{Proposition} $\kappa$ is constant on orbits of $\xi$. \\

\paragraph{Proof} Let $t$ be tangent to $\mathcal{N}$.  Then, since 
(\ref{eq:normal_dagger}) is valid everywhere on $\mathcal{N}$
\bea
t\cdot\partial\kappa^2 & = & -\left.\left(D^{\mu}\xi^{\nu}\right)
t^{\rho}D_{\rho}D_{\mu}\xi_{\nu}\right|_{\mathcal{N}} \\
 & = & -\left(D^{\mu}\xi^{\nu}\right)t^{\rho}
R_{\nu\mu\rho}^{\I\I\I\I\sigma}\xi_{\sigma} \qquad \mbox{(using Lemma)}
\eea
Now, $\xi$ is tangent to $\mcN$ (in addition to being normal to it).  
Choosing $t=\xi$ we have
\bea
\xi\cdot \partial\kappa^2 & = & -\left(D^{\mu}\xi^{\nu}\right) 
R_{\nu\mu\rho\sigma}\xi^{\rho}\xi^{\sigma}  \\
 & = & 0 \qquad \mbox{(since $R_{\nu\mu\rho\sigma}=-R_{\nu\mu\sigma\rho}$)}
\eea
so $\kappa$ is constant on orbits of $\xi$.

\subsubsection{Non-degenerate Killing horizons $(\kappa\neq 0)$}

Suppose $\kappa\neq 0$ on one orbit of $\xi$ in $\mathcal{N}$.  Then this 
orbit coincides with only \emph{part} of a null generator of $\mathcal{N}$. To
see this, choose coordinates on $\mathcal{N}$ such that 
\be
\xi=\pd{}{\alpha} \qquad  \mbox{(except at points where $\xi=0$)}
\ee
i.e. such that the group parameter $\alpha$ is one of the coordinates.  Then 
if $\alpha=\alpha(\lambda)$ on an orbit of $\xi$ with an affine parameter
$\lambda$
\be
\left.\xi\right|_{\subtext{orbit}} =\frac{d\lambda}{d\alpha}\frac{d}
{d\lambda}=fl \quad \left\{\begin{array}{rcl} f & = & \bgfrac{d\lambda}{d\alpha}
\\ \\ l & = & \bgfrac{d}{d\lambda} =
\bgfrac{dx^{\mu}(\lambda)}{d\lambda}\partial_{\mu} \end{array}\right.
\ee
Now 
\be
\pd{}{\alpha}\ln\left|f\right|=\kappa
\ee
where $\kappa$ is \emph{constant} for orbit on $\mcN$. For such orbits, 
$f=f_0e^{\kappa\alpha}$ for arbitrary constant $f_0$.  Because of freedom to
shift $\alpha$ by a constant we can choose $f_0 = \pm \kappa$ without loss of
generality, i.e. 
\be
\frac{d\lambda}{d\alpha}=\pm\kappa e^{\kappa\alpha} \quad \Rightarrow 
\quad \lambda = \pm e^{\kappa\alpha}+\mbox{constant}
\ee
Choose constant $=0$
\bebox{ \lambda=\pm e^{\kappa\alpha} }
As $\alpha$ ranges from $-\infty$ to $\infty$ we cover the $\lambda>0$ or 
the $\lambda<0$ portion of the generator of $\mathcal{N}$ (geodesic in $\mcN$
with normal $l$).  The bifurcation point\index{bifurcation!point} $\lambda=0$ is
a fixed point of $\xi$, which can be shown to be a 2-sphere, called the
bifurcation 2-sphere\index{bifurcation!2-sphere}, (BK-axis for Kruskal).
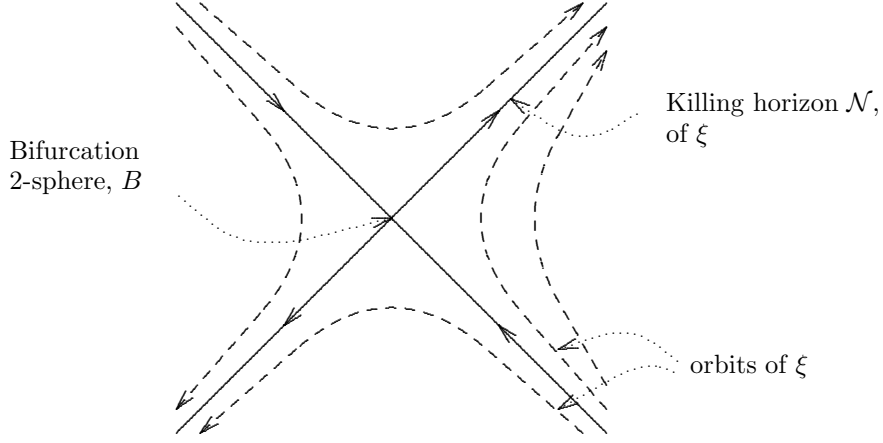
\begin{center}\input{p31-1.pictex}\end{center}
This is called a \emphin{bifurcate Killing 
horizon}\index{Killing!horizon!bifurcate}.

\paragraph{Proposition} If $\mathcal{N}$ is a bifurcate Killing 
horizon of $\xi$, with bifurcation 2-sphere, $B$, then $\kappa^2$ is 
constant on $\mathcal{N}$.

\paragraph{Proof}  $\kappa^2$ is constant on each orbit of $\xi$.  The value 
of this constant is the value of $\kappa^2$ at the limit point of the orbit on
$B$, so $\kappa^2$ is constant on $\mcN$ if it is constant on $B$.  But we saw
previously that
\bea
t\cdot\partial \kappa^2 & = & -\left.\left(D^{\mu}\xi^{\nu}\right)
t^{\rho}R_{\nu\mu\rho}^{\I\I\I\I\sigma}\xi_{\sigma}\right|_{\mathcal{N}}  \\
 & = & 0 \quad  \mbox{on $B$ since $\left.\xi_{\sigma}\right|_B=0$ }
\eea
Since $t$ can be any tangent to $B$, $\kappa^2$ is constant on $B$, and 
hence on $\mcN$.

\paragraph{Example} $\mathcal{N}$ is $\left\{U=0\right\}
\cup\left\{V=0\right\}$ of Kruskal spacetime, and $\xi=k$, the 
time-translation Killing vector field.

On $\mcN$,
\be
k=\left\{ \begin{array}{ccc} \bgfrac{1}{4M}V\pd{}{V} & \mbox{on} 
& \{U=0\} \\ \\
	-\bgfrac{1}{4M}U\pd{}{U} & \mbox{on} & \{V=0\} \end{array}\right\}=fl
\ee
where
\be
f=\left\{ \begin{array}{ccc} \bgfrac{1}{4M}V & \mbox{on} & \{U=0\} \\ \\
	-\bgfrac{1}{4M}U & \mbox{on} & \{V=0\} \end{array}\right\}, \quad  
l=\left\{ \begin{array}{ccc} \bgpd{}{V} & \mbox{on} & \{U=0\} \\ \\ \bgpd{}{U} &
\mbox{on} & \{V=0\} \end{array} \right\} 
\ee
Since $l$ is normal to $\mathcal{N}$, \emph{$\mathcal{N}$ is a Killing 
horizon of $k$}.  Since $l\cdot Dl=0$, the surface gravity is 
\bea
\kappa=k\cdot\partial\ln\left|f\right| & = & \left\{ \begin{array}{ccc} 
\bgfrac{1}{4M}V\pd{}{V}\ln\left|V\right| & \mbox{on} & U=0 \\ \\
-\bgfrac{1}{4M}U\pd{}{U}\ln\left|U\right| & \mbox{on} & V=0 \end{array} 
\right. \\
 & = & \left\{ \begin{array}{ccc}
\bgfrac{1}{4M} & \mbox{on} & \{U=0\} \\ \\ -\bgfrac{1}{4M} & 
\mbox{on} & \{V=0\} \end{array} \right.
\eea
So $\kappa^2=1/(4M)^2$ is indeed a constant on $\mathcal{N}$.  Note that 
orbits of $k$ lie either entirely in $\{U=0\}$ or in $\{V=0\}$ or are fixed
points on $B$, which allows a difference of sign in $\kappa$ on the two branches
of $\mcN$.

[N.B. Reinstating factors of $c$ and $G$, $\left|\kappa\right|=
\bgfrac{c^3}{4GM}$]

\subsubsection{Normalization of $\kappa$}

If $\mcN$ is a Killing horizon of $\xi$ with surface gravity $\kappa$, 
then it is also a Killing horizon of $c\xi$ with surface gravity $c^2\kappa$
[from formula (\ref{eq:normal_dagger}) for $\kappa$] for any constant $c$.  
Thus surface gravity is not a property of $\mcN$ \emph{alone}, it also 
depends on the normalization of $\xi$. 

There is no natural normalization of $\xi$ on $\mcN$ since $\xi^2=0$ there, 
but in an asymptotically flat spacetime there is a natural normalization at
spatial infinity, e.g. for the time-translation Killing vector field $k$ we
choose
\be
k^2\to -1 \quad \mbox{as} \quad r\to\infty
\ee
This fixes $k$, and hence $\kappa$, up to a sign, and the sign of $\kappa$ 
is fixed by requiring $k$ to be future-directed.

\subsubsection{Degenerate Killing Horizon ($\kappa=0$)}

In this case, the group parameter on the horizon is also an affine parameter, 
so there is no bifurcation 2-sphere.  More on this case later.

\subsection{Rindler spacetime}\index{Rindler!spacetime}

Return to Schwarzschild solution
\be
ds^2=-\left(\Schr\right)dt^2+\left(\Schr\right)^{-1}dr^2+r^2d\Omega^2 
\label{eq:schw_dagger}
\ee
and let
\be
r-2M=\frac{x^2}{8M}
\ee
Then
\bea
\Schr & = & \frac{(\kappa x)^2}{1+(\kappa x)^2} \qquad 
\left(\kappa=\frac{1}{4M}\right) \\
 & \approx & (\kappa x)^2 \qquad \mbox{near $x=0$} \\
dr^2 & = & (\kappa x)^2 dx^2
\eea
so for $r\approx 2M$ we have
\be
ds^2 \approx \underbrace{ -(\kappa x)^2dt^2+dx^2 }_{\begin{array}{c} 
\subtext{2-dim Rindler} \\ \subtext{spacetime} \end{array} } + \underbrace{
\frac{1}{4\kappa^2}d\Omega^2 }_{\begin{array}{c} \subtext{2-sphere of} \\
\subtext{radius $1/(2\kappa)$} \end{array} }
\ee
so we can expect to learn something about the spacetime near the Killing 
horizon at $r=2M$ by studying the 2-dimensional \emph{Rindler spacetime}
\be
ds^2 = -(\kappa x)^2dt^2+dx^2 \qquad (x >0) 
\ee
This metric is singular at $x=0$, but this is just a coordinate singularity.  
To see this, introduce the Kruskal-type coordinates
\be
U'=-xe^{-\kappa t},\quad V'=xe^{\kappa t} 
\ee
in terms of which the Rindler metric\index{Rindler!metric} becomes
\be
ds^2=-dU'\,dV'
\ee
Now set
\be
U'=T-X, \quad V'=T+X
\ee
to get
\be
ds^2=-dT^2 +dX^2
\ee
i.e. the \emph{Rindler spacetime is just 2-dim Minkowski in unusual 
coordinates}.  Moreover, the Rindler coordinates with $x>0$ cover only the
$U'<0,\;V'>0$ region of 2d Minkowski
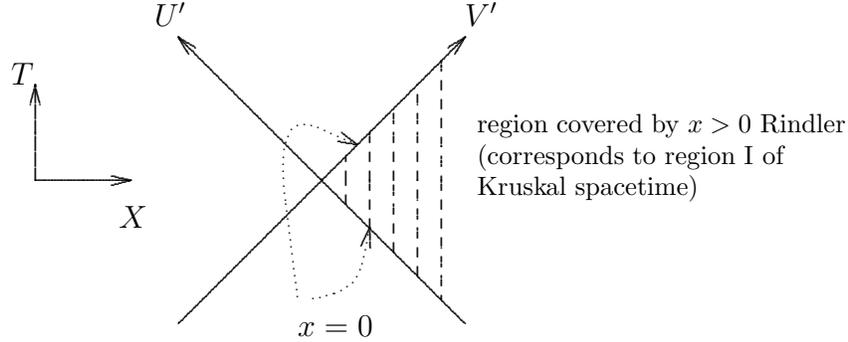
\begin{center}\input{p35-1.pictex}\end{center}
From what we know about the surface $r=2M$ of Schwarzschild it follows that 
the lines $U'=0,\;V'=0$, i.e. $x=0$ of Rindler is a Killing horizon of
$k=\fltpd{}{t}$ with surface gravity $\pm\kappa$.

\paragraph{Exercise}
\newcounter{Rindler_ex}
\begin{list}{(\roman{Rindler_ex})}
{\usecounter{Rindler_ex}}
\item Show that $U'=0$ and $V'=0$ are \emph{null curves}.

\item Show that 
\be
k = \kappa\left(V'\pd{}{V'}-U'\pd{}{U'}\right)
\ee
and that $\left.k\right|_{U'=0}$ is normal to $U'=0$. (So $\{U'=0\}$ is a 
Killing horizon).

\item
\be
\left.(k\cdot Dk)^{\mu}\right|_{U'=0} =\left.\kappa k^{\mu}\right|_{U'=0}
\ee

\end{list}

Note that $k^2=-(\kappa x)^2\to -\infty$ as $x\to\infty$, so \emph{there is 
no natural normalization of $k$ for Rindler}.

i.e. In contrast to Schwarzschild only the fact that $\kappa\neq 0$ is a 
property of the Killing horizon itself - the actual value of $\kappa$ depends on
an arbitrary normalization of $k$ --- so what is the meaning of the value of
$\kappa$?

\subsubsection{Acceleration Horizons}

\paragraph{Proposition}  The proper acceleration of a particle at $x=a^{-1}$ in 
Rindler spacetime (i.e. on an orbit of $k$) is constant and equal to $a$. 

\paragraph{Proof}  A particle on a timelike orbit $X^{\mu}(\tau)$ of a Killing 
vector field $\xi$ has 4-velocity
\be
u^{\mu}=\frac{\xi^{\mu}}{\left(-\xi^2\right)^{1/2}} \qquad 
\mbox{(since $u\propto \xi$ and $u\cdot u=-1$)}
\ee
Its proper 4-acceleration is
\bea
a^{\mu} & = & D_{(\tau)}u^{\mu}=u\cdot Du^{\mu} \\
 & = & \frac{\xi\cdot D\xi^{\mu}}{-\xi^2}+
\frac{\left(\xi\cdot\partial\xi^2\right)\xi^{\mu}}{2\xi^2}
\eea
But $\xi\cdot\partial\xi^2=2\xi^{\mu}\xi^{\nu}D_{\mu}\xi_{\nu}=0$ 
for Killing vector field, so
\be
a^{\mu}=\frac{\xi\cdot D\xi^{\mu}}{-\xi^2}
\ee
and `proper acceleration' is magnitude $|a|$ of $a^{\mu}$.  \\

For Rindler with $\xi=k$ we have (\bold{Exercise})
\be
a^{\mu}\partial_{\mu}=\frac{1}{U'}\pd{}{V'}+\frac{1}{V'}\pd{}{U'}
\ee
so
\bea
|a| & \equiv & \left(a^{\mu}a^{\nu}g_{\mu\nu}\right)^{1/2}=
\left(-\frac{1}{U'V'}\right)^{1/2} \\
 & = & \frac{1}{x}
\eea
so for $x=a^{-1}$ (constant) we have $|a|=a$, i.e. \emph{orbits of $k$ 
in Rindler are worldlines of constant proper acceleration}.  The acceleration
increases \emph{without bound} as $x\to 0$, so the Killing horizon at $x=0$ is
called an \emphin{acceleration horizon}.
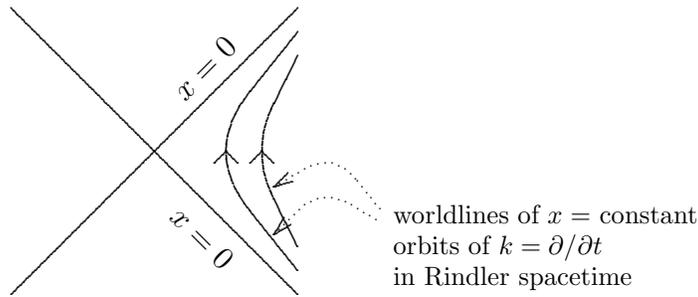
\begin{center}\input{p38-1.pictex}\end{center}
Although the \emph{proper acceleration} of an $x=$ constant worldline 
diverges as $x\to 0$ its acceleration as measured by another $x=$ constant
observer will remain finite.  Since
\be
d\tau^2=(\kappa x)^2dt^2 \qquad \mbox{(for $x=a^{-1}$, constant)}
\ee
the acceleration as measured by \emph{an observer whose proper time is $t$} is
\be
\left(\frac{d\tau}{dt}\right)\times \frac{1}{x} = (\kappa x)\times 
\frac{1}{x}=\kappa
\ee
which has a \emph{finite} limit, $\kappa$, as $x\to 0$. \\

In Rindler spacetime such an observer is one with constant proper 
acceleration $\kappa$, but these observers are \emph{in no way `special`}
because the normalization of $t$ was arbitrary.
\be
t\to \lambda t \quad \Rightarrow \quad \kappa \to \lambda^{-1}\kappa , 
\quad (\lambda\in \R)
\ee
For Schwarzschild, however,
\be
d\tau^2=dt^2 \quad \Rightarrow \quad \left\{\begin{array}{ll} 
r=\mbox{constant} \to \infty \\ \theta,\phi\;\mbox{constant} \end{array}\right.
\ee
i.e. an observer whose proper time is $t$ is one at spatial $\infty$.  Thus

\fbox{\parbox{6in}{\emphin{surface gravity} is the acceleration of a static 
particle near the horizon as measured at spatial infinity}} \\

This explains the term `surface gravity' for $\kappa$.

\subsection{Surface Gravity and Hawking Temperature}

We can study the behaviour of QFT in a black hole spacetime using 
\emph{Euclidean path integrals}.  In Minkowski spacetime this involves setting
\be
t=i\tau
\ee
and continuing $\tau$ from imaginary to real values.  Thus $\tau$ 
is `imaginary time'\index{imaginary time} here (\ul{not} proper time on some
worldline). \\

In the black hole spacetime this leads to a continuation of the 
Schwarzschild metric to the \emph{Euclidean Schwarzschild metric}.
\be
ds^2_{\subtext{E}}=\left(\Schr\right)d\tau^2+\frac{dr^2}
{\left(\Schr\right)}+r^2d\Omega^2
\ee
This is singular at $r=2M$.  To examine the region \emph{near $r=2M$} we set
\be
r-2M=\frac{x^2}{8M}
\ee
to get
\be
ds^2_{\subtext{E}} \approx 
\underbrace{ (\kappa x)^2d\tau^2+dx^2 }_{\subtext{Euclidean Rindler}}
+\frac{1}{4\kappa^2}d\Omega^2
\ee
Not surprisingly, the metric near $r=2M$ is the product of the metric on 
$\mbox{S}^2$ and the Euclidean Rindler
spacetime\index{Rindler!spacetime!Euclidean}
\be
ds^2_{\subtext{E}} = dx^2+x^2d(\kappa\tau)^2
\ee
This is just $\bb{E}^2$ in plane polar coordinates if we make the 
\emph{periodic identification}
\be
\tau \sim \tau +\frac{2\pi}{\kappa}
\ee
i.e. the singularity of Euclidean Schwarzschild at $r=2M$ (and of 
Euclidean Rindler at $x=0$) is just a coordinate singularity provided that
imaginary time coordinate $\tau$ is periodic with period $2\pi/\kappa$.
This means that the Euclidean functional integral must be taken over fields 
$\Phi(\vec{x},\tau)$ that are periodic in $\tau$ with period $2\pi/\kappa$ 
[Why this is so is not self-evident, which is presumably why the Hawking
temperature was not first found this way. Closer analysis shows that the
non-singularity of the Euclidean metric is required for equilibrium].

Now, the Euclidean functional integral is
\be
Z =\int \left[\mcD\Phi\right]e^{-S_{\subtext{E}}\left[\Phi\right]}
\ee
where 
\be
S_{\subtext{E}} = \int dt\left(-i p\dot{q}+H\right)
\ee
is the Euclidean action. If the functional integral is taken over fields $\Phi$
that are periodic in imaginary time with period $\hbar \beta$ then it can be
written as (see QFT course)
\be
Z=\tr\, e^{-\beta H}\, ,
\ee
which is the partition function for a quantum mechanical system with 
Hamiltonian $H$ at temperature $T$ given by $\beta=(k_B T)^{-1}$ where $k_B$ is
Boltzman's constant. \\

But we just saw that $\hbar\beta=2\pi/\kappa$ for Schwarzschild, so we deduce
that a QFT  can be in equilibrium with a black hole only at the \emph{Hawking
temperature}\index{Hawking!temperature}
\be
T_H   =   \frac{\kappa}{2\pi}\frac{\hbar}{k_B} \\
\ee
i.e. in units for which $\hbar=1$, $k_B=1$
\bebox{
T_H=\frac{\kappa}{2\pi}
}
\ul{N.B.}
\newcounter{hawkingtempNB}
\begin{list}{(\roman{hawkingtempNB})}
{\usecounter{hawkingtempNB}}
\item At any other temperature, Euclidean Schwarzschild has a conical 
singularity\index{singularity!conical} $\rightarrow$ no equilibrium.

\item Equilibrium at Hawking temperature is unstable since if the black 
hole absorbs radiation its mass increases and its temperature \emph{decreases},
i.e. the black hole has \emph{negative} specific heat.

\end{list}

\subsection{Tolman Law - Unruh Temperature}\index{Tolman law}
\index{Unruh!temperature}

\paragraph{Tolman Law}  The local temperature $T$ of a static 
self-gravitating system in thermal equilibrium satisfies 
\be
\left(-k^2\right)^{1/2}T=T_0
\ee
where $T_0$ is constant and $k$ is the timelike Killing vector 
field $\fltpd{}{t}$.  If $\left(k^2\right)\to -1$ asymptotically we can identify
$T_0$ as the temperature `as seen from infinity'.  For a Schwarzschild black
hole we have
\be
T_0=T_H=\frac{\kappa}{2\pi}
\ee
Near $r=2M$ we have, in Rindler coordinates,
\be
(\kappa x)T=\frac{\kappa}{2\pi}
\ee
so
\be
T=\frac{x^{-1}}{2\pi}
\ee
is the temperature measured by a static observer (on orbit of $k$) near the 
horizon.  But $x=a^{-1}$, constant, for such an observer, where $a$ is proper
acceleration.  So
\be
T=\frac{a}{2\pi}
\ee
is the local (Unruh) temperature.  It is a general feature of quantum mechanics
(Unruh effect\index{Unruh!effect}) that an observer accelerating in Minkowski
spacetime appears to be in a heat bath at the Unruh temperature. \\

In Rindler spacetime the Tolman law states that
\be
(\kappa x)T=T_0
\ee
Since $T=x^{-1}/(2\pi)$ for $x=$ constant, we deduce that 
$T_0=\kappa/(2\pi)$, as in Schwarzschild, but this is now just the temperature
of the observer with constant acceleration $\kappa$, who is of no particular
significance.  Note that in Rindler spacetime
\be
T=\frac{x^{-1}}{2\pi} \to 0 \quad \mbox{as} \; x\to \infty
\ee
so the Hawking temperature (i.e. temperature as measured at spatial 
$\infty$) is actually zero. 

This is expected because Rindler is just Minkowski in unusual coordinates, there is nothing inside which could radiate.  But for a black hole
\be
T_{\subtext{local}}\to T_H \quad \mbox{at infinity}
\ee
$\Rightarrow$ the black hole must be radiating at this temperature. We shall
confirm this later.

\section{Carter-Penrose Diagrams}

\subsection{Conformal Compactification}

A black hole is a ``region of spacetime from which no signal can escape to 
infinity'' (Penrose).  This is unsatisfactory because `infinity' is not part of
the spacetime.  However the `definition' concerns the \emph{causal structure} of
spacetime which is unchanged by \emphin{conformal compactification}
\be
ds^2\to d\tilde{s}^2 = \Lambda^2(\vec{r},t)ds^2, \quad \Lambda\neq 0
\ee
We can choose $\Lambda$ in such a way that all points at $\infty$ in the 
original metric are at \emph{finite} affine parameter in the new metric.  For
this to happen we must choose $\Lambda$ s.t.
\be
\Lambda(\vec{r},t) \to 0 \quad \mbox{as } |\vec{r}|\to \infty \quad 
\mbox{and/or } |t|\to\infty
\ee
In this case `infinity' can be identified as those points $(\vec{r},t)$ for 
which $\Lambda(\vec{r},t)=0$.  These points are \ul{not} part of the original
spacetime but they can be added to it to yield a \emph{conformal
compactification} of the spacetime.

\subsubsection{Example 1}  Minkowski space
\be
ds^2=-dt^2+dr^2+r^2d\Omega^2
\ee
Let
\be
\left\{ \begin{array}{rcl} u & = & t-r \\ v & = & t+r \end{array}\right\} 
\to ds^2=-du\,dv +\frac{(u-v)^2}{4}d\Omega^2
\ee
Now set
\be
\left\{ \begin{array}{rclcc} u & = & \tan\tilde{U} & \quad & 
-\pi/2<\tilde{U}<\pi/2 \\ v & = & \tan\tilde{V} & \quad & -
\pi/2<\tilde{V}<\pi/2
\end{array} \right\} \begin{array}{l} \mbox{with $\tilde{V} \ge \tilde{U}$} \\
\mbox{since $r\ge0$} \end{array}
\ee
In these coordinates,
\be
ds^2=\left(2\cos\tilde{U}\cos\tilde{V}\right)^{-2}
\left[-4d\tilde{U}\,d\tilde{V}+\sin^2\left(\tilde{V}-
\tilde{U}\right)d\Omega^2\right]
\ee
To approach $\infty$ in this metric we must take 
$\left|\tilde{U}\right|\to \pi/2$ or $\left|\tilde{V}\right|\to \pi/2$, so by
choosing
\be
\Lambda = 2\cos\tilde{U}\cos \tilde{V}
\ee
we bring these points to finite affine parameter in the new metric
\be
d\tilde{s}^2=\Lambda ds^2=-4d\tilde{U}d\tilde{V}+
\sin^2\left(\tilde{V}-\tilde{U}\right)d\Omega^2 
\ee
We can now add the `points at infinity'.  Taking the restriction 
$\tilde{V}\ge \tilde{U}$ into account, these are
\bdm
\begin{array}{cccccl}
\left.\begin{array}{rcl}  \tilde{U} & = & -\pi/2 \\ \tilde{V} & = & \pi/2
\end{array} \right\} &
\Leftrightarrow & 
\left\{\begin{array}{rcl} u & \to & -\infty \\ v & \to & \infty 
\end{array}\right\} &
\Leftrightarrow & 
\left\{ \begin{array}{c} r\to\infty \\ \mbox{$t$ finite} 
\end{array} \right\} 
\mbox{\emph{spatial} $\infty$, $i_0$} \\ \\
\left.\begin{array}{rcl}  \tilde{U} & = & \pm\pi/2 \\ 
\tilde{V} & = & \pm\pi/2 \end{array} \right\} &
\Leftrightarrow & 
\left\{\begin{array}{rcl}
u & \to & \pm\infty \\ v & \to & \pm\infty \end{array}\right\} &
\Leftrightarrow & 
\left\{ \begin{array}{c} t\to\pm\infty \\ \mbox{$r$ finite} 
\end{array} \right\}
\begin{array}{l} \mbox{past and future} \\ 
\mbox{\emph{temporal} $\infty$, $i_{\pm}$} 
\end{array} \\ \\
\left.\begin{array}{rcl} \tilde{U} & = & -\pi/2 \\ 
|\tilde{V}| & \neq & \pi/2 \end{array} \right\} &
\Leftrightarrow & 
\left\{\begin{array}{c} u\to -\infty \\ 
\mbox{$v$ finite} \end{array}\right\} &
\Leftrightarrow & 
\left\{ \begin{array}{c} r\to\infty \\ t\to -\infty \\ 
\mbox{$r+t$ finite} \end{array} \right\}   
\begin{array}{l} \mbox{past null $\infty$} \\ \scri^- \end{array} \\ \\
\left.\begin{array}{rcl} |\tilde{U}| & \neq & \pi/2 \\ 
\tilde{V} & = & \pi/2 \end{array} \right\} &
\Leftrightarrow & 
\left\{\begin{array}{c} \mbox{$u$ finite} \\ 
v\to\infty \end{array}\right\} & 
\Leftrightarrow & 
\left\{ \begin{array}{c} r\to\infty \\ t\to\infty \\ 
\mbox{$r-t$ finite} \end{array} \right\} 
\begin{array}{l} \mbox{future null $\infty$} \\ \scri^+ \end{array}
\end{array}
\edm
Minkowski spacetime is conformally embedded in the new spacetime with 
metric $d\tilde{s}^2$ with boundary at $\Lambda=0$. \\

%\be
%d\tilde{s}^2=-4d\tilde{U}\,d\tilde{V}+
%\sin^2\left(\tilde{V}-\tilde{U}\right)d\Omega^2
%\ee

Introducing the new time and space coordinates $\tau,\chi$ by 
\be
\tau = \tilde{V}+\tilde{U}, \quad \chi=\tilde{V}-\tilde{U}
\ee
we have
\bebox{
\begin{array}{rcl}
d\tilde{s}^2 & = & \Lambda ds^2 = -d\tau^2+d\chi^2+\sin^2\chi d\Omega^2  \\ \\
\Lambda & = & \cos \tau+\cos\chi \end{array}
}
$\chi$ is an angular variable which must be identified modulo $2\pi$, 
$\chi\sim \chi+2\pi$.  If no other restriction is placed on the ranges of $\tau$
and $\chi$, then this metric $d\tilde{s}^2$ is that of the \emphin{Einstein
Static Universe}, of topology $\R$ (time) $\times$ $\mbox{S}^3$ (space). \\

The 2-spheres of constant $\chi\neq 0,\pi$ have radius 
$\left|\sin\chi\right|$ (the points $\chi=0,\pi$ are the poles of a 3-sphere). 
If we represent each 2-sphere of constant $\chi$ as a point the E.S.U. can be
drawn as a cylinder.
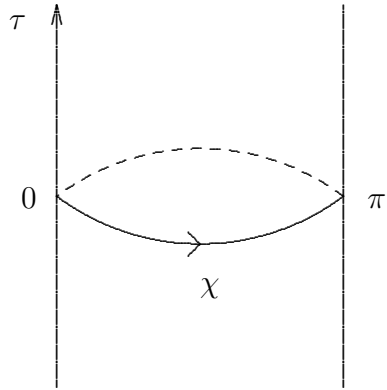
\begin{center}\input{p46-1.pictex}\end{center}
But compactified Minkowski spacetime is conformal to the triangular region
\be
-\pi \le \tau \le \pi, \quad 0 \le \chi \le \pi
\ee
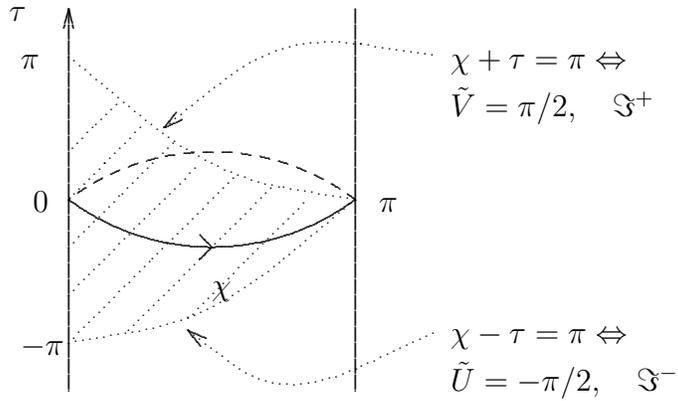
\begin{center}\input{p47-1.pictex}\end{center}
Flatten the cylinder to get the \emphin{Carter-Penrose diagram} of 
Minkowski spacetime.
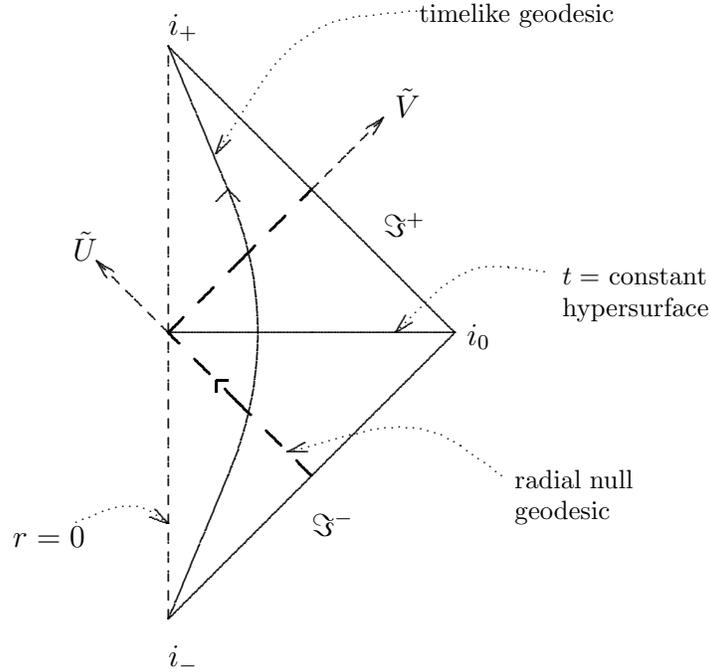
\begin{center}\input{p47-2.pictex}\end{center}
Each point represents a 2-sphere, except points on $r=0$ and $i_0,i_{\pm}$.  
Light rays travel at $45^{^0}$ from $\scri^-$ through $r=0$ and then out to
$\scri^+$.  [$\scri^{\pm}$ are null hypersurfaces]. \\

\emph{Spatial sections} of the compactified spacetime are topologically 
$\mbox{S}^3$ because of the addition of the point $i_0$.  Thus, they are not
only compact, but also have no boundary.  This is not true of the whole
spacetime.  Asymptotically it is possible to identify points on the boundary of
compactified spacetime to obtain a compact manifold without boundary (the group
U(2); see Question I.6).  More generally, this is not possible because $i_{\pm}$
are singular points that cannot be added (see \bold{Example 3: Kruskal}).  
%This is illustrated by ... Kruskal 

\subsubsection{Example 2: Rindler Spacetime}

\be
ds^2=-dU'\,dV'
\ee
Let
\be
\left.\begin{array}{rcl} U' & = & \tan \tilde{U} \\ V' & = & \tan \tilde{V} 
\end{array} \right\} \begin{array}{c} -\pi/2<\tilde{U} <\pi/2 \\
-\pi/2<\tilde{V}<\pi/2 \end{array} 
\ee
Then
\bea
ds^2 & = & -\left(\cos\tilde{U}\cos\tilde{V}
\right)^{-2}d\tilde{U}\,d\tilde{V} \\
 & = & \Lambda^{-2}d\tilde{s}^2 \qquad 
\left(\Lambda=\cos\tilde{U}\cos\tilde{V}\right)
\eea
i.e. conformally compactified spacetime with metric 
$d\tilde{s}^2=-d\tilde{U}\,d\tilde{V}$ is same as before but with the above
\emph{finite} ranges for coordinates $\tilde{U},\tilde{V}$.

The points at infinity are those for which $\Lambda=0$, 
$\left|\tilde{U}\right|=\pi/2$, $\left|\tilde{V}\right|=\pi/2$. 
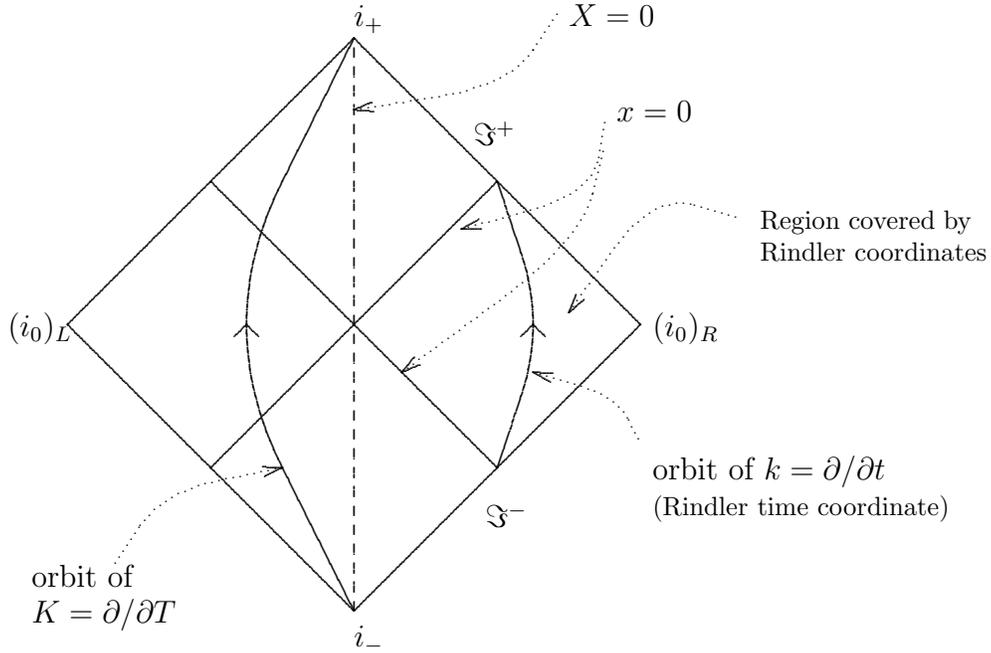
\begin{center}\input{p50-1.pictex}\end{center}
Similar to 4-dim Minkowski, but $i_0$ is now two points.

\subsubsection{Example 3:  Kruskal Spacetime}

\be
ds^2=-\left(\Schr\right)du\,dv +r^2d\Omega^2 \qquad \mbox{in region I}
\ee
Let
\be
\left\{\begin{array}{rcl} u=\tan \tilde{U} \\ 
v=\tan \tilde{V} \end{array} \quad   
\begin{array}{c} -\pi/2<\tilde{U} <\pi/2 \\
-\pi/2<\tilde{V}<\pi/2 \end{array}  \right\}
\ee
Then
\be
ds^2=\left(2\cos\tilde{U}\cos\tilde{V}\right)^{-2}
\left[-4\left(\Schr\right)d\tilde{U}d\tilde{V}+
r^2\cos^2\tilde{U}\cos^2\tilde{V}d\Omega^2\right]
\ee
Using the fact that
\be
r^* = \half (v-u)=\frac{\sin\left(\tilde{V}-
\tilde{U}\right)}{2\cos\tilde{U}\cos\tilde{V}} 
\ee
we have
\be
d\tilde{s}^2 = \Lambda^2 ds^2=-4\left(\Schr\right)
d\tilde{U}d\tilde{V}+\left(\frac{r}{r^*}\right)^2
\sin^2\left(\tilde{V}-\tilde{U}\right)d\Omega^2
\ee
Kruskal is an example of an \emph{asymptotically flat spacetime}.  
It approaches the metric of compactified Minkowski spacetime as $r\to\infty$
(with or without fixing $t$) so $i_0$, and $\scri^{\pm}$ can be added as
before.  Near $r=2M$ we can introduce KS-type coordinates to pass through the
horizon.  In this way one can deduce that the CP diagram for the Kruskal
spacetime is 
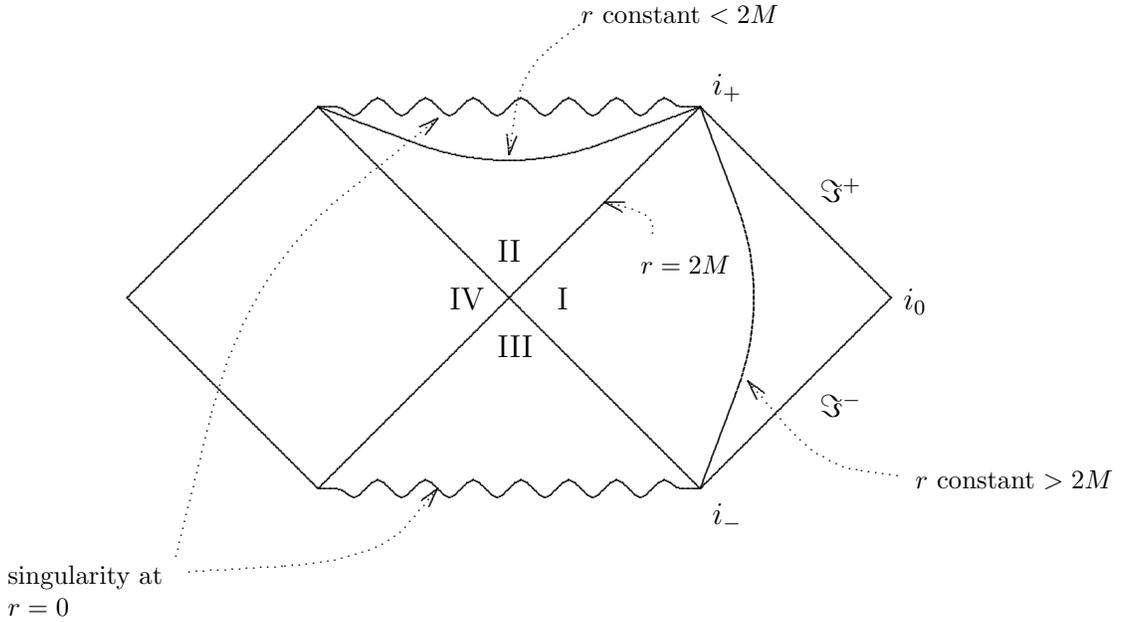
\begin{center}\input{p49-1.pictex}\end{center}
\paragraph{Note}
\newcounter{Kruskal_note}
\begin{list}{(\roman{Kruskal_note})}
{\usecounter{Kruskal_note}}
\item All $r=$ constant hypersurfaces meet at $i_+$ including 
the $r=0$ hypersurface, which is singular, so $i_+$ is a singular point. 
Similarly for $i_-$, so these points cannot be added.

\item We can adjust $\Lambda$ so that $r=0$ is represented by a straight line.

\end{list}
 
In the case of a collapsing star, only that part of the CP diagram of 
Kruskal that is exterior to the star is relevant.  The details of the interior
region depend on the physics of the star.  For pressure-free, spherical 
collapse, all parts of the star not initially at $r=0$ reach the singularity at
$r=0$ \emph{simultaneously}, so the CP diagram is
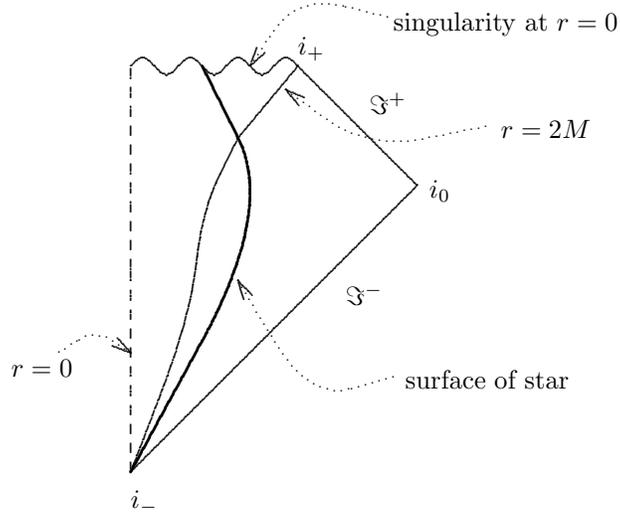
\begin{center}\input{p49-2.pictex}\end{center}

\section{Asymptopia}

A spacetime $(M,g)$ is \emph{asymptotically simple}
\index{asymptotically!simple} if $\exists$ a manifold
$(\widetilde{M},\tilde{g})$ with boundary $\partial\widetilde{M}=\overline{M}$
and a continuous embedding $f(M):M\to\widetilde{M}$ s.t.
\newcounter{asymsimple}
\begin{list}{(\roman{asymsimple})}
{\usecounter{asymsimple}}
\item $f(M)=\widetilde{M}-\partial \widetilde{M}$ 

\item $\exists$ a smooth function $\Lambda$ on $\widetilde{M}$ with 
$\Lambda >0$ on $f(M)$ and $\tilde{g}=\Lambda^2f(g)$.

\item $\Lambda=0$ but $d\Lambda\neq 0$ on $\partial\widetilde{M}$.

\item Every null geodesic in $M$ acquires 2 endpoints on $\partial M$.

\paragraph{Example} $M=$ Minkowski, $\widetilde{M}=$ compactified 
Minkowski.  \\

Condition (iv) excludes black hole spacetime.  This motivates the 
following definition: \\

A \emph{weakly asymptotically simple}\index{asymptotically!simple!weakly} 
spacetime $(M,g)$ is one for which $\exists$ an open set $U\subset M$ that is
isometric to an open neighborhood of $\partial\widetilde{M}$, where
$\widetilde{M}$ is the `conformal compactification' of some asymptotically
simple manifold. 

\paragraph{Example}  $M=$ Kruskal, $\widetilde{M}$ its conformal 
`compactification'. 

\paragraph{Note}
\newcounter{was}
\begin{list}{(\roman{was})}
{\usecounter{was}}
\item $\widetilde{M}$ is not actually compact because 
$\partial\widetilde{M}$ excludes $i_{\pm}$.  

\item $M$ is not asymptotically simple because geodesics that enter 
$r<2M$ cannot end on $\scri^+$.
\end{list}

\subsubsection{Asymptotic flatness}

An \emph{asymptotically flat} spacetime is one that is both weakly asymptotically simple and is \emph{asymptotically empty}\index{asymptotically!empty} in the sense that

\item $R_{\mu\nu}=0$ in an open neighborhood of $\partial M$ in $\overline{M}$.

\end{list}

This excludes, for example, anti-de Sitter space.  It also excludes 
spacetimes with long range electromagnetic fields that we don't wish to exclude
so condition (v) requires modification to deal with electromagnetic fields. \\

Asymptotically flat spacetimes have the same type of structure for 
$\scri^{\pm}$ and $i_0$ as Minkowski spacetime.
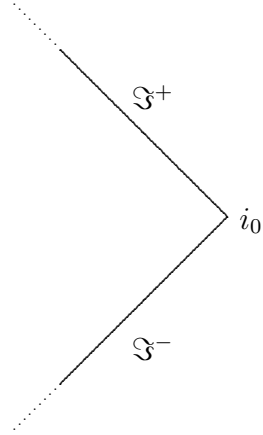
\begin{center}\input{p52-1.pictex}\end{center}
In particular they admit vectors that are asymptotic to the Killing 
vectors of Minkowski spacetime near $i_0$, which \emph{allows a definition of
total mass, momentum and angular momentum on spacelike hypersurfaces}.  The
asymptotic symmetries on $\scri^{\pm}$ are much more complicated (the `BMS'
group, which will not be discussed in this course).

\section{The Event Horizon}

Assume spacetime $M$ is weakly asymptotically flat.  Define
\bdm
J^-(U)
\edm
to be the \emph{causal past} of a set of points $U\subset M$ and 
\bdm
\overline{J}^-(U)
\edm
to be the topological closure of $J^-$, i.e. including limit points.  Define 
the \emph{boundary} of $\overline{J}^{-}$ to be
\be
\dot{J}^-(U)= \overline{J}^-(U)-J^-(U)
\ee
The \emphin{future event horizon} of $M$ is
\be
\mcH^+=\dot{J}^-\left(\scri^+\right)
\ee
i.e. the \emph{boundary of the closure of the causal past of $\scri^+$}.

\paragraph{Example}  Spacetime of a spherically-symmetric collapsing star
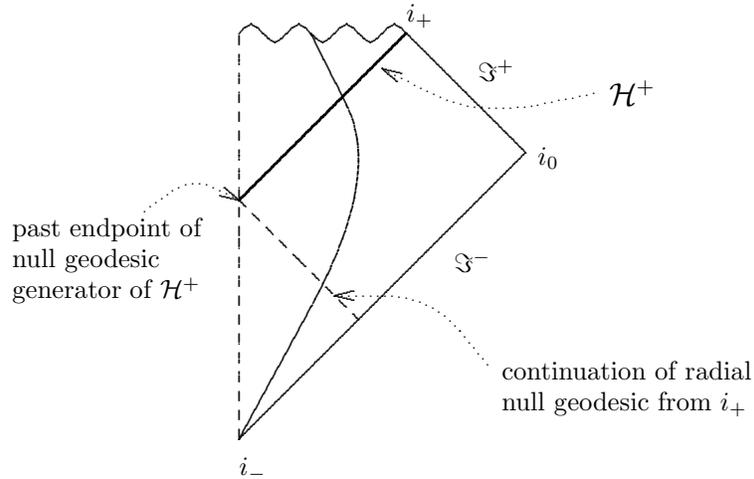
\begin{center}\input{p53-1.pictex}\end{center}

\subsubsection{Properties of the Future Event Horizon, $\mcH^+$}
\newcounter{propfeh}
\begin{list}{(\roman{propfeh})}
{\usecounter{propfeh}}
\item $i_0$ and $\scri^-$ are contained in $J^-\left(\scri^+\right)$, so 
they are \ul{not} part of $\mcH^+$.

\item $\mcH^+$ is a null hypersurface.

\item No two points of $\mcH^+$ are timelike separated.  For nearby points 
this follows from (ii) but is also true globally.  Suppose that $\alpha$ and
$\beta$ were two such points with $\alpha\in J^-(\beta)$.  The timelike curve
between them could then be deformed to a nearby timelike curve between $\alpha'$
and $\beta'$ with $\beta'\in J^-\left(\scri^+\right)$ but $\alpha'\not\in
J^-\left(\scri^+\right)$
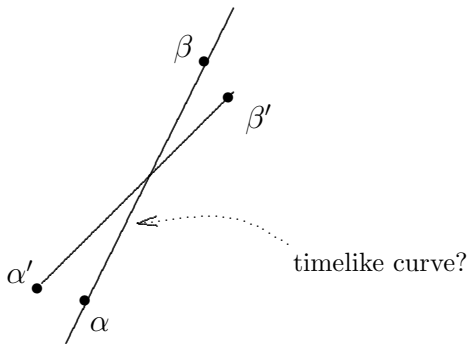
\begin{center}\input{p54-1.pictex}\end{center}
But $\alpha'\in J^-(\beta)\in J^-\left(\scri^+\right)$, so we have a 
contradiction.  The timelike curve between $\alpha$ and $\beta$ cannot exist.

\item The null geodesic generators of $\mcH^+$ may have \emph{past endpoints} 
in the sense that the continuation of the geodesic further into the past is no
longer in $\mcH^+$, e.g. at $r=0$ for a spherically symmetric star, as shown in
diagram above.

\item If a generator of $\mcH^+$ had a future endpoint, the future continuation 
of the null geodesic beyond a certain point would leave $\mcH^+$.  This cannot
happen.

\end{list}

\fbox{\parbox{6in}{
\bold{Theorem} (Penrose) The generators of $\mcH^+$ have no future endpoints}}

\paragraph{Proof} Consider the causal past $J^-(S)$ of some set $S$.
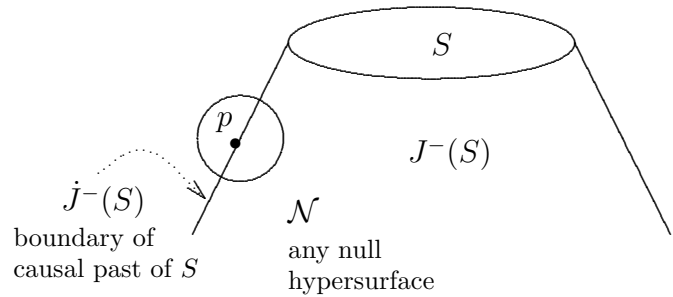
\begin{center}\input{p55-0.pictex}\end{center}
Consider a point $p\in\dot{J}^-(S)$, $p\not\in S,\overline{S}$.
% and a compact region $U$ containing $p$ but not containing any past 
  Endpoints of the null geodesic in $\dot{J}^-(S)$ through $p$.  
Consider also an infinite sequence of timelike curves 
$\left\{\gamma_i\right\}$ from $p_i\in$ neighborhood of $p$ and 
$\in J^-(S)$ to
$S$
%$\scri^+$ to points $\left\{p_n\in U\right\}$ 
s.t. $p$ is the limit point of $\left\{p_i\right\}$ on $\dot{J}^-(S)$.
%$\mcH^+$.  Let $q_n$ be the point at which $\gamma_i$ intersects the 
%boundary of $U$.
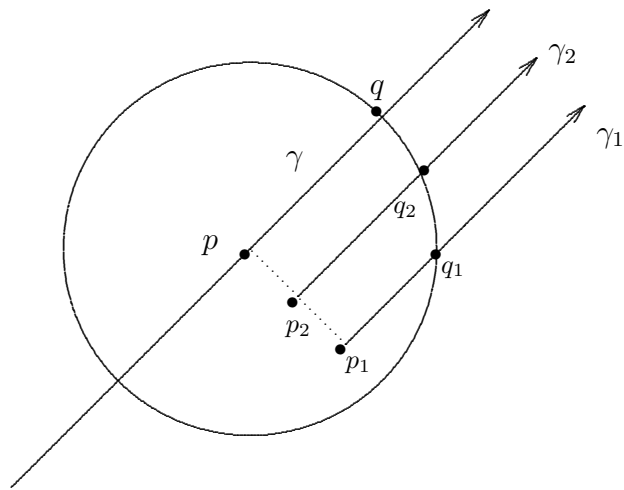
\begin{center}\input{p55-1.pictex}\end{center}
The points $\left\{q_i\right\}$ must have a limit point $q$ on $\dot{J}^-(S)$.
%$\mcH^+$.  
Being the limit of timelike curves, the curve $\gamma$ from $p$ to $q$ 
cannot be spacelike, but can be null (lightlike).  It cannot be timelike either
from property (iii) above, so it is a segment of the null geodesic generator of
$\mcN$
%$\mcH^+$ 
through $p$.  The argument can now be repeated with $p$ replaced by $q$ to 
find another segment from $q$ to a point, 
%$r\in\mcH^+$
$r\in\mcN$, but further in the future.  It must be a segment of the 
\emph{same} generator because otherwise there exists a deformation to a timelike
curve in $\mcN$ separating $p$ and $r$.
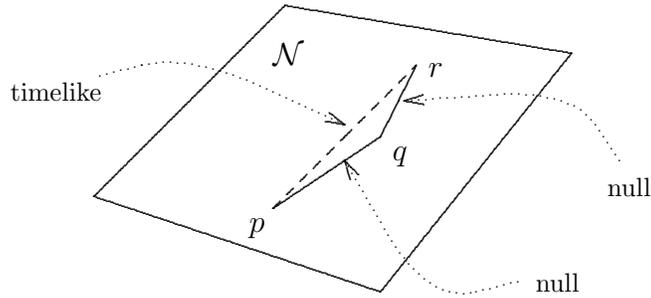
\begin{center}\input{p55-2.pictex}\end{center}
Choosing $S=\scri^+$, then gives Penrose's Theorem. \\

Properties (iv) and (v) show that \emph{null geodesics may enter $\mcH^+$ 
but cannot leave it}. \\

This result may appear inconsistent with time-reversibility, but is not.  
The time-reverse statement is that null geodesics may leave but cannot enter the
\emph{past event horizon}, $\mcH^-$.  $\mcH^-$ is defined as for $\mcH^+$ with
$J^-\left(\scri^+\right)$ replaced by $J^+\left(\scri^-\right)$, i.e. the causal
future of $\scri^-$.  The time-symmetric Kruskal spacetime has both a future and
a past event horizon.
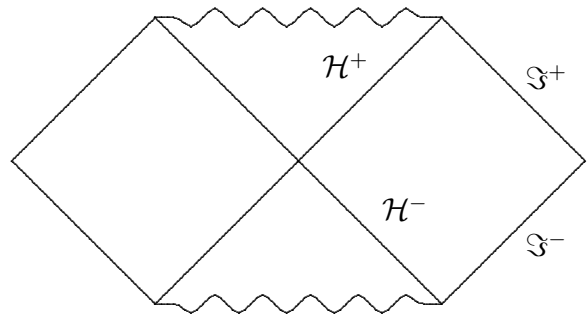
\begin{center}\input{p56-1.pictex}\end{center}
The location of the event horizon $\mcH^+$ generally requires knowledge of the 
\emph{complete} spacetime.  Its location cannot be determined by observations
over a finite time interval. \\

However if we wait until the black hole settles down to a stationary spacetime 
we can invoke: \\

\fbox{\parbox{6in}{
\bold{Theorem} (Hawking)  The event horizon of a stationary asymptotically 
flat spacetime is a Killing horizon (but \ul{not} \emph{necessarily} of
$\fltpd{}{t}$).}}

This theorem is the essential input needed in the proof of the uniqueness
theorems for stationary black holes, to be considered later.

\section{Black Holes vs. Naked Singularities}

The singularity at $r=0$ that occurs in spherically symmetric collapse is 
hidden in the sense that no signal from it can reach $\scri^+$.  This is not
true of the Kruskal spacetime manifold since a signal from $r=0$ in the white
hole region \emph{can} reach $\scri^+$.
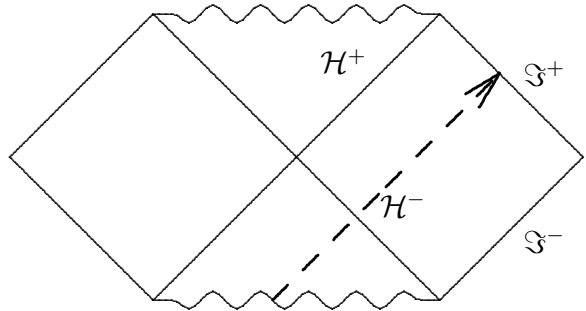
\begin{center}\input{p57-1.pictex}\end{center}
This singularity is \emph{naked}\index{naked singularity}.  Another example 
of a naked singularity is the $M<0$ Schwarzschild solution
\be
ds^2=-\left(1+\frac{2|M|}{r}\right)dt^2+
\frac{1}{\left(1+\frac{2|M|}{r}\right)}dr^2+r^2d\Omega^2
\ee
This solves Einstein's equations so we have no a priori reason to exclude 
it.  The CP diagram is
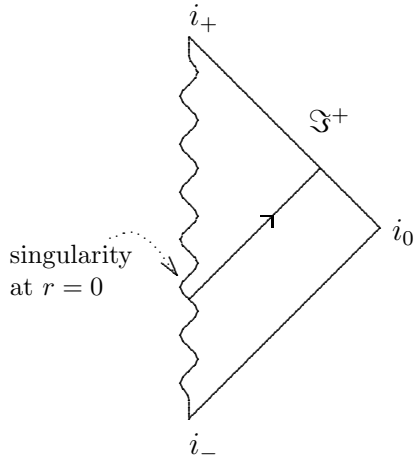
\begin{center}\input{p57-2.pictex}\end{center}
Neither of these examples is relevant to gravitational collapse, but 
consider the CP diagram:
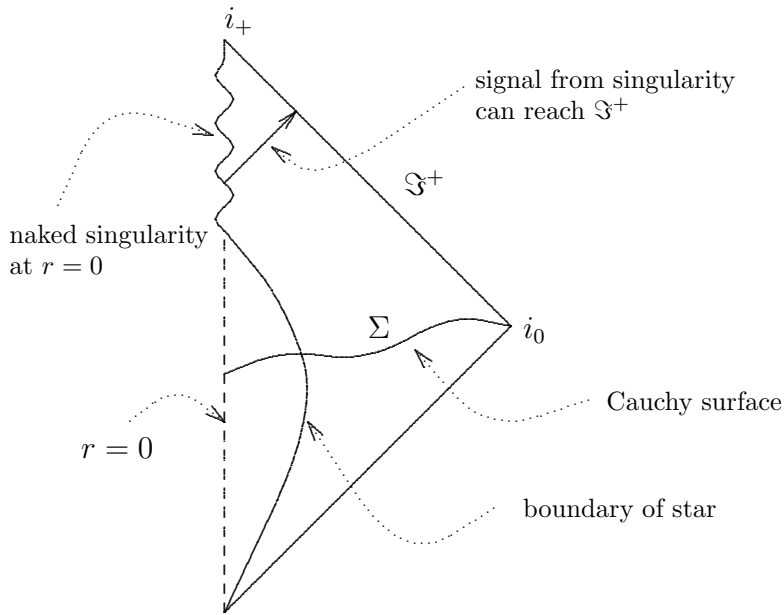
\begin{center}\input{p58-1.pictex}\end{center}
At late times the spacetime is $M<0$ Schwarzschild but at earlier times 
it is non-singular.  Under these circumstances it can be shown that $M\ge 0$ for
physically reasonable matter (the `positive energy' theorem) so the possibility
illustrated by the above CP diagram (formation of a naked singularity in
\emph{spherically-symmetric} collapse) cannot occur.  There remains the
possibility that naked singularities could form in non-spherical collapse.  If
this were to happen the future would eventually cease to be predictable from
data given on an initial spacelike hypersurface ($\Sigma$ in CP diagram above). 
There is considerable evidence that this possibility cannot be realized for
physically reasonable matter, which led Penrose to suggest the:

\paragraph{Cosmic Censorship Conjecture}  `Naked singularities cannot form 
from gravitational collapse in an asymptotically flat spacetime that is
non-singular on some initial spacelike hypersurface (Cauchy surface).' \\

\paragraph{Notes}
\newcounter{ccc}
\begin{list}{(\roman{ccc})}
{\usecounter{ccc}}
\item Certain types of `trivial' naked singularities must be excluded.

\item  Initial, cosmological, singularities are excluded.

\item There is no proof.  This is the major unsolved problem in classical G.R.
\end{list}

\chapter{Charged Black Holes}

\section{Reissner-Nordstr\"om}

Consider the Einstein-Maxwell action
\be
S=\frac{1}{16\pi G}\int \dx{4}{x}\sqrt{-g}\left[ R-F_{\mu\nu}F^{\mu\nu}\right], 
\qquad \left(R=R_{\mu\nu}^{\I\I\I\mu\nu}\right)
\ee
The unusual normalization of the Maxwell term means that the magnitude of the 
Coulomb force between point charges $Q_1,Q_2$ at separation $r$ (large) in flat
space is
\be
\frac{G\left|Q_1Q_2\right|}{r^2} \quad \mbox{(`geometrized' units of charge)}
\ee
The source-free Einstein-Maxwell equations are
\bea
G_{\mu\nu} & = & 2\left(F_{\mu\lambda}F_{\nu}^{\I\lambda}-
\frac{1}{4}g_{\mu\nu}F_{\rho\sigma}F^{\rho\sigma}\right) \\
D_{\mu}F^{\mu\nu} & = & 0 
\eea
They have the \emph{spherically-symmetric Reissner-Nordstr\"om}
\index{Reissner-Nordstr\"om solution} (RN) solution (which generalizes
Schwarzschild)
\bea
ds^2 & = & -\left(1-\frac{2M}{r}+\frac{Q^2}{r^2}\right)dt^2+\frac{dr^2}
{\left(1-\frac{2M}{r}+\frac{Q^2}{r^2}\right)}+r^2d\Omega^2  \\
A & = & \frac{Q}{r}dt \quad \mbox{(Maxwell 1-form potential $F=dA$)} 
\eea
The parameter $Q$ is clearly the \emph{electric charge}.

The RN metric can be written as 
\be
ds^2=-\frac{\Delta}{r^2}dt^2+\frac{r^2}{\Delta}dr^2+r^2d\Omega^2 
\ee 
where
\be
\Delta = r^2-2Mr+Q^2 = \left(r-r_+\right)\left(r-r_-\right)
\ee
where $r_{\pm}$ are not necessarily real
\be
r_{\pm} = M\pm \sqrt{M^2-Q^2}
\ee

There are therefore \ul{3 cases} to consider:
\newcounter{RNcases}
\begin{list}{\roman{RNcases})}
{\usecounter{RNcases}}

\item \ul{ $M<|Q|$} \\

$\Delta$ has no real roots so there is no horizon and the singularity at 
$r=0$ is naked. 

This case is similar to $M<0$ Schwarzschild.  According to the cosmic 
censorship hypothesis\index{cosmic censorship hypothesis} this case could not
occur in gravitational collapse.  As confirmation, consider a shell of matter of
charge $Q$ and radius $R$ in Newtonian gravity but incorporating

a) Equivalence of inertial mass $M$ with total energy, from special relativity.

b) Equivalence of inertial and gravitational mass from general relativity.

\be
\underbrace{M_{\subtext{total}}}_{\stackrel{\uparrow}{\mbox{total energy}}} = 
\underbrace{M_0}_{\stackrel{\uparrow}{\mbox{rest mass
energy}}}+\underbrace{\frac{GQ^2}{R}}_{\stackrel{\uparrow}{\mbox{Coulomb
energy}}}-\underbrace{\frac{GM^2}{R}}_{\stackrel{\uparrow}
{\stackrel{\mbox{grav.
binding energy}}{\mbox{($M$=total mass)}}}}
\ee
This is a quadratic equation for $M$.  The solution with $M\to M_0$ as 
$R\to\infty$ is
\be
M(R) = \frac{1}{2G}\left[\left(R^2+4GM_0R+4G^2Q^2\right)^{1/2}-R\right]
\ee
The shell will only undergo gravitational collapse iff $M$ decreases with 
decreasing $R$ (so allowing K.E. to increase).  Now
\be
M'=\frac{ G\left(M^2-Q^2\right)}{ 2MGR+R^2}
\ee
so collapse occurs only if $M>|Q|$ as expected. \\

Now consider $M(R)$ as $R\to 0$.
\be
M \longrightarrow |Q| \quad 
\mbox{\emph{independent of $M_0$}}
\ee
So GR resolves the infinite self-energy problem of point particles in 
classical EM.  A point particle becomes an extreme ($M=|Q|$) RN black 
hole (case (iii) below).

\paragraph{Remark}  The electron has $M\ll |Q|$ (at least when probed at 
distances $\gg GM/c^2$) because the gravitational attraction is negligible
compared to the Coulomb repulsion.  But the electron is \emph{intrinsically
quantum mechanical}, since its Compton wavelength $\gg$ Schwarzschild radius. 
Clearly the applicability of GR requires
\be
\frac{\mbox{Compton wavelength}}{\mbox{Schwarzschild radius}} = 
\frac{ \hbar/Mc}{MG/c^2} = \frac{\hbar c}{M^2 G} \ll 1 
\ee
i.e.
\be
M \gg \left(\frac{\hbar c}{G}\right)^{1/2} \equiv M_P \quad 
\mbox{(Planck mass)}
\ee
This is satisfied by any macroscopic object but not by elementary particles.

More generally the domains of applicability of classical physics QFT and 
GR are illustrated in the following diagram.
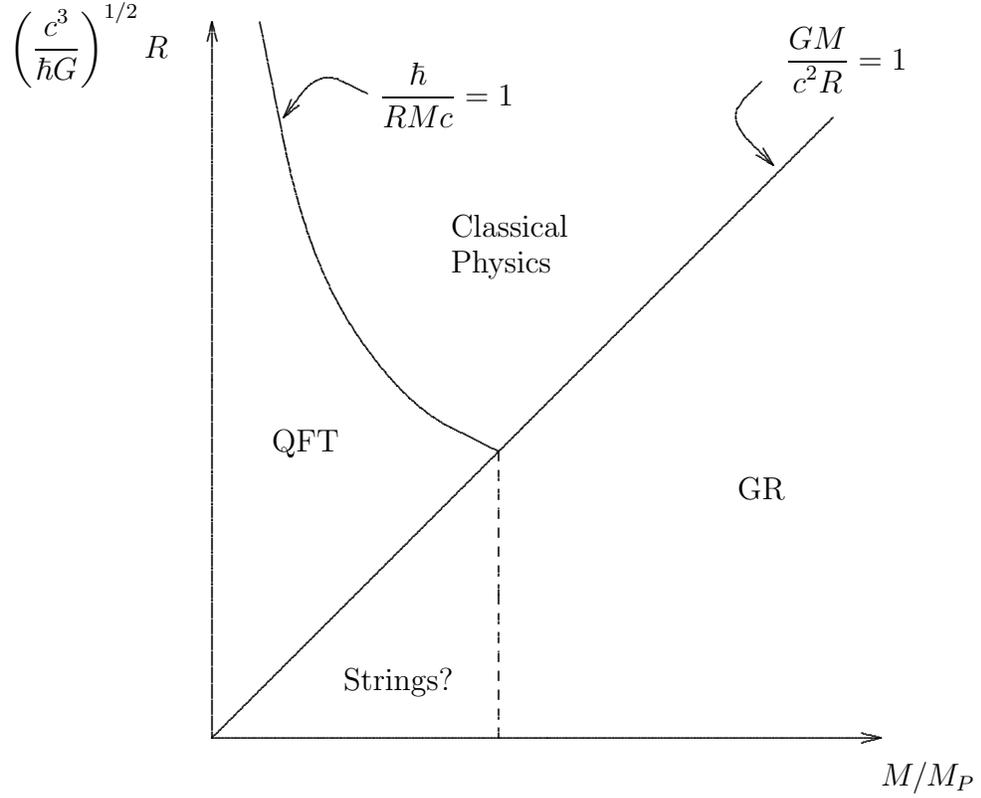
\begin{center}\input{p62-1.pictex}\end{center}

\item \ul{$M>|Q|$}  \\

$\Delta$ vanishes at $r=r_+$ and $r=r_-$ real, so metric is singular there, 
but these are coordinate singularities.  To see this we proceed as for $r=2M$ in
Schwarzschild.  Define $r^*$ by
\bea
dr^* & = &\frac{r^2}{\Delta}dr= \frac{dr}{\left(1-\frac{2M}{r}+
\frac{Q^2}{r^2}\right)} \\
\Rightarrow\; r^* & = & r+\frac{1}{2\kappa_+}\ln 
\left(\frac{ \left|r-r_+\right|}{r_+}\right) +
\frac{1}{2\kappa_-}\ln\left(\frac{ \left|r-r_-\right|}{r_-}\right)+\mbox{const}
\eea
where
\bebox{\kappa_{\pm}=\frac{ \left( r_{\pm}-r_{\mp} \right)}{ 2r_{\pm}^2} }
We then introduce the radial null coordinates $u,v$ as before
\be
v=t+r^*, \quad u = t-r^*
\ee
The RN metric in ingoing Eddington-Finkelstein coordinates $(v,r,\theta,\phi)$ 
is
\be
ds^2=-\frac{\Delta}{r^2}dv^2+2dv\,dr+r^2d\Omega^2
\ee
which is non-singular everywhere except at $r=0$.  Hence the $\Delta=0$ 
singularities of RN were coordinate singularities.  The hypersurfaces of
constant $r$ are null when $g^{rr}=\Delta/r^2=0$, i.e. when $\Delta=0$, so
$r=r_{\pm}$ are null hypersurfaces, $\mathcal{N}_{\pm}$. \\

\fbox{\parbox{6in}{
\paragraph{Proposition} The null hypersurfaces $\mcN_{\pm}$ of RN are 
Killing horizons of the Killing vector field $k=\partial/\partial v$ (the
extension of $\partial/\partial t$ in RN coordinates) with surface gravities
$\kappa_{\pm}$.}}

\paragraph{Proof}  The normals to $\mcN_{\pm}$ are 
\be
l_{\pm} = \left.f_{\pm}\left(g^{rr}\pd{}{r}+g^{vr}\pd{}{v}
\right)\right|_{\mcN_{\pm}} = f_{\pm}\pd{}{v}
\ee
(note $g^{rr}=0$ on $\mcN_{\pm}$ and $g^{vr}=1$) for some arbitrary 
functions $f_{\pm}$ which we can choose s.t. \fbox{$l_{\pm}Dl_{\pm}^{\mu}=0$}
(tangent to an affinely parameterized geodesic) so
\be
\pd{}{v} = f_{\pm}^{-1}l_{\pm}
\ee
which shows that $\mcN_{\pm}$ are Killing horizons of $\pd{}{v}$ 
(This is Killing because in EF coordinates the metric is $v$-independent). We
can interpret the LHS of this equation as a derivative w.r.t the group
parameter, and the RHS as a derivative w.r.t the affine parameter.  Now
\bea
(k\cdot Dk)^r & = & \Gamma^r_{\I vv}=-\half g^{rr}g_{vv,r} = 0 
\quad \mbox{on }\mcN_{\pm} \\
\left.(k\cdot Dk)^v\right|_{r=r_{\pm}} & = & \Gamma^v_{\I vv}= 
-\half g^{vr}g_{vv,r} = \left.\frac{1}{2r^2}\pd{}{r}\Delta\right 
|_{r=r_{\pm}} \\
 & = & \frac{1}{2r_{\pm}^2}\left(r_{\pm}-r_{\mp}\right) 
\quad \mbox{on }\mcN_{\pm} \\ 
& = & \kappa_{\pm}
\eea
\bebox{
\thdots \quad k\cdot Dk^{\mu}  =  \kappa_{\pm}k^{\mu}}
Since $k=\partial/\partial t$ in static coordinates we have $k^2\to -1$ as 
$r\to \infty$.  So we identify $\kappa_{\pm}$ as the surface gravities of
$\mcN_{\pm}$.
%\end{list}

Each of the Killing horizons $\mcN_{\pm}$ will have a bifurcation 2-sphere 
in the neighborhood of which we can introduce the KS-type coordinates
\be
U^{\pm}=-e^{-\kappa_{\pm}u},\quad V^{\pm}=e^{\kappa_{\pm}v}
\ee
For the $+$ sign we have
\be
ds^2=-\frac{r_+r_-}{\kappa_+^2}\frac{e^{-2\kappa_+r}}{r^2}
\left(\frac{r_-}{r-r_-}\right)^{\left(\frac{\kappa_+}
{\kappa_-}-1\right)}dU^+\,dV^++r^2d\Omega^2
\ee
where $r\left(U^+,V^+\right)$ is determined implicitly by
\be
U^+V^+=-e^{2\kappa_+r}\left(\frac{r-r_+}{r_+}\right)
\left(\frac{r-r_-}{r_-}\right)^{\kappa_+/\kappa_-}
\ee
This metric covers four regions of the maximal analytic extension of RN,
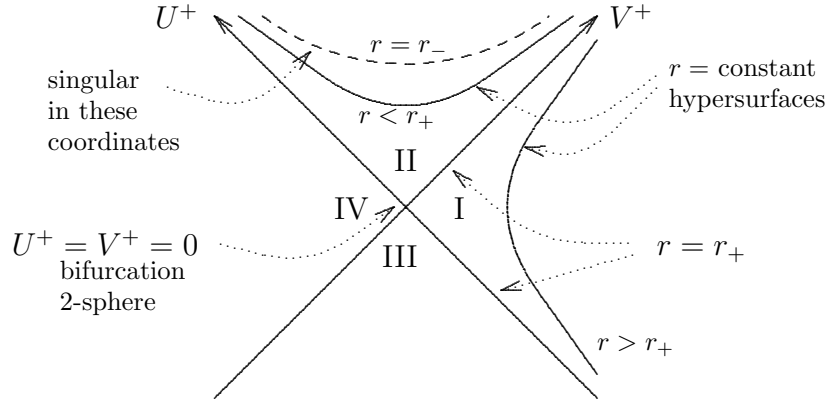
\begin{center}\input{p65-1.pictex}\end{center}
These coordinates do not cover $r\le r_-$ because of the coordinate 
singularity at $r=r_-$ (and $U^+V^+$ is complex for $r<r_-$), but $r=r_-$ and a
similar four regions are covered by the $\left(U^-V^-\right)$ KS-type
coordinates to this case (Exercise).
\bea
ds^2 & = & -\frac{r_+r_-}{\kappa_-^2}\frac{e^{-2\kappa_-r}}{r^2}
\left(\frac{r_+}{r_+-r}\right)^{\frac{\kappa_-}
{\kappa_+}-1}dU^-\,dV^-+r^2d\Omega^2
\\ U^-V^- & = &
-e^{-2\kappa_-r}\left(\frac{r_--r}{r_-}\right)
\left(\frac{r_+-r}{r_+}\right)^{\kappa_-/\kappa_+}
\eea
This metric covers four regions around $U^-=V^-=0$.
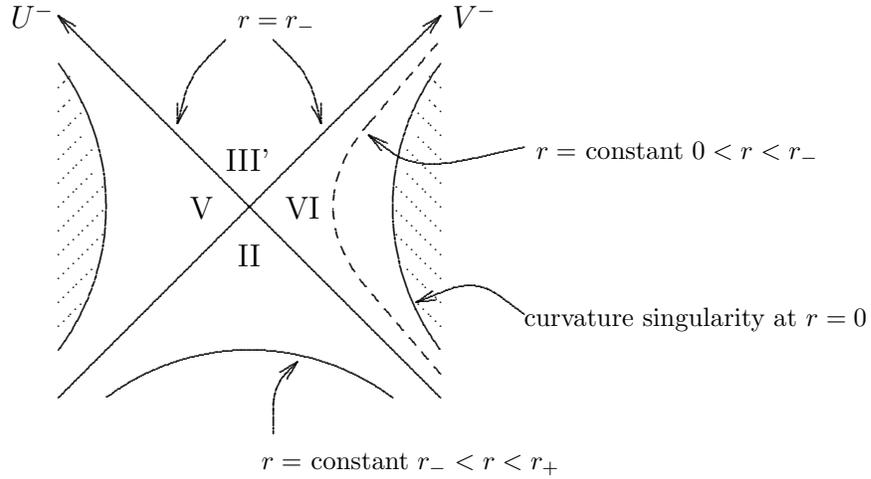
\begin{center}\input{p66-1.pictex}\end{center}
Region II is the same as the region II covered by the 
$\left(U^+,V^+\right)$ coordinates.  The other regions are new.  Regions V and
VI contain the curvature singularity at $r=0$, which is \emph{timelike} because
the normal to $r=$ constant is spacelike for $\Delta>0$, e.g. in $r<r_-$. \\

We know that region II of the diagram is connected to an exterior spacetime 
in the past (regions I, III, and IV),  by time-reversal invariance, region III'
must be connected to another exterior region (isometric regions I', II', and
IV').
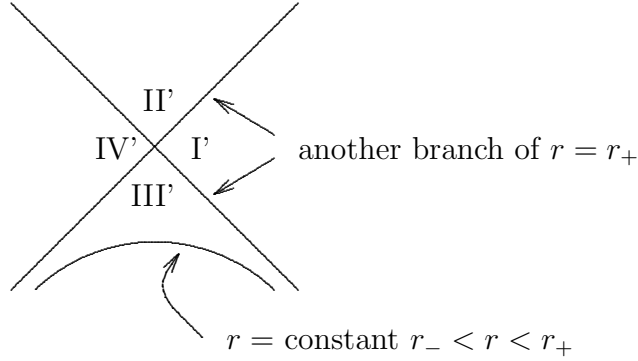
\begin{center}\input{p66-2.pictex}\end{center}
Regions I' and IV' are new asymptotically flat `exterior' spacetimes.  
Continuing in this manner we can find an infinite sequence of them.

\subsubsection{Internal Infinities} 

Consider a path of constant $r,\theta,\phi$ in any region for which 
$\Delta <0$, e.g. region II. In ingoing EF coordinates
\bea
ds^2 & = & -\frac{\Delta}{r^2}dv^2 \\
 & = & \frac{|\Delta|}{r^2}dv^2 \quad \mbox{since are considering 
$\Delta<0$ by hypothesis}
\eea
Since $ds^2>0$ the path is spacelike.  The distance along it from $v=0$ 
to $v=-\infty$ (i.e. to $V^+=0$ or $V^-=0$) is
\bea
s & = & \int^0_{-\infty}\frac{|\Delta|^{1/2}}{r}dv = 
\frac{|\Delta|^{1/2}}{r}\int^0_{-\infty}dv \quad 
\mbox{since $r$ is constant} \\
 & = & \infty
\eea
So there is an `internal' spatial infinity behind the $r=r_+$ horizon.  
(Note that one can still reach $V^{\pm}=0$ in finite proper time on a
\emph{timelike} path, so the null hypersurfaces $V^{\pm}=0$ are part of the
spacetime). \\

If all points at $\infty$, external and internal, are brought to finite 
affine parameter by a conformal transformation, one finds the following CP
diagram, which can be infinitely extended in both directions:
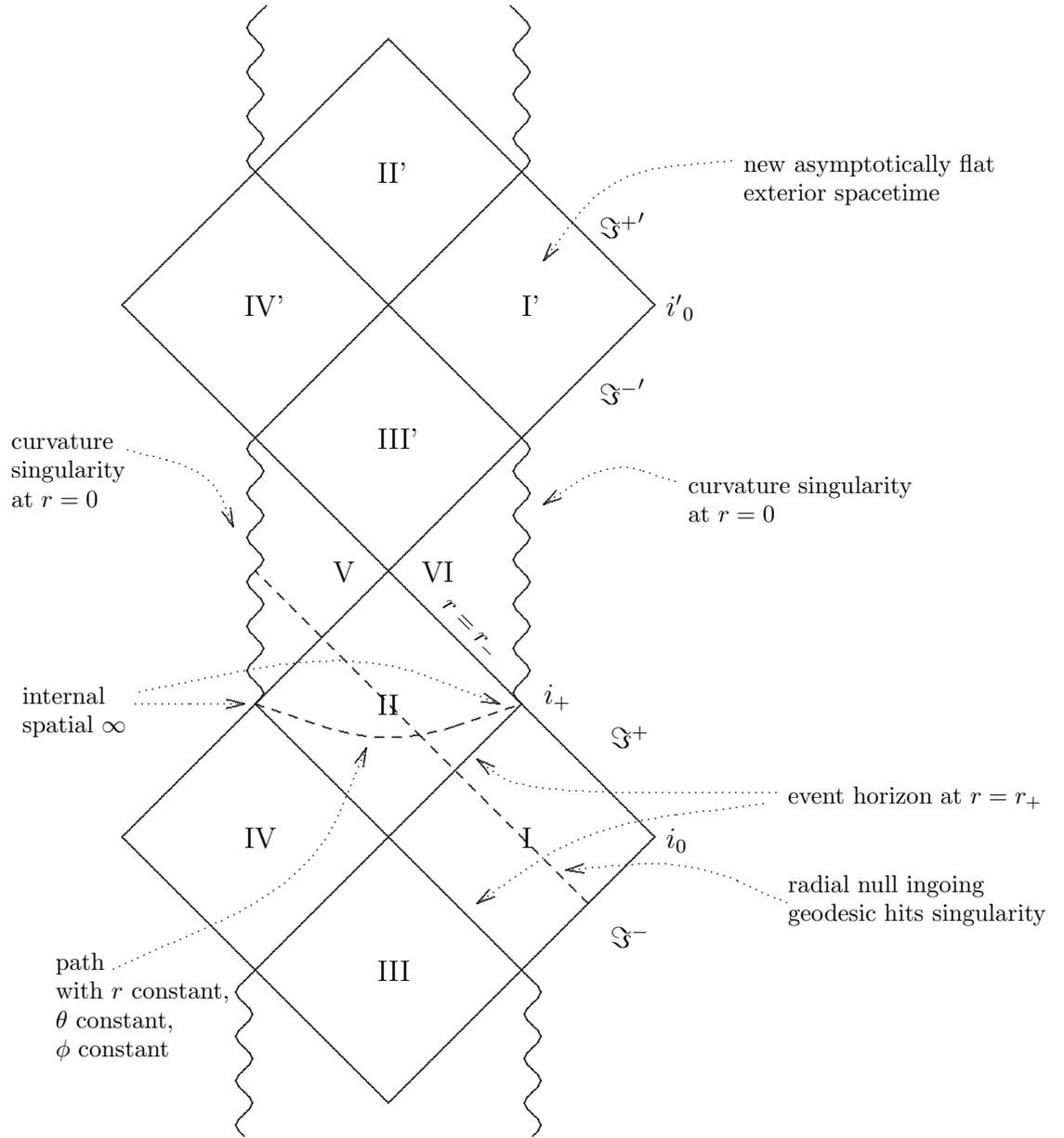
\begin{center}\input{p67-1.pictex}\end{center}
%\end{list} wait for third case 

\section{Pressure-Free Collapse to RN}

Consider a spherical dust ball for which each particle of dust has 
charge/mass ratio 
\be
\gamma=\frac{Q}{M},\quad |\gamma|<1
\ee
where $Q$ is the total charge and $M$ is the total mass.  The exterior 
metric is $M>|Q|$ RN.  The trajectory of a particle at the surface is the same
as that of a radially infalling particle of charge/mass ratio $\gamma$ in the RN
spacetime.  This is \ul{not} a geodesic because of the additional electrostatic
repulsion.  From the result of Question II.4, we see that the trajectory of a
point on the surface obeys
\be
\left(\frac{dr}{d\tau}\right)^2 = \varepsilon^2-V_{\subtext{eff}},
\quad (\varepsilon <1)
\ee
where
\be
V_{\subtext{eff}} = 1-\left(1-\varepsilon \gamma^2\right)\frac{2M}{r}+
\left(1-\gamma^2\right)\frac{Q^2}{r^2}
\ee
\begin{center}\input{p68-1.pictex}\end{center}
\be
r_0=\frac{ \left(1-\gamma^2\right)}{\left(1-\varepsilon\gamma^2\right)}
\frac{Q^2}{M}=\frac{
\gamma^2\left(1-\gamma^2\right)}{\left(1-\varepsilon\gamma^2\right)}M
\ee
The collapse will therefore be halted by the electrostatic repulsion.  
All timelike curves that enter $r<r_+$ must continue to $r<r_-$, so the `bounce'
will occur in region V.  The dust ball then enters region III', explodes as a
white hole into region I' and then recollapses and re-expands indefinitely.

This is illustrated by the following CP diagram
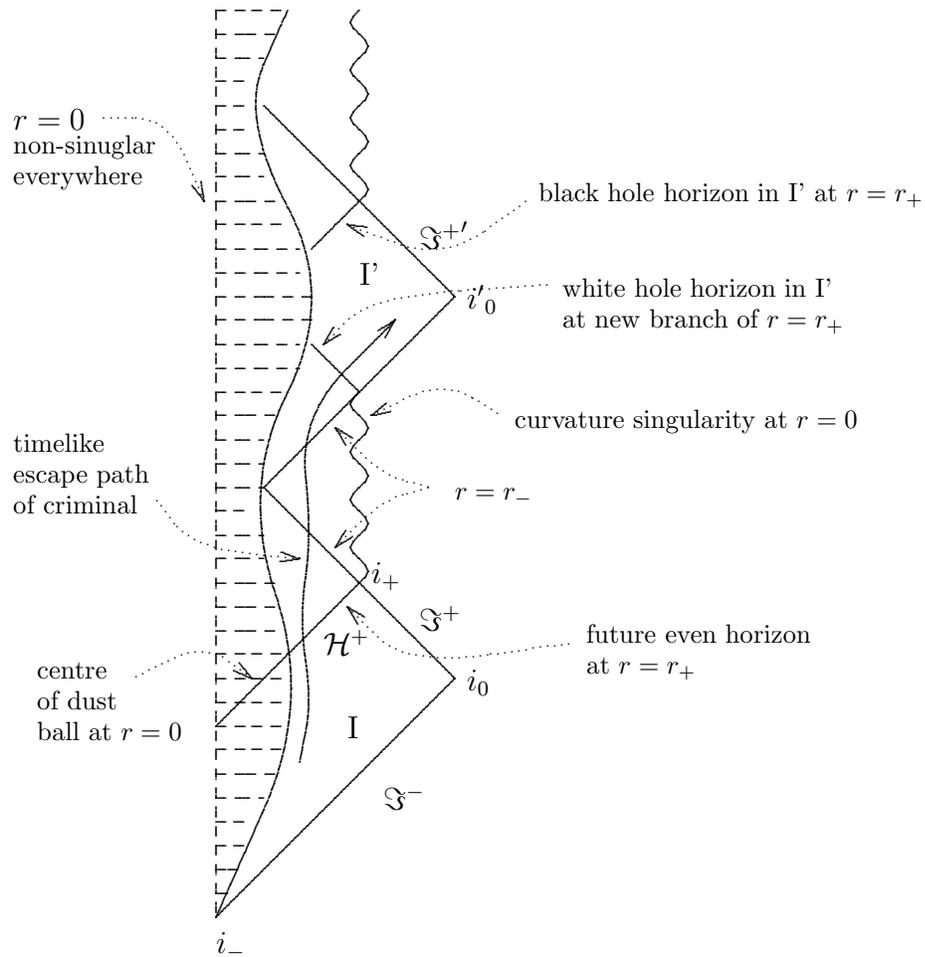
\begin{center}\input{p69-1.pictex}\end{center}
\ul{Notes}
\newcounter{RNnotes}
\begin{list}{\roman{RNnotes})}
{\usecounter{RNnotes}}
\item No singularity is visible from $\scri^+$, in agreement with cosmic 
censorship.
\item Although the dust ball never collapses to zero size and its interior 
is completely non-singular, there is nevertheless a singularity behind
$\mathcal{H}^+$ on another branch of $r=0$, in agreement with the singularity
theorems.
\item It seems that a criminal could escape justice in universe I by escaping 
on a timelike path into universe I'.  Is this science fiction?
\end{list}

\section{Cauchy Horizons}

A particle on an ingoing radial geodesic of RN (e.g. surface of 
collapsing star) will `hit' the singularity at $r=0$, but once in region V or VI
it can accelerate away from the singularity then enter the new exterior region
via the white hole region III'.  However, there is no way to ensure in advance
of entering the black hole (e.g. by programming of rockets) that it will do so
because to get to region I' it must cross a \emph{Cauchy
horizon}\index{Cauchy!horizon}, a concept that will now be elaborated. 

\paragraph{Definition}  A \emph{partial Cauchy surface}
\index{Cauchy!surface!partial}, $\Sigma$, for a spacetime $M$ is a hypersurface
which no causal curve intersects more than once. 

\paragraph{Definition}  A causal curve is \emph{past-inextendable} if 
it has no past endpoint in $M$.

\paragraph{Definition}  The \emph{future domain of dependence}, 
$D^+(\Sigma)$ of $\Sigma$, is the set of points $p\in M$ for which every
past-inextendable causal curve through $p$ intersects $\Sigma$.
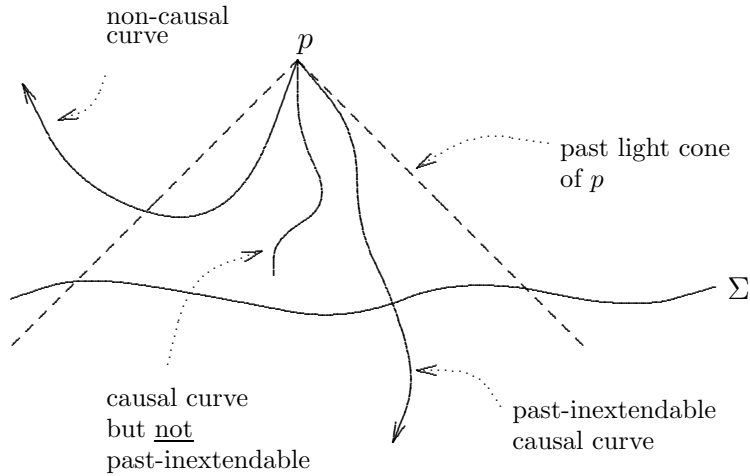
\begin{center}\input{p70-1.pictex}\end{center}
The significance of $D^+(\Sigma)$ is that the behavior of solutions of 
hyperbolic PDE's \emph{outside} $D^+(\Sigma)$ is not determined by initial data
on $\Sigma$. \\

The past domain of dependence, $D^-(\Sigma)$ of $\Sigma$, is defined 
similarly and $\Sigma$ is said to be a \emphin{Cauchy surface} for $M$ if
\be
D^+(\Sigma)\cup D^-(\Sigma) = M
\ee
If $M$ has a Cauchy surface it is said to be \emph{globally hyperbolic}.  
Examples of globally hyperbolic spacetimes are
\newcounter{GHSex}
\begin{list}{\arabic{GHSex})}
{\usecounter{GHSex}}
\item Spherical, pressure-free collapse (Schwarzschild)
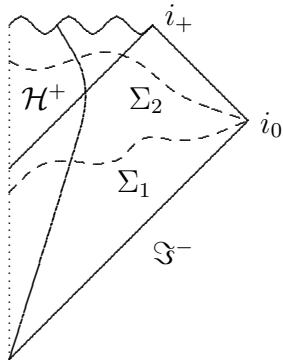
\begin{center}\input{p71-1.pictex}\end{center}
$\Sigma_1$ and $\Sigma_2$ are both Cauchy surfaces. \\

\item Kruskal
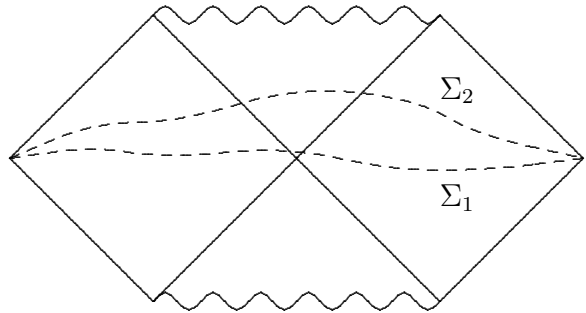
\begin{center}\input{p71-2.pictex}\end{center}
$\Sigma_1$ and $\Sigma_2$ are both Cauchy surfaces. \\
\end{list}
If $M$ is not globally hyperbolic then $D^+(\Sigma)$ or $D^-(\Sigma)$ will 
have a boundary in $M$, called the \emph{future or past Cauchy horizon}.

% end of p71
\subsubsection{Examples}
\newcounter{CHex}
\begin{list}{(\roman{CHex})}
{\usecounter{CHex}}
\item Gravitationally-collapsed charged dust ball.
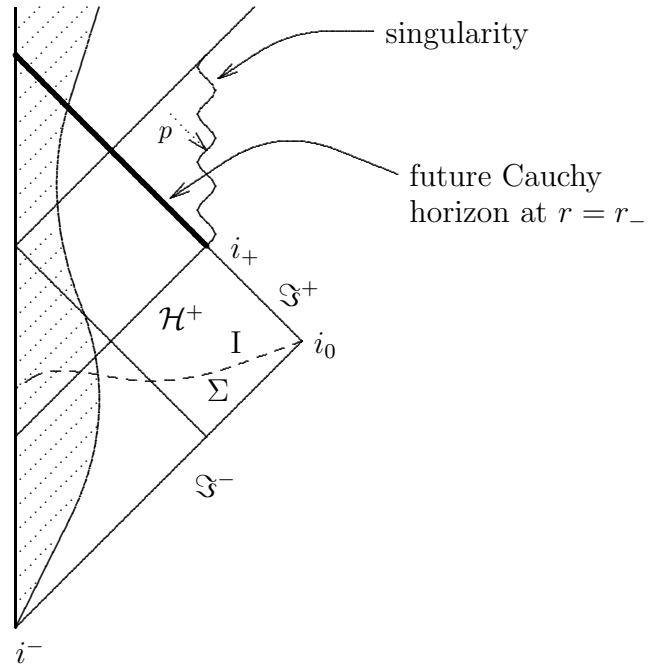
\begin{center}\input{p72-1.pictex}\end{center}
\item Maximal analytic extension of RN
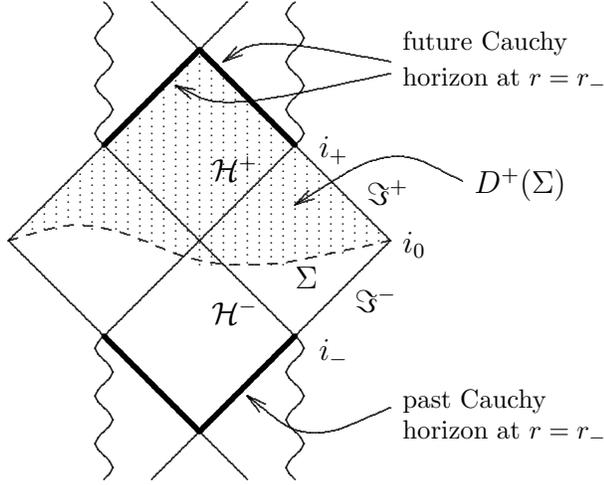
\begin{center}\input{p72-2.pictex}\end{center}
\end{list}

In example (i) a strange feature of the future Cauchy horizon is that the 
entire infinite history of the external spacetime in region I is in its causal
past, i.e. visible, so signals from I must undergo an infinite blueshift as they
approach the Cauchy horizon.  For this reason, the Cauchy horizon usually
becomes singular when subjected to any perturbation, no matter how small.  For
any physically realistic collapse, the Cauchy horizon is a \emph{singular null
hypersurface} for which new physics beyond GR is needed.

\section{Isotropic Coordinates for RN}

Let 
\be
r=\rho+M+\frac{M^2-Q^2}{4\rho}
\ee
Then
\bea
ds^2 & = &-\frac{\Delta dt^2}{r^2(\rho)}+\frac{r^2(\rho)}{\rho^2}
\underbrace{ \left(d\rho^2+\rho^2d\Omega^2\right)}_{\subtext{flat space metric}}
\\
\Delta & = &  \left[ \rho-\frac{\left(M^2-Q^2\right)}{4\rho}\right]^2 
\eea
is RN metric in isotropic coordinates $(t,\rho,\theta,\phi)$.  As in $Q=0$ case, there are \emph{two} values $\rho$ for every value of $r>r_+$, but $\rho$ is complex for $r<r_+$.
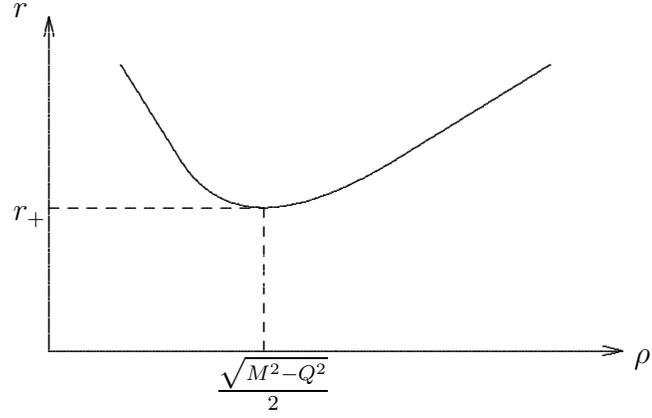
\begin{center}\input{p73-1.pictex}\end{center}
This new metric covers \emph{two} isometric regions (I\&IV) exchanged by the 
geometry.
\be
\rho \to \frac{M^2-Q^2}{4\rho}
\ee
The fixed points set at $\rho=\sqrt{M^2-Q^2}/2$ (i.e. $r=r_+$) is a minimal 
2-sphere of an ER bridge as in the $Q=0$ case.
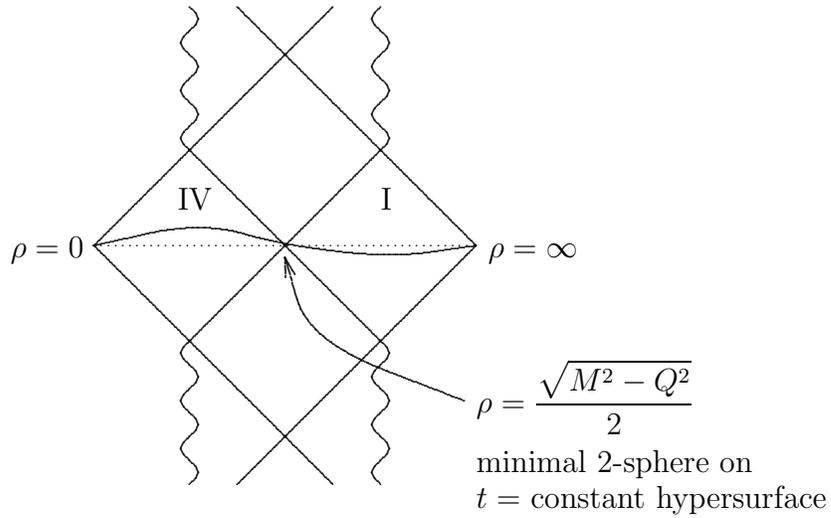
\begin{center}\input{p74-1.pictex}\end{center}
The distance to the horizon at $r=r_+$ along a curve of constant 
$t,\theta,\phi$ from $r=R$ is
\bea
s & = & \int^R_{r_+} \frac{dr}{\sqrt{ \left(1-\frac{r_+}{r}\right)
\left(1-\frac{r_-}{r}\right)}}  \\
 & \to & \infty \quad \mbox{as }r_+-r_-\to 0,\mbox{ i.e. as }M-|Q|\to 0
\eea
so the ER bridge separating regions I \& IV becomes \emph{infinitely long} 
in the limit as $|Q|\to M$.  In this limit, the spatial sections look like:
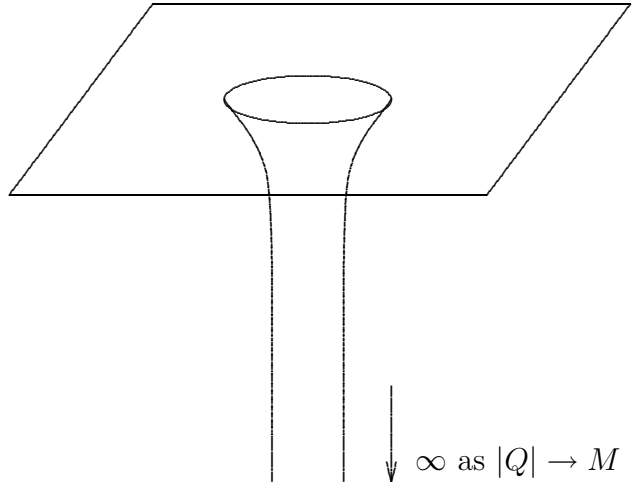
\begin{center}\input{p74-2.pictex}\end{center}

\item \ul{$M=|Q|$ `Extreme' RN ($r_{\pm}=M$)}
\be
ds^2 = -\left(1-\frac{M}{r}\right)^2dt^2+\frac{dr^2}
{\left(1-\frac{M}{r}\right)^2}+r^2d\Omega^2
\ee
This is singular at $r=M$ so define the Regge-Wheeler coordinate
\be
r^*=r+2M\ln\left|\frac{r-M}{M}\right|-\frac{M^2}{r-M} \quad 
\Rightarrow \quad dr^*=\frac{dr}{1-\frac{M}{r}}
\ee
and introduce ingoing EF coordinates as before.  Then
\be
ds^2=-\left(1-\frac{M}{r}\right)^2 dv^2+2dv\,dr+r^2d\Omega^2
\ee
This is non-singular on the null hypersurface $r=M$.

\paragraph{Proposition}  $r=M$ is a \emph{degenerate} (i.e. surface 
gravity $\kappa=0$) Killing horizon\index{Killing!horizon!degenerate} of the
Killing vector field $k=\partial/\partial v$.

\paragraph{Proof}  From the previous calculation $l=f\partial/\partial v$ 
so $r=M$ is a Killing horizon of $k$, and $k\cdot Dk=0$ when $r_+=r_-=M$. 

Since the orbits of $k$ on $r=M$ are affinely parameterized they must go to 
infinite affine parameter in both directions $\Rightarrow$ \emph{internal
$\infty$}.  This is the same internal $\infty$ that we find down the 
infinite ER
bridge.

Note that $k$ is null on $r=2M$, but \emph{timelike everywhere else}, 
so region II has disappeared and region I now leads directly to region V.  
The CP diagram is 
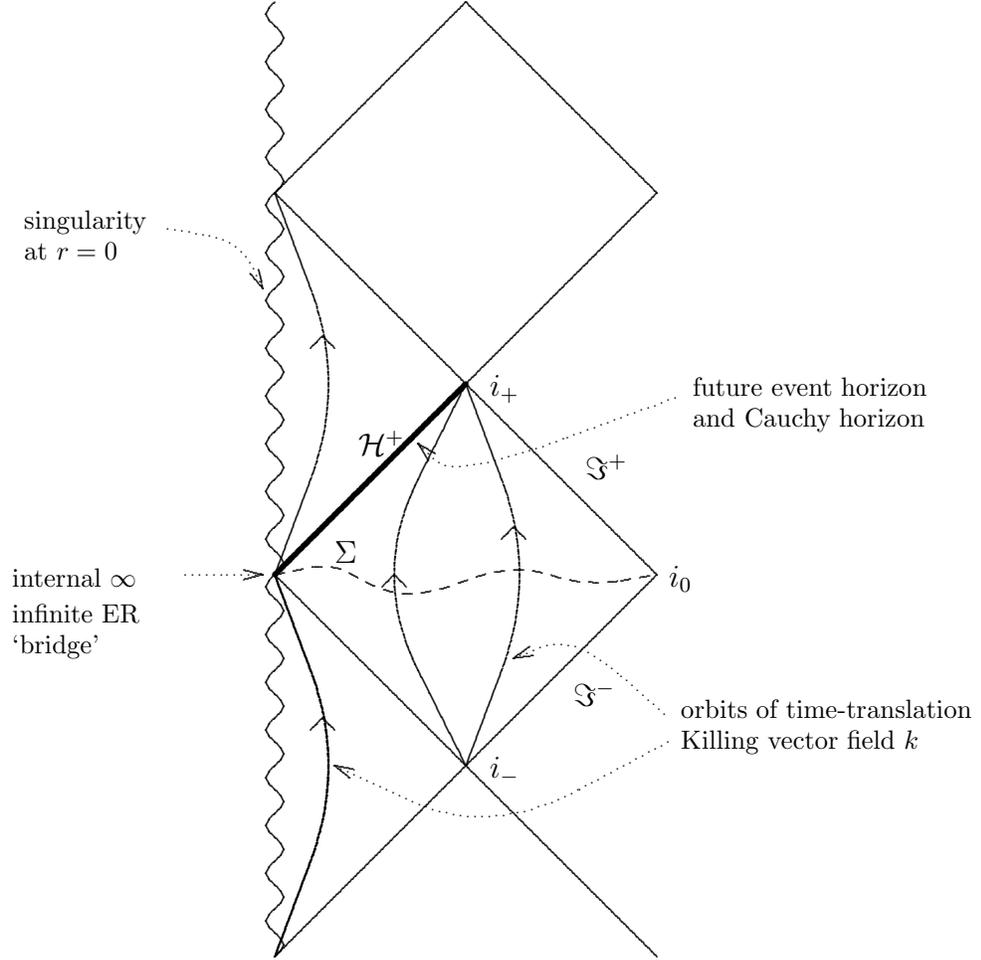
\begin{center}\input{p76-1.pictex}\end{center}

\subsection{Nature of Internal $\infty$ in Extreme RN}

The asymptotic metric as $r\to\infty$ is Minkowski.  To determine the 
asymptotic metric as $r\to M$ we introduce the new coordinate $\lambda$ by
$r=M(1+\lambda)$ and keep only the leading terms in $\lambda$, to get
\bea
F & \sim & d\lambda\wedge d t \\
ds^2 & \sim & \underbrace{ \left(-\lambda^2 dt^2+M^2
\lambda^{-2}d\lambda^2\right)}_{adS_2} +
\underbrace{M^2d\Omega^2}_{\stackrel{\subtext{2-sphere}}{\subtext{ of radius
$M$}}}
\eea
This is the Robinson-Bertotti metric.  It is a kind of `Kaluza-Klein' 
vacuum\index{Kaluza-Klein vacuum} in which two directions are compactified and
the `effective' spacetime is the two-dimensional `anti-de Sitter' ($adS_2$)
spacetime of constant negative curvature.  (See Q.II.7).

\end{list}

\subsection{Multi Black Hole Solutions}

The extreme RN in isotropic coordinates is
\be
ds^2=V^{-2}dt^2+V^2\left(d\rho^2+\rho^2d\Omega^2\right)
\ee
where 
\be
V=1+\frac{M}{\rho}
\ee
This is a special case of the multi black hole solution
\be
ds^2=V^{-2}dt^2+V^2d\vec{x}\cdot d\vec{x}
\ee
where $d\vec{x}\cdot d\vec{x}$ is the Euclidean 3-metric and $V$ is any 
solution of $\nabla^2 V=0$.  In particular,
\be
V=1+\sum_{i=1}^N \frac{M_i}{\left| \vec{x}-\vec{\bar{x}}^i\right|}
\ee
yields the metric for $N$ extreme black holes of masses $M_i$ at positions 
$\bar{x}_i$.  Note that the `points' $\bar{x}_i$ are actually minimal
2-spheres.  There are no $\delta$-function singularities at $x=\bar{x}_i$
because the lines of force continue indefinitely into the asymptotically RB
regions (`charge without charge'). \\

Note that a static multi black hole solution is possible only when there 
is an exact balance between the gravitational attraction and the electrostatic
repulsion.  This occurs only for $M=|Q|$.

\chapter{Rotating Black Holes}
%\section{Rotating Black Holes}

\section{Uniqueness Theorems}\index{uniqueness theorems}

\subsection{Spacetime Symmetries}

\paragraph{Definition}  An asymptotically flat spacetime is 
\emphin{stationary} if and only if there exists a Killing vector field, $k$,
that is timelike near $\infty$ (where we may normalize it s.t. $k^2\to -1$).

i.e. outside a possible horizon, $k=\partial/\partial t$ where $t$ is a 
time coordinate.  The general stationary metric in these coordinates is 
therefore
\be
ds^2= g_{00}(\vec{x})dt^2 + 2g_{0i}(\vec{x})dt\,dx^i+g_{ij}(\vec{x})dx^i\,dx^j
\ee
A stationary spacetime is \emphin{static} at least near $\infty$ if it 
is also \emph{invariant under time-reversal}. This requires $g_{0i}=0$, so the
general static metric can be written as
\be
ds^2= g_{00}(\vec{x})dt^2+g_{ij}(\vec{x})dx^i\,dx^j
\ee
for a static spacetime outside a possible horizon.

\paragraph{Definition}  An asymptotically flat spacetime is 
\emphin{axisymmetric} if there exists a Killing vector field $m$ (an
`axial' Killing vector field) that is spacelike near $\infty$ and for which
\emph{all orbits are closed}.  

We can choose coordinates such that
\be
m=\pd{}{\phi}
\ee
where $\phi$ is a coordinate \emph{identified modulo $2\pi$}, such that
$m^2/r^2\rightarrow 1$ as $r\rightarrow\infty$. Thus, as for $k$, there is
a natural choice of normalization for an axial Killing vector field in an
asymptotically flat spacetime.

%\subsection{Uniqueness Theorems}\index{uniqueness theorems}

\paragraph{Birkhoff's theorem}\index{Birkhoff's theorem} says that any spherically
symmetric vacuum solution is static, which effectively implies that it must be
Schwarzschild.  A generalization of this theorem to the Einstein-Maxwell
system shows that the only spherically symmetric solution is RN. \\

But suppose we know only that the metric exterior to a star is static.  
Unfortunately static $\not\Rightarrow$ spherical symmetry.  However, if the
`star' is actually a black hole we have:

\paragraph{Israel's theorem}\index{Israel's theorem} If $(M,g)$ is an 
asymptotically-flat, \emph{static}, \emph{vacuum} spacetime that is non-singular
on and outside an event horizon, then $(M,g)$ is Schwarzschild. \\

Even more remarkable is the:

\paragraph{Carter-Robinson theorem}\index{Carter-Robinson theorem} If 
$(M,g)$ is an asymptotically-flat \emph{stationary} and \emph{axi-symmetric}
vacuum spacetime that is non-singular on and outside an event horizon, then
$(M,g)$ is a member of the two-parameter Kerr family (given later).  The
parameters are the mass $M$ an the angular momentum $J$. \\

The assumption of axi-symmetry has since been shown to be unnecessary, i.e. 
\emph{for black holes, stationarity $\Rightarrow$ axisymmetry} (Hawking, Wald).
\\

Stationarity $\Leftrightarrow$ equilibrium, so we expect the final state of 
gravitational collapse to be a stationary spacetime.  The uniqueness theorems
say that if the collapse is to a black hole then this spacetime is uniquely
determined by its mass and angular momentum (cf. state of matter in thermal
equilibrium).  Thus, \emph{all multipole moments of the gravitational field are
radiated away} in the collapse to a black hole, except the monopole and dipole
moments (which can't be radiated away because the graviton\index{graviton} has
spin 2). \\

These theorems can be generalized to `vacuum' Einstein-Maxwell equations.  
The result is that a stationary black hole spacetimes must belong to the
3-parameter \emphin{Kerr-Newman family}.  In \emphin{Boyer-Linquist
coordinates}  the KN metric is
\bebox{
\begin{array}{rcl}
ds^2 & = & -\bgfrac{ \left(\Delta -a^2\sin^2\theta\right)}{\Sigma}dt^2 - 2 a 
\sin^2\theta \bgfrac{ \left(r^2+a^2-\Delta\right)}{\Sigma}dt\,d\phi \\
 & & +\left( \bgfrac{ \left(r^2+a^2\right)^2-\Delta a^2\sin^2\theta}{\Sigma}
\right)\sin^2\theta d\phi^2 +\bgfrac{\Sigma}{\Delta}dr^2+\Sigma d\theta^2 
\end{array} }
where
\bebox{
\begin{array}{rcl}
\Sigma & = & r^2+a^2\cos^2\theta \\
\Delta & = & r^2-2Mr+a^2+e^2 \end{array}}
The three parameters are $M$, $a$, and $e$.  It can be shown that
\be
a=\frac{J}{M}
\ee
where $J$ is the total angular momentum, while
\be
e = \sqrt{ Q^2+P^2}
\ee
where $Q$ and $P$ are the electric and magnetic (monopole) charges, 
respectively.  The Maxwell 1-form of the KN solution is 
\bebox{
A= \bgfrac{ Qr\left(dt-a\sin^2\theta d\phi\right)-P\cos\theta
\left[a dt-\left(r^2+a^2\right)d\phi\right] }{\Sigma} }
\subsubsection{Remarks}
\newcounter{KNremarks}
\begin{list}{(\roman{KNremarks})}
{\usecounter{KNremarks}}
\item When $a=0$ the KN solution reduces to the RN solution.
\item Taking $\phi \to -\phi$ effectively changes the sign of $a$, so we 
may choose $a \ge 0$ without loss of generality.
\item  The KN solution has the discrete isometry
\be
t\to -t, \quad \phi \to -\phi
\ee
\end{list}

\section{The Kerr Solution}

This is obtained from KN by setting $e=0$.  Then
\bea
\Delta & = & r^2-2Mr+a^2 \\
(\Sigma & = & r^2+a^2\cos^2\theta ) 
\eea
The Kerr metric\index{Kerr metric} is important astrophysically since it 
is a good \emph{approximation} to the metric of a rotating star at large
distances where all multipole moments except $l=0$ and $l=1$ are unimportant. 
The only known solution of Einstein's equations for which Kerr is \emph{exact}
for $r>R$ is the Kerr solution itself (for which $T_{\mu\nu}=0$), i.e. it has
not been matched to any known non-vacuum solution that could represent the
interior of a star, in contrast to the Schwarzschild solution which is
guaranteed by Birkhoff's theorem to be the exact exterior spacetime that
matches on to the interior solution for any spherically symmetric star. \\

The Kerr metric in BL coordinates has \emph{coordinate} singularities at
\newcounter{Kerrsing}
\begin{list}{(\alph{Kerrsing})}
{\usecounter{Kerrsing}}
\item $\theta=0$ (i.e on axis of symmetry)
\item $\Delta =0$
\end{list}
Write
\be
\Delta=\left(r-r_+\right)\left(r-r_-\right) 
\ee
where
\be
r_{\pm}=M\pm \sqrt{M^2-a^2}
\ee
There are 3 cases to consider
\newcounter{Kerrcases}
\begin{list}{(\roman{Kerrcases})}
{\usecounter{Kerrcases}}
\item \ul{$M^2<a^2$}:  $r_{\pm}$ are complex, so $\Delta$ has no real zeroes, 
and there are no coordinate singularities there.  The metric still has a
coordinate singularity at $\theta=0$.  More significantly, it has a
\emph{curvature singularity} at $\Sigma=0$, i.e.
\be
r=0,\quad \theta = \pi/2
\ee
The nature of this singularity is best seen in Kerr-Schild 
coordinates\index{Kerr-Schild coordinates} $(\tilde{t},x,y,z)$ (which also
removes the coordinate singularity at $\theta=0$).  These are defined by
\bea
x+iy & = & (r+ia)\sin\theta \exp \left[i\int\left(d\phi+
\frac{a}{\Delta}dr\right)\right] \\
z & = & r\cos\theta \\
\tilde{t} & = & \int \left(dt+\frac{r^2+a^2}{\Delta}dr\right)-r
\eea
which implies that $r=r(x,y,z)$ is given implicitly by
\be
r^4-\left(x^2+y^2+z^2-a^2\right)r^2-a^2z^2=0
\ee
In these coordinates the metric is
\bea
 ds^2 &  = & -d\tilde{t}^2+dx^2+dy^2+dz^2  \\
 &  + & \frac{2Mr^3}{r^4+a^2z^2} \left[ \frac{ r(x\,dx+y\,dy)-
a(x\,dy-y\,dx) }{r^2+a^2} + \frac{zdz}{r}+d\tilde{t}\right]^2  \nn
\eea
which shows that the spacetime is flat (Minkowski) when $M=0$.  \\

The surfaces of constant $\tilde{t},r$ are confocal ellipsoids which 
degenerate at $r=0$ to the disc $z=0,\;x^2+y^2\le a^2$.
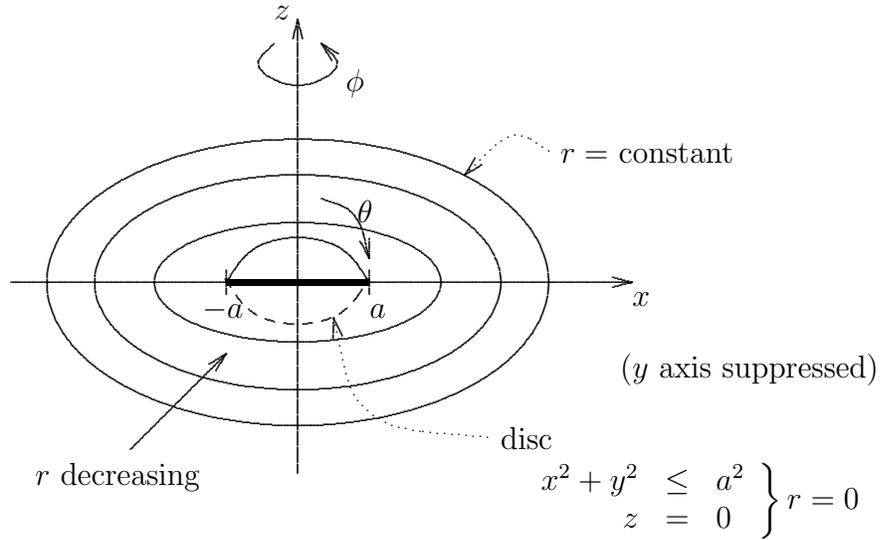
\begin{center}\input{p83-1.pictex}\end{center}
$\theta=\pi/2$ corresponds to the boundary of the disc at $x^2+y^2=a^2$ 
so the curvature singularity occurs on the boundary of the disc, i.e. on
the `ring'
\be
x^2+y^2=a^2,\quad z=0
\ee
There is no reason to restrict $r$ to be positive.  The spacetime can be 
analytically continued through the disc to another asymptotically flat region
with $r<0$.

%\end{list} %(Kerrcases)

\subsubsection{Causal structure}  

Because we now have only axial symmetry we really need a 3-dim spacetime 
diagram to encode the causal structure, but the $\theta=0,\pi/2$ submanifolds
are \emphin{totally-geodesic}, i.e. a geodesic that is initially tangent to the
submanifold remains tangent to it, so we can draw 2-dim CP diagrams for them.
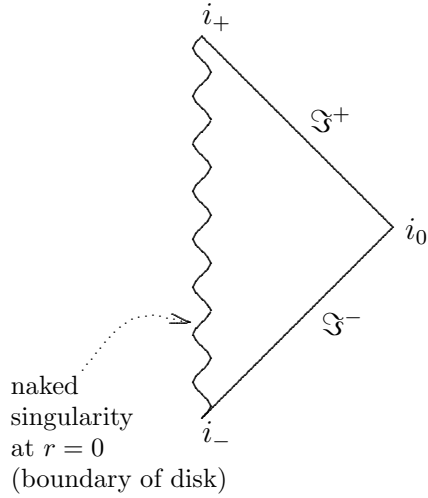
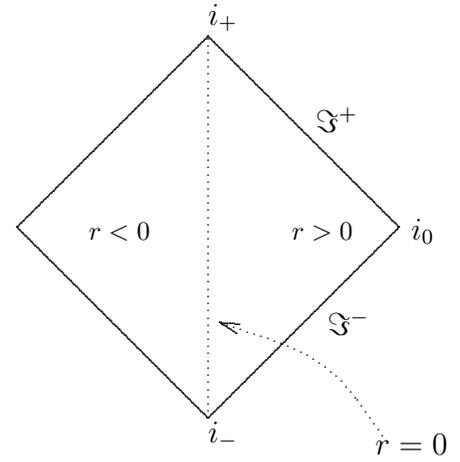
\begin{center}\input{p84-1.pictex}\end{center}
For $\theta=\pi/2$ each point in the diagram represents a circle 
$(0\le\phi< 2\pi)$.  Each ingoing radial geodesic hits the ring singularity at
$r=0$, which is clearly \emph{naked}.  For $\theta=0$ we are considering only
geodesics on the axis of symmetry.  Ingoing radial null geodesics pass through
the disc at $r=0$ into the other region with $r<0$.  We can summarize both
diagrams by the single one.
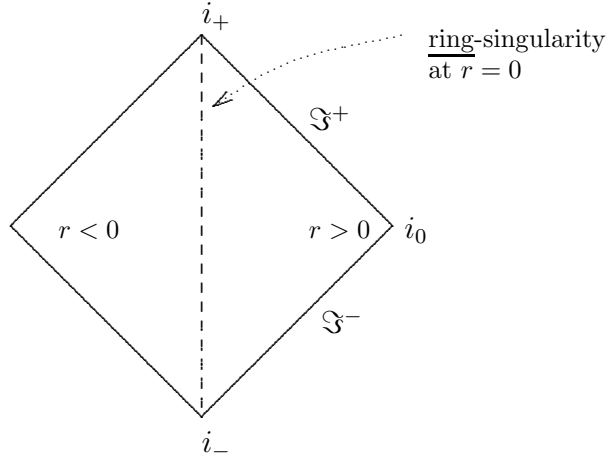
\begin{center}\input{p84-2.pictex}\end{center}
The spacetime is unphysical for another reason.  Consider the norm of the 
Killing vector field $m=\partial/\partial\phi$:
\be
m^2= g_{\phi\phi} =a^2\sin^2\theta\left(1+\frac{r^2}{a^2}\right)+
\frac{Ma^2}{r}\left( \frac{2\sin^4\theta}{1+\frac{a^2}{r^2}\cos^2\theta} \right)
\ee
Let $r/a=\delta$ (small) and consider $\theta=\pi/2+\delta$.  Then
\bea
m^2 & = & a^2+\frac{Ma}{\delta}+\mathcal{O}(\delta), \quad \mbox{for 
$\delta \ll 1$}  \\
 & < & 0 \quad \mbox{for sufficiently small negative $\delta$} \nn
\eea
So \emph{$m$ becomes timelike near the ring-singularity on the $r<0$ branch}.  
But the orbits of $m$ are \emph{closed}, so the spacetime admits closed timelike
curves (CTCs).  This constitute a \emphin{global violation of causality}. \\

Moreover because of the absence of a horizon these CTCs may be deformed to 
pass through \emph{any point} of the spacetime (Carter).  They also miss the
singularity by a distance $\sim M$, for $M\sim a$, and $M$ can be arbitrarily
large.  Since the ring singularity would be naked for $M^2<a^2$, then even if
the white hole region is replaced by a collapsing star, we can invoke cosmic
censorship to rule out $M^2<a^2$.

\item \ul{$M^2>a^2$}.  We still have a ring-singularity but now the metric 
(in BL coordinates) is singular at $r=r_+$ and $r=r_-$.  These are coordinate
singularities.  To see this we define new coordinates $v$ and $\chi$ by
\bea
dv & = & dt+\frac{\left(r^2-a^2\right)}{\Delta}dr \\
d\chi & = & d\phi+\frac{a}{\Delta} dr 
\eea
This yields the Kerr solution in Kerr coordinates $(v,r,\theta,\chi)$ which 
are analogous to ingoing EF for Schwarzschild:
\bebox{
\begin{array}{rcl}
ds^2 & = & -\bgfrac{ \left(\Delta-a^2\sin^2\theta\right)}{\Sigma}dv^2 +
2dv\,dr-\bgfrac{2a\sin^2\theta\left(r^2+a^2-\Delta\right)}{\Sigma} dv\,d\chi \\
 & & -2a\sin^2\theta d\chi\,dr+\bgfrac{ \left[\left(r^2+a^2\right)^2-
\Delta a^2\sin^2\theta \right]}{\Sigma} \sin^2\theta d\chi^2+\Sigma d\theta^2 
\end{array}
}
This metric is non singular when $\Delta = 0$, i.e. when $r=r_+$ or $r=r_-$.

\fbox{\parbox{6in}{
\paragraph{Proposition}  The hypersurfaces $r=r_{\pm}$ are Killing horizons of 
the Killing vector fields
\be
\xi_{\pm}=k+\left(\frac{a}{r_{\pm}^2+a^2}\right)m
\ee
with surface gravities
\be
\kappa_{\pm} = \frac{ r_{\pm}-r_{\mp} }{2\left(r_{\pm}^2+a^2\right)}
\ee 
}} % end parbox

\paragraph{Proof}  Let $\mcN_{\pm}$ be the hypersurfaces $r=r_{\pm}$.  The 
normals are
\bea
l_{\pm} & = & \left. f_{\pm} g^{\mu r}\right|_{\mcN_{\pm}}\partial_{\mu}, 
\quad \mbox{for some non-zero functions $f_{\pm}$} \\
 & = & -\left(\frac{r_{\pm}^2+a^2}{r_{\pm}^2+a^2\cos^2\theta}\right)f_{\pm} 
\underbrace{ \left( \pd{}{v}+\frac{a}{r_{\pm}^2+a^2}\pd{}{\chi} \right)
}_{\displaystyle \xi_{\pm}} \quad \mbox{(Exercise)}
\eea
First
\be
l_{\pm}^2 \propto \left. \left(g_{vv}+\frac{2a}{r^2+a^2}g_{v\chi}+
\frac{a^2}{\left(r^2+a^2\right)^2}g_{\chi\chi}\right) \right|_{\Delta=0} = 0 
\ee
so $\mcN_{\pm}$ are null hypersurfaces.  Since 
$\left.\xi_{\pm}\right|_{\mcN_{\pm}}\propto l_{\pm}$, they are Killing horizons
of $\xi_{\pm}$.  It remains to compute $\xi_{\pm}D\xi_{\pm}^{\mu}$.  This gives
the result for $\kappa_{\pm}$ (Exercise). \\

This result can be used to find KS type coordinates that cover 4 
regions around a BK axis of each Killing horizon, and the $\theta=0$ and
$\theta=\pi/2$ CP diagram of the maximal analytic extension of 
$M^2 > a^2$ Kerr can be found. Note that the diagram can be extended infinitely
in both time directions.
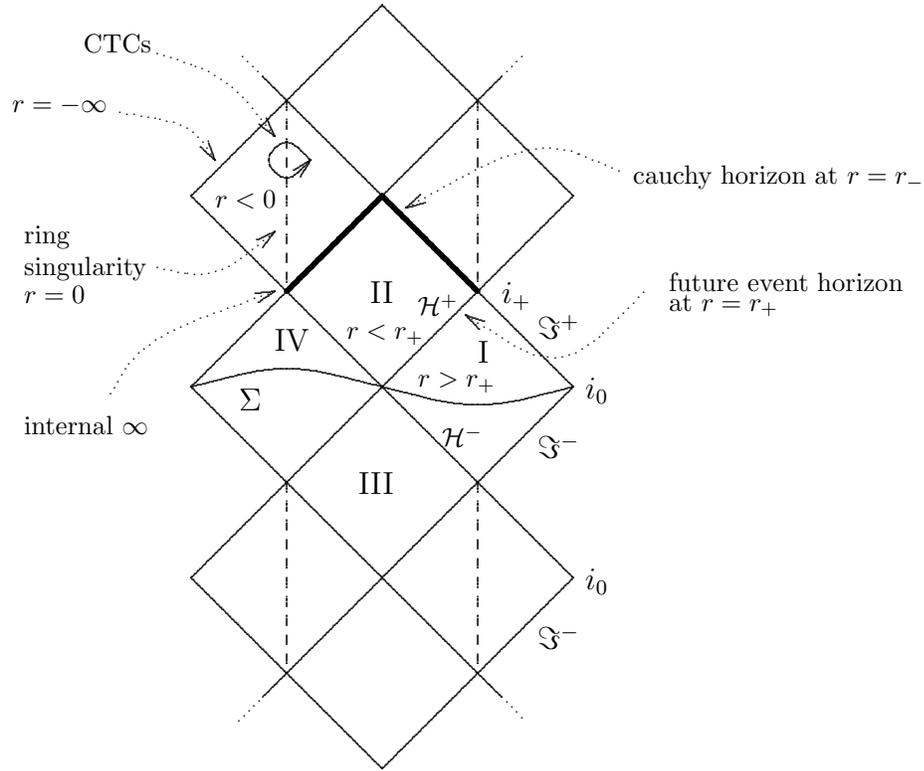
\begin{center}\input{p87-1.pictex}\end{center}

\subsection{Angular Velocity of the Horizon}

The event horizon is a Killing horizon of
\be
\xi=k+\Omega_Hm
\ee
where
\be
\Omega_H = \frac{a}{r_{+}^2+a^2}= \frac{J}{2M\left[M^2+\sqrt{M^4-J^2}\right]}
\ee
In coordinates for which $k=\partial/\partial t$ and 
$m=\partial/\partial\phi$ we have that
\be
\xi^{\mu}\partial_{\mu}\left(\phi-\Omega_H t\right)=0
\ee
i.e. $\phi=\Omega_H t+\mbox{constant}$, on orbits of $\xi$, whereas $\phi$ 
is constant on orbits of $k$.  Note that $k$ is \emph{unique}.  Consider 
\be
(k+\alpha m)^2=g_{tt}+2\alpha g_{t\phi}+\alpha^2g_{\phi\phi}
\ee
As long as $g_{t\phi}$ is finite and $g_{\phi\phi}\sim r^2$ as $r\to\infty$, 
we have $(k+\alpha m)^2\sim \alpha^2r^2>0$ (if $\alpha\neq 0$) as $r\to\infty$. 
So there can be only \emph{one} Killing vector $k$ that is timelike at $\infty$
and normalized s.t. $k^2\to -1$ as $r\to \infty$.

Thus particles on orbits of $\xi$ rotate with angular velocity $\Omega_H$ 
relative to static particles, those on orbits of $k$, and hence relative to a
stationary frame at $\infty$.  Since the null geodesic generators of the horizon
follow orbits of $\xi$ the black hole is rotating with angular velocity
$\Omega_H$.

\paragraph{Lemma} $\xi\cdot k=0$ on  a Killing horizon, $\mcN$, of $\xi$.

\paragraph{Proof} 
\bea
\left. \xi\cdot k\right|_{\mcN} & = &  \left.\xi^2\right|_{\mcN}-\left. 
\Omega_H \xi \cdot m\right|_{\mcN}  \\
 & = & -\left.\Omega_H\xi\cdot m\right|_{\mcN} \quad \mbox{ (since 
$\xi^2=0$ on $\mcN$)}
\eea
Now, $\mcN$ is a fixed point set of $m$, since $m$ is Killing  (Choose 
coordinates s.t. $m=\partial/\partial\phi$.  The metric is $\phi$ independent,
so the position of the horizon is independent of $\phi$).  So $m$ must be
tangent to $\mcN$ or $l\cdot m=0$ where $l$ is normal to $\mcN$.  But
$\xi\propto l$ on $\mcN$, so $\left.\xi\cdot m\right|_{\mcN}=0$.  Hence result.

\subsubsection{Consistency checks (See Question III.3)}

$\xi^2=0$ implies that
\be
k^2+2\Omega_H m\cdot k-m^2\Omega_H = 0, \quad \mbox{on $\mcN$}
\ee
But $\xi\cdot k=0$ implies that
\be
k^2+\Omega_H m\cdot k=0, \quad \mbox{on $\mcN$}
\ee
Consistency requires 
\be
\left.D\equiv (k\cdot m)^2-k^2m^2\right|_{\mcN}=0
\ee
For Kerr, $D=\Delta\sin^2\theta =0$ on $\mcN$ $\Box$.

Also
\bea
\Omega_H & = & -\frac{k^2}{m\cdot k} = \left. -
\frac{g_{tt}}{g_{t\phi}}\right|_{\mcN} \quad \mbox{in BL coordinates} \\
 & = & \frac{-a^2\sin^2\theta}{-2a\sin^2\theta\left(r^2_++a^2\right)} \\
 & = & \frac{a}{r_+^2+a^2} \quad \Box.
\eea

\item \ul{$M^2=a^2$ Extreme Kerr}

In this case we have a \emph{degenerate} ($\kappa=0$) Killing horizon 
at $r=M$ of the Killing vector field
\be
\xi = k+\Omega_H m, \quad \Omega_H=\frac{a}{2M}
\ee 
The CP diagram is
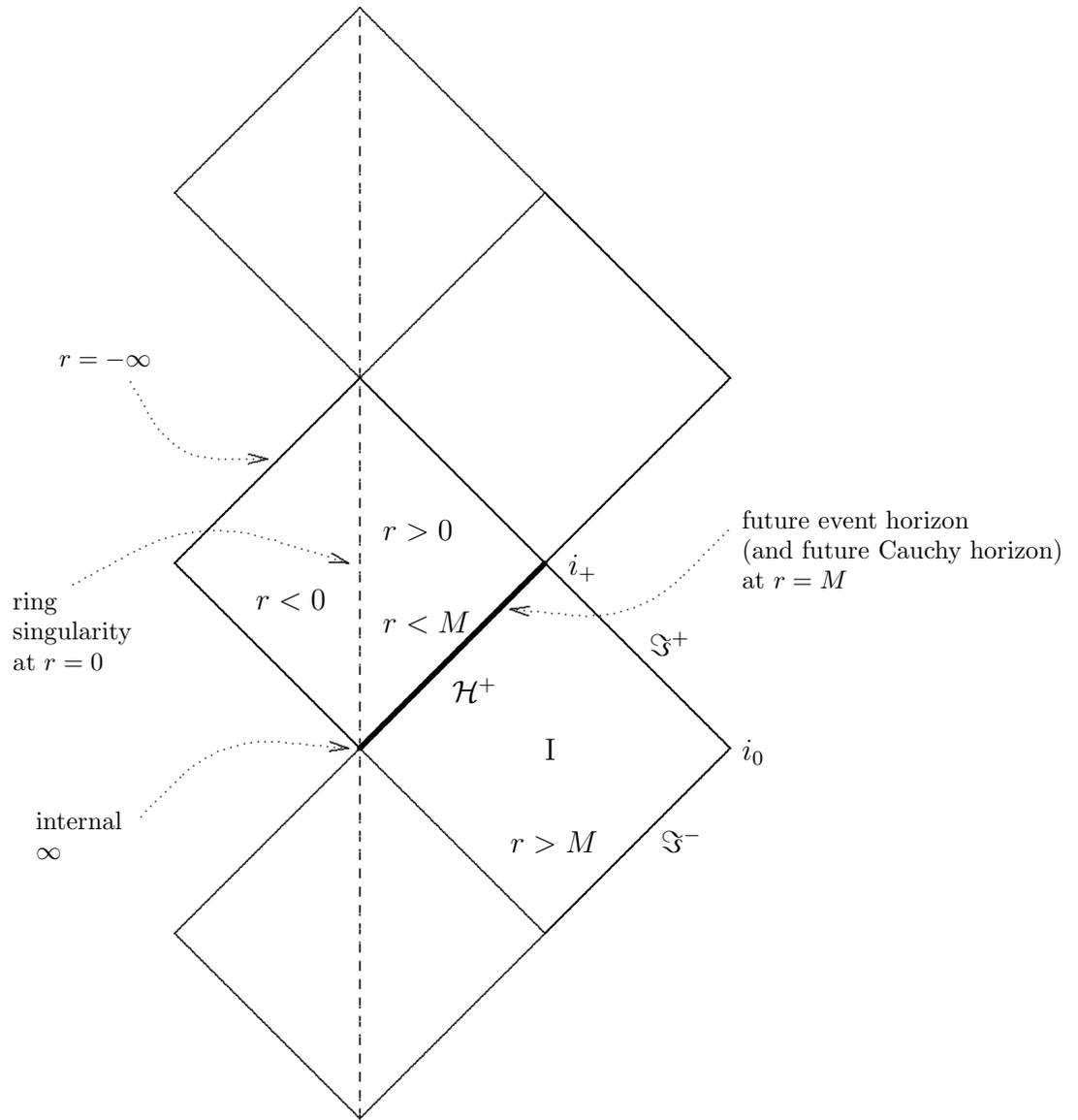
\begin{center}\input{p90-1.pictex}\end{center}

%\paragraph{Uniqueness of $k$} 
%\be
%(k+\alpha m)^2=g_{tt}+2\alpha g_{t\phi}+\alpha^2g_{\phi\phi}
%\ee
%As long as $g_{t\phi}$ is finite and $g_{\phi\phi}\sim r^2$ as $r\to\infty$, 
%we have $(k+\alpha m)^2\sim \alpha^2r^2>0$ (if $\alpha\neq 0$) as $r\to\infty$. 
So there can be only \emph{one} Killing vector $k$ for which $k\cdot k\to -1$ as
$r\to \infty$.

\end{list} %(Kerrcases)

N.B. If you change the sign of $r$ in the Kerr metric this effectively 
changes the sign of $M$.

\section{The Ergosphere}\index{ergosphere}

Although $k$ is timelike at $\infty$ it need not be timelike everywhere 
outside the horizon.  For Kerr,
\be
k^2=g_{tt}=-\frac{ \left(\Delta-a^2\sin^2\theta\right) }
{\Sigma}=-\left( 1-\frac{ 2Mr }{r^2+a^2\cos^2\theta }\right)
\ee
so $k$ is timelike provided that
\be
r^2+a^2\cos^2\theta -2Mr > 0 
\ee
For $M^2 \gg a^2$ this implies that
\be
r>M+\sqrt{M^2-a^2\cos^2\theta}
\ee
(or $r< M-\sqrt{M^2-a^2\cos^2\theta}$, but this is not physically relevant).

The boundary of this region, i.e. the hypersurface 
\be
r=M+\sqrt{M^2-a^2\cos^2\theta}
\ee
is the \emph{ergosphere}.  The ergosphere intersects the event horizon at 
$\theta=0,\pi$, but it lies \emph{outside} the horizon for other values of
$\theta$.  Thus, \emph{$k$ can become spacelike in a region outside the event
horizon}.  This is called the \emphin{ergoregion}.
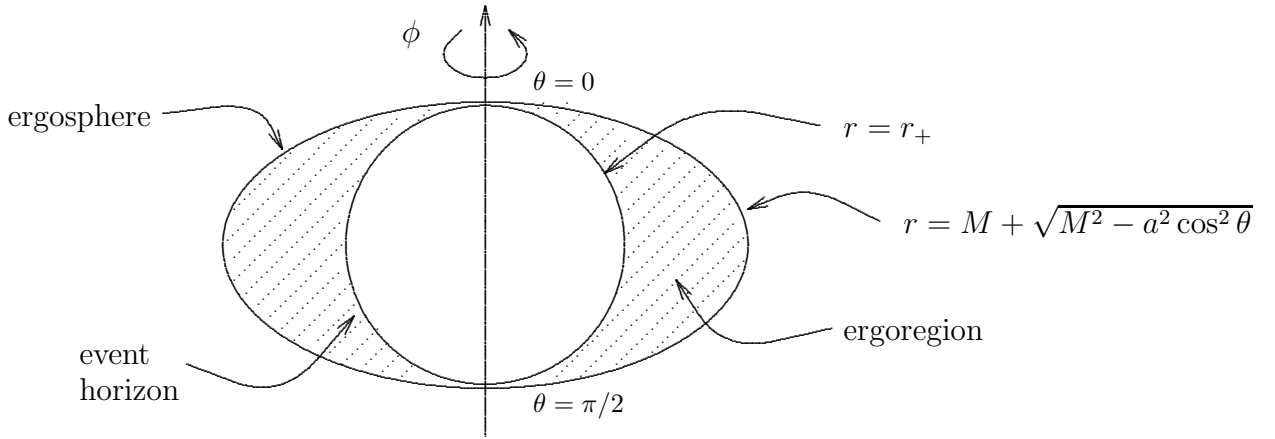
\begin{center}\input{p91-1.pictex}\end{center}

\section{The Penrose Process}\index{Penrose!process}

Suppose that a particle approaches a Kerr black hole along a geodesic.  If 
$p$ is its 4-momentum we can identify the constant of the motion
\be
E=-p\cdot k
\ee
as its energy (since $E=p^0$ at $\infty$).  Now suppose that the particle 
decays into two others, one of which falls into the hole while the other escapes
to $\infty$.
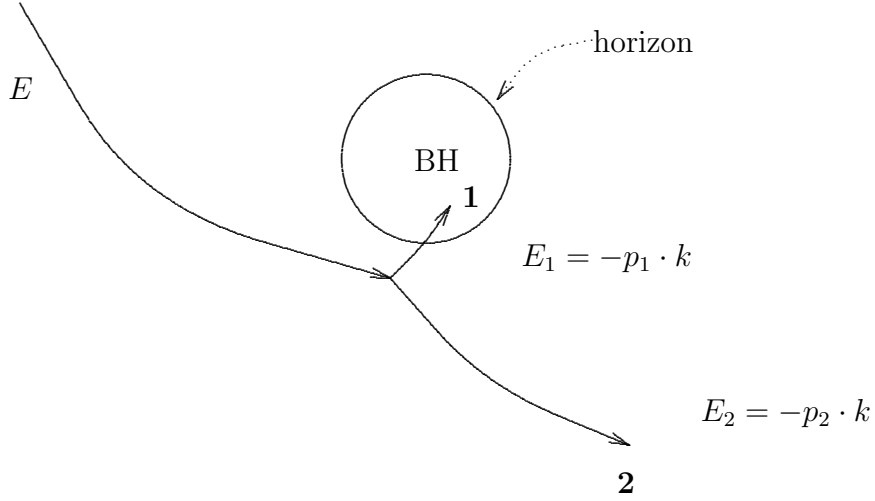
\begin{center}\input{p92-1.pictex}\end{center}
By conservation of energy
\be
E_2=E-E_1
\ee
Normally $E_1>0$ so $E_2 < E$, but in this case
\be
E_1=-p_1\cdot k
\ee
which is \emph{not necessarily positive in the ergoregion} since $k$ may 
be spacelike there.  Thus, if the decay takes place in the ergoregion we may
have $E_2>E$, so \emph{energy has been extracted from the black hole}.

\subsection{Limits to Energy Extraction}

For particles passing through the horizon at $r=r_+$ we have
\be
-p\cdot \xi \ge 0
\ee 
Since $\xi$ is future-directed null on the horizon and $p$ is 
future-directed timelike or null.  Since $\xi = k+\Omega_H m$,
\be
E-\Omega_H L \ge 0 
\ee
where $L=p\cdot m$ is the component of the particle's angular momentum in 
the direction defined by $m$ (only this component is a constant of the motion). 
Thus
\be
L \le \frac{E}{\Omega_H}
\ee
If $E$ is negative, as it is for particle \bold{1} in the Penrose process 
then $L$ is also negative, so the hole's angular momentum is reduced.  We end up
with a hole of mass $M+\delta M$ and angular momentum $J+\delta J$ where $\delta
M=E$ and $\delta J=L$ so
\be
\delta J \le \frac{\delta M}{\Omega_H} = 
\frac{ 2M\left( M^2+\sqrt{ M^4-J^2}\right) }{J} \delta M
\ee 
from formula for $\Omega_H$.  This is equivalent to (Exercise)
\be
\delta\left( M^2+\sqrt{ M^4-J^2}\right) \ge 0 
\ee
(This quantity must increase in the Penrose process).

\paragraph{Lemma}  $A=8\pi\left[M^2+\sqrt{M^4-J^2}\right]$ is the `area of 
the event horizon', of a Kerr black hole (i.e. area of intersection of $\mcH^+$
with partial Cauchy surface, e.g. area of $v=$ constant, $r=r_+$ in Kerr
coordinates (See Question III.5).

\paragraph{Corollary}  Energy extraction by Penrose process is limited by 
the requirement that $\delta A \ge 0$.  This is a special case of the second 
law of black hole mechanics.

\subsection{Super-radiance}\index{super-radiance}

The Penrose process has a close analogue in the scattering of radiation by a 
Kerr black hole.  For simplicity, consider a massless scalar field $\Phi$.  Its
stress tensor is
\be
T_{\mu\nu} = \partial_{\mu}\Phi\partial_{\nu}\Phi-
\half g_{\mu\nu}(\partial\Phi)^2
\ee
Since $D_{\mu}T^{\mu}_{\I\I \nu}=0$ we have
\be
D_{\mu}\left(T^{\mu}_{\I\I\nu}k^{\nu}\right) = T^{\mu\nu}D_{\mu}k_{\nu}=0
\ee
so we can consider
\be
j^{\mu}=-T^{\mu}_{\I\I\nu}k^{\nu} = -\partial^{\mu}\Phi k\cdot 
\partial\Phi+\half k^{\mu}(\partial\Phi)^2
\ee
as the future directed ($-k\cdot J>0$) energy flux 4-vector of $\Phi$. 
Now consider the following region, $S$, of spacetime with a null hypersurface
$\mcN \subset \mcH^+$ as one boundary.
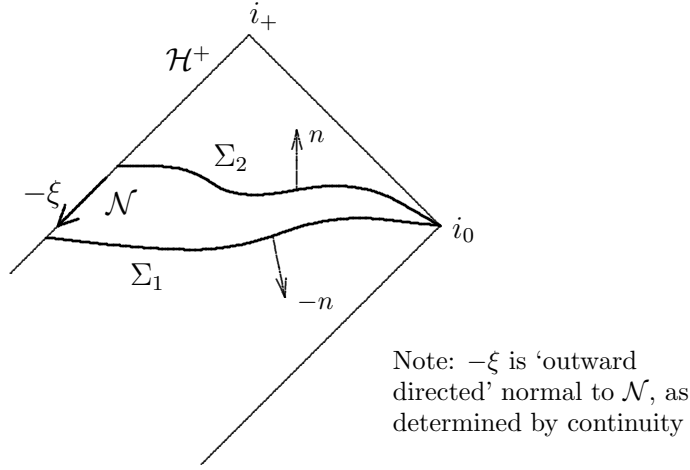
\begin{center}\input{p94-1.pictex}\end{center}
Assume that $\partial\Phi=0$ at $i_0$.  Since $D_{\mu}j^{\mu}=0$ we have
\bea
0 & = & \int_S \dx{4}{x} \sqrt{-g}D_{\mu}j^{\mu}=\int_{\partial S} 
dS_{\mu}\,j^{\mu} \\
 & = & \int_{\Sigma_2} dS_{\mu}\,j^{\mu}-\int_{\Sigma_1} 
dS_{\mu}\,j^{\mu} -\int_{\mcN} dS_{\mu}\,j^{\mu} \\
 & = & E_2 -E_1 - \int_{\mcN} dS_{\mu}\,j^{\mu}
\eea
where $E_i$ is the energy of the scalar field on the spacelike 
hypersurface $\Sigma_i$.  The energy going through the horizon is therefore
\bea
\Delta E = E_1-E_2 & = & -\int_{\mcN}dS_{\mu}\, j^{\mu} \\
 & = & -\int dA\,dv\,\xi_{\mu}j^{\mu}, \quad \mbox{($v$ is Kerr coordinate)}
\eea
The energy flux lost/unit time (power) is therefore
\bea
P & = & -\int dA\, \xi_{\mu}j^{\mu}=\int dA\, 
(\xi\cdot\partial\Phi)(k\cdot D\Phi) \\
\lefteqn{ \mbox{(since $\xi\cdot k=0$ on horizon by previous Lemma)} } 
\nn \\
 & = & \int dA\, \left(\pd{}{v}\Phi+\Omega_H \pd{}{\chi}\Phi\right)
\left(\pd{\Phi}{v}\right)
\eea
For a wave-mode of angular-frequency $\omega$
\be
\Phi=\Phi_0 \cos\left(\omega v- \nu\chi\right), \quad \nu\in\Z \quad 
\mbox{(angular quantum no.)}
\ee
The time average power lost across the horizon is
\be
P=\half \Phi_0^2 A\omega(\omega-\nu\Omega)
\ee
where $A$ is the area of the horizon. 

$P$ is positive for most values of $\omega$, but for $\omega$ in the range
\be
0 < \omega < \nu\Omega_H
\ee
it is negative, i.e. a wave-mode with $\omega,\nu$ satisfying the 
inequality is \emph{amplified} by the black hole.

\subsubsection{Remarks}
\newcounter{superradrem}
\begin{list}{\roman{superradrem})}
{\usecounter{superradrem}}
\item Process is positive only for $\nu\neq 0$ because the amplified field 
must also take away angular momentum from the hole.

\item Process is similar to stimulated emission in atomic physics, which
suggests the possibility of a spontaneous emission effect. This can be shown to
occur in the quantum theory so any black with an ergoregion cannot be stable
quantum mechanically.

\item  We have neglected the back-reaction of $\Phi$ on the metric.  When 
corrected for back-reaction the metric can be stationary only if
$\partial\Phi/\partial\phi=0$, but then $j^{\mu}=0$ and the black hole energy
doesn't change, i.e. strictly speaking super-radiance is incompatible with
stationarity.

\end{list}

\chapter{Energy and Angular Momentum}

\section{Covariant Formulation of Charge Integral}

In the usual Minkowski space formulation with charge density $\rho(\vec{x},t)$, 
the charge in a volume $V$ is written as
\bea
Q  & = &  \int_V dV\,\rho = \int_V dV\, \vec{\nabla}\cdot \vec{E} \quad \mbox{by Maxwell's eqs.}  \\
Q  & = &  \oint_{\partial V} d\vec{S}\cdot \vec{E} \quad \mbox{by Gauss' law}
\eea
where surface integral is over boundary of $V$.  Note that,
\be
\vec{\nabla}\cdot\vec{E} = \frac{1}{\sqrt{\Dim{3}g}}
\partial_i\sqrt{\Dim{3}g}E^i, \quad dV =\dx{3}{x}\sqrt{\Dim{3}g}
\ee
where $\Dim{3}g$ is the determinant of the 3-metric, so
\be
\int dV\vec{\nabla}\cdot \vec{E}=\int \dx{3}{x}\partial_i
\left(\sqrt{\Dim{3}g}E^i\right) = \int dS_i E^i\, .
\ee
The Lorentz covariant formulation uses the similar result 
\be
\frac{1}{\sqrt{-\Dim{4}g}}\partial_{\mu}\left(\sqrt{-\Dim{4}g}
F^{\mu\nu}\right) = D_{\mu}F^{\mu\nu}\, .
\ee
The volume $V$ is replaced by an arbitrary spacelike hypersurface $\Sigma$
(partial Cauchy surface) with boundary $\partial\Sigma$.  The volume element on
$\Sigma$ is a \emph{non-spacelike} co-vector (1-form) $dS_{\mu}$.  Given the
current density 4-vector $j^{\mu}(x)$ we write
\be
Q=\int_{\Sigma}dS_{\mu}j^{\mu}
\ee
We can choose $\Sigma$ (at least locally) to be $t=$ constant, in which 
case $dS_{\mu}=(dV,\vec{0})$.  Since $j^0=\rho$, we recover the previous
expression for $Q$.  Now use Maxwell's equations.  $D_{\nu}F^{\mu\nu}=j^{\mu}$
to rewrite $Q$ as
\bea
Q & = & \int_{\Sigma} dS_{\mu} D_{\nu}F^{\mu\nu} \\
 & = & \half \oint_{\partial\Sigma} dS_{\mu\nu}F^{\mu\nu} \quad 
\mbox{by Gauss' law}
\eea
where $dS_{\mu\nu}$ is the area element of $\partial\Sigma$.  When $\Sigma$ 
is $t=$ constant the only non-vanishing components of $dS_{\mu\nu}$ are
\be
dS_{0i}=-dS_{i0}\equiv dS_i
\ee
in which case
\be
Q= \oint_{\partial\Sigma}dS_i\, F^{0i}
\ee
But $F^{0i}=-F^{i0}=E^i$, so we recover the previous formula.

\section{ADM energy}\index{ADM energy}

We cannot define energy in the same way because this is associated with a 
conserved \emph{symmetric tensor} $T^{\mu\nu}$, rather than a vector.  This is
not unexpected because a \emph{locally conserved energy can exist only in a
spacetime admitting a timelike Killing vector field}. \\

[Unlike photons, which do \ul{not} carry charge, gravitons\index{graviton} 
\emph{do} carry energy $\Rightarrow$ possibility of energy exchange between
matter and its gravitational field.] \\

We can still define a \emph{total} energy in asymptotically flat spacetimes 
as a surface integral at infinity because $\partial/\partial t$ is
asymptotically Killing in such spacetimes.  In this case
\be
g_{\mu\nu} \to \eta_{\mu\nu} \quad \mbox{as }r\to\infty \quad 
(\mbox{$\eta_{\mu\nu}$ Minkowski metric})
\ee
We shall assume that, \emph{in Cartesian coordinates},
\be
h_{\mu\nu}=g_{\mu\nu}-\eta_{\mu\nu}=\mathcal{O}\left(\frac{1}{r}\right)
\ee
which will justify a linearization of Einstein's equations near $\infty$.

\paragraph{Exercise}  Show that $G_{\mu\nu}=8\pi GT_{\mu\nu}$ becomes the 
Pauli-Fierz equation\index{Pauli-Fierz equation}
\bebox{
\Box h_{\mu\nu}+h_{,\mu\nu}-2h_{(\mu,\nu)}=-16\pi G\left(T_{\mu\nu}-\half 
\eta_{\mu\nu}T\right)
\label{eq:ADMdagger}
}
where
\bea
\Box & = & \eta^{\mu\nu}\partial_{\mu}\partial_{\nu} \\
h & = & \eta^{\mu\nu} h_{\mu\nu} \\
h_{\mu} & = & \eta^{\nu\rho}h_{\rho\mu,\nu} = h^{\nu}_{\I \mu,\nu} \\
T & = &  \eta^{\mu\nu}T_{\mu\nu}
\eea
Take the trace to get
\bebox{
\Box h - h^{\mu}_{\I,\mu} = 8\pi GT
\label{eq:ADMstar}
}

We shall first consider a \emphin{weak static dust}  source
\be
T_{\mu\nu} = \left( \begin{array}{c|ccc}
\rho & & 0 & \\ \hline
 & & & \\
0 & & 0 & \\
 & & & \end{array} \right) \quad \mbox{zero pressure for `dust'}
\ee
\bean
\begin{array}{ccl}
\dot{\rho}  =  0 \quad & \mbox{for} & \mbox{\emph{static}}  \\
\left.\begin{array}{rcl}
4\pi G\rho & \ll & 1 \\
T_{0i} & = & 0 \end{array} \right\} & \mbox{for} & \mbox{\emph{weak}} 
\end{array}
\eean
Since source is static we may assume static $h_{\mu\nu}$, i.e. 
$\dot{h}_{\mu\nu}=0$.  Then $\mu=\nu=0$ component of (\ref{eq:ADMdagger})
becomes 
\be
\nabla^2h_{00}=-8\pi G T_{00} 
\label{eq:ADM1}
\ee 
while (\ref{eq:ADMstar}) becomes
\be
-\nabla^2h_{00}+\underbrace{ \nabla^2h_{jj} - 
h_{ij,ij}}_{\displaystyle \partial_i\left(\partial_i h_{jj}-\partial_j
h_{ij}\right)} = -8\pi GT_{00}
\label{eq:ADM2}
\ee
Add (\ref{eq:ADM1}) and (\ref{eq:ADM2}) to get
\be
T_{00}=\frac{1}{16\pi G}\partial_i\left(\partial_jh_{ij}-
\partial_ih_{jj}\right) \quad \mbox{(Cartesian coordinates)}
\ee
Since the source is weak we can assume that the spacetime is almost 
Minkowski, i.e. we treat $h_{\mu\nu}$ as a field on Minkowski spacetime. The 
total energy is now found by integrating $T_{00}$ over all space.
\be
E = \int_{\stackrel{\subtext{$t=$ constant}}{\subtext{all space}}} 
\dx{3}{x} T_{00}
\ee
  Using Gauss' law we can rewrite result as the surface integral
\be
E=\frac{1}{16\pi G}\oint_{\infty}dS_i \left(\partial_j h_{ij}-
\partial_i h_{jj}\right) \quad \mbox{(Cartesian coordinates)}
\ee
But this depends \emph{only} on the asymptotic data, so we may now change 
the source in any way we wish in the interior without changing $E$, provided
that the asymptotic metric is unchanged.  So \emph{formula for $E$ is valid in
general}. \\

This is the ADM formula for the energy of asymptotically flat spacetimes.

\subsection{Alternative Formula for ADM Energy}

Subtract (\ref{eq:ADM2}) from (\ref{eq:ADM1}) to get
\be
\partial_i\left(\partial_j h_{ij}-\partial_i h_{jj}\right)=-2\nabla^2 h_{00}
\ee This allows us to rewrite ADM formula as
\be
E=-\frac{1}{8\pi G} \oint_{\infty} dS_i\, \partial_i h_{00}
\ee
But (Exercise)
\be
g^{ij}\Gamma_{0j}^{\I\I 0} = -\half \partial_i h_{00} +\mathcal{O}
\left(\frac{1}{r^3}\right) \quad \mbox{($\Gamma=$ affine connection)}
\ee
and hence
\bea
E & = & \frac{1}{4\pi G}\oint_{\infty} dS_i\, g^{ij}\Gamma_{0j}^{\I\I 0} \\
 & = & \frac{1}{4\pi G}\oint_{\infty} dS_{0i}\, D^ik^0 \quad \mbox{where } 
k=\pd{}{t},\; dS_i\equiv dS_{0i}
\eea
But $k$ is asymptotically Killing, i.e.
\be
D^{\mu}k^{\nu}+D^{\nu}k^{\mu} = \mathcal{O}\left(\frac{1}{r^3}\right)
\ee
so
\be
E= -\frac{1}{8\pi G}\oint_{\infty} dS_{\mu\nu}D^{\mu}k^{\nu}
\ee

\section{Komar Integrals}\index{Komar integrals}

Let $V$ be a volume of spacetime on a spacelike hypersurface $\Sigma$, with 
boundary $\partial V$.  To every \emph{Killing} vector field $\xi$ we can
associate the Komar integral
\be
Q_{\xi}(V) = \frac{c}{16\pi G}\oint_{\partial V}dS_{\mu\nu}D^{\mu}\xi^{\nu} 
\ee
for some constant $c$.  Using Gauss' law
\be
Q_{\xi}(V) = \frac{c}{8\pi G}\int_V dS_{\mu}D_{\nu}D^{\mu}\xi^{\nu}
\ee
\fbox{\parbox{6in}{
\paragraph{Lemma} $D_{\nu}D_{\mu}\xi^{\nu} = R_{\mu\nu}\xi^{\nu}$ for 
Killing vector field $\xi$.
\paragraph{Proof} By contraction of previous `Killing vector Lemma.'
}} \\

Using Lemma,
\bea
Q_{\xi}(V) & = & \frac{c}{8\pi G} \int_V dS_{\mu}R^{\mu}_{\I\nu}\xi^{\nu}  \\
 & = & c\int dS_{\mu}\left(T^{\mu}_{\I\I\nu}\xi^{\nu}-\half 
T\xi^{\mu}\right) \quad \mbox{(by Einstein's eqs.)} \\
 & = & \int dS_{\mu}J^{\mu}(\xi)
\eea
where
\be
J^{\mu}(\xi) = c\left(T^{\mu}_{\I\I\nu}\xi^{\nu}-\half T\xi^{\mu}\right)
\ee
\paragraph{Proposition} $\partial_{\mu}J^{\mu}(\xi)=0$.
\paragraph{Proof} Using $D_{\mu}T^{\mu\nu}=0$ we have
\bea
D_{\mu}J^{\mu} & = & c\underbrace{ \left(T^{\mu\nu}D_{\mu}\xi_{\nu}-
\half TD_{\mu}\xi^{\mu}\right) }_{\displaystyle \mbox{0 for Killing vector
$\xi$}} -\frac{c}{2}\xi\cdot \partial T \\
 & = & \frac{c}{2}\xi \cdot\partial R \quad \mbox{(by Einstein's eqs.)} \\
 & = & 0 \quad \mbox{for Killing vector field $\xi$}
\eea
(In this last step, choose coordinates s.t. $\xi\cdot\partial=
\partial/\partial\alpha$, then the metric is $\alpha$-independent
($\fltpd{g_{\mu\nu}}{\alpha}=0$), so $R$ is too ($\fltpd{R}{\alpha}=0$)). \\

Since $J^{\mu}(\xi)$ is a `conserved current', the charge $Q_{\xi}(V)$ is 
time-independent provided $J^{\mu}(\xi)$ vanishes on $\partial V$, 
\emph{just as
for electric charge}.
\paragraph{Exercise} $\xi=k$ (time-translation Killing vector field)
\bebox{
E(V)=-\frac{1}{8\pi G}\oint_{\partial V}dS_{\mu\nu}D^{\mu}k^{\nu}
}
i.e. $c=-2$, is fixed by comparison with previous formula derived for total 
energy, i.e. by choosing $V=$ 2-sphere at spatial $\infty$.

\fbox{\parbox{6in}{
\paragraph{Exercise} Verify that $E(V)=M$ for Schwarzschild, for any $V$ with 
$\partial V$ in exterior $(r>2M)$ spacetime.
}}

\subsection{Angular Momentum in Axisymmetric Spacetimes}

Return to Komar integral.  Let $\xi=m=\fltpd{}{\phi}$ and choose $c=1$ to get
\bebox{
J(V)=\frac{1}{16\pi G}\oint_{\partial V}dS_{\mu\nu}D^{\mu}m^{\nu}
}
\emph{Note here factor of $-1/2$ relative to Komar integral for the energy}.

To check coefficient, use Gauss' law to write $J(V)=\int_V dS_{\mu}J^{\mu}(m)$ 
where
\be
J^{\mu}(m)=T^{\mu}_{\I\I\nu}m^{\nu}-\half Tm^{\mu}
\ee
If we choose $V$ to be on $t=$ constant hypersurface, and 
$m=\partial/\partial\phi$, then $dS_{\mu}m^{\mu}=0$, so
\be
J(V) = \int_{V}dV T^0_{\I\I\nu}m^{\nu} = \int_V 
dV\left(T^0_{\I\I 2}x^1-T^0_{\I\I 1}x^2 \right)
\ee 
in Cartesian coordinates $\left\{x^i; i=1,2,3\right\}$ where 
\be
m=x^1\pd{}{x^2}-x^2\pd{}{x^1}
\ee
For a \emph{weak source}, $g\approx \eta$ and
\be
J(V) \approx \varepsilon_{3jk}\int_V \dx{3}{x}x^j T^{k0}
\ee
which is result for $3^{\subtext{rd}}$ component of angular momentum of field 
in Minkowski spacetime with stress tensor $T_{\mu\nu}$. 

So the \emph{total} angular momentum of an asymptotically flat spacetime is 
found by taking $\partial V$ to be a 2-sphere at spatial infinity
\bebox{
J=\frac{1}{16\pi G}\oint_{\infty}dS_{\mu\nu}D^{\mu}m^{\nu}
}

\section{Energy Conditions}

\fbox{\parbox{6in}{
$T_{\mu\nu}$ satisfies the \emphin{dominant energy condition} if for 
\emph{all} future-directed timelike vector fields $v$, the vector field
\be
j(v) \equiv -v^{\mu}T_{\mu}^{\I\nu}\partial_{\nu}
\ee
is future-directed non-spacelike, or zero.
}} \\

All physically reasonable matter satisfies this condition, e.g. for massless 
scalar field $\Phi$ (with $T_{\mu\nu}=
\partial_{\mu}\Phi\partial_{\nu}\Phi-\half
g_{\mu\nu}(\partial\Phi)^2$):
\bea
j^{\mu}(v) & = & -v\cdot \partial\Phi 
\partial^{\mu}\Phi+\half v^{\mu}(\partial\Phi)^2 \\
j^2(v) & = & \frac{1}{4}v^2\underbrace{ 
\left((\partial\Phi)^2\right)^2}_{\ge 0} \le 0 \quad \mbox{if $v^2<0$}
\eea
so $j(v)$ is timelike or null if $v$ is timelike.  Since $v$ is 
assumed future-directed, $j(v)$ will be too if $-v\cdot j >0$.  Allowing for
$j=0$ means that we have to prove that $-v\cdot j\ge 0$.  Now
\bea
-v\cdot j & = & (v\cdot\partial\Phi)^2-\half v^2(\partial\Phi)^2 \\
 & = & \half(v\cdot \partial\Phi)^2+\half\left(-v^2\right)
\left(\partial\Phi-\frac{v(v\cdot\partial\Phi)}{v^2}\right)^2 
\eea
But $(v\cdot\partial\Phi)^2\ge 0$ and $-v^2>0$ for timelike $v$, so we 
have to prove that 
\be
\left(\partial\Phi-\frac{v(v\cdot\partial\Phi)}{v^2}\right)^2 \ge 0
\ee
i.e. that $\left(\partial\Phi-\frac{v(v\cdot\partial\Phi)}{v^2}\right)$ is 
spacelike or zero.  This follows from
\be
v\cdot \left(\partial\Phi-\frac{v(v\cdot\partial\Phi)}{v^2}\right) = 0
\ee
since $v\cdot V < 0$ for any non-zero timelike or null vector for timelike 
$v$ (choose coordinates s.t. $v=(1,\vec{0})$).  So if $v\cdot V=0$ then $V$
cannot be timelike or null. \\

Since $-v\cdot j=v^{\mu}v^{\nu}T^{\mu\nu}$, the dominant energy condition 
implies that $v^{\mu}v^{\nu}T_{\mu\nu} \ge 0$ for all timelike $v$.  By
continuity it also implies the

\fbox{\parbox{6in}{
\emph{Weak energy condition}\index{weak energy condition}
\be
v^{\mu}v^{\nu}T_{\mu\nu}\ge 0 \quad \forall\; \mbox{non-spacelike $v$}
\ee
}}\\

There is also the 

\fbox{\parbox{6in}{
\emph{Strong energy condition}\index{strong energy condition}
\be
v^{\mu}v^{\nu}\left(T_{\mu\nu}-\half g_{\mu\nu}T\right)\ge 0 \quad 
\forall\;\mbox{non-spacelike $v$}
\ee
}}\\

Note, \emph{Dominant $\not\Leftrightarrow$ Strong}. \\

The strong energy condition is needed to prove the singularity theorems, but 
the dominant energy condition is the physically important one.  (An inflationary
universe violates the strong energy condition).  For example it is needed for 
the

\subsubsection{Positive Energy Theorem (Shoen \& Yau, Witten)}
\index{positive energy theorem}

The ADM energy of an asymptotically-flat spacetime satisfying 
$G_{\mu\nu}=8\pi GT_{\mu\nu}$ is positive semi-definite, and vanishes
\emph{only} for Minkowski spacetime with $T_{\mu\nu}=0$, provided that 
\newcounter{positenergy}
\begin{list}{\roman{positenergy})}
{\usecounter{positenergy}}
\item $\exists$ an initially non-singular Cauchy surface (otherwise 
$M<0$ Schwarzschild would be a counter-example).

\item $T_{\mu\nu}$ satisfies the dominant energy condition (clearly, 
\emph{some} condition on $T_{\mu\nu}$ is necessary).

\item Some other technical assumptions which we ignore here.
\end{list}

\chapter{Black Hole Mechanics}

\section{Geodesic Congruences}

\paragraph{Definition}  A congruence\index{congruence} is a family of curves 
such that precisely one curve of the family passes through each point.  It is a
geodesic congruence\index{congruence!geodesic}\index{geodesic!congruence} if the
curves are geodesics. \\

The equations of a geodesic congruence may be written as 
$x^{\mu}=x^{\mu}\left(y^{\alpha},\lambda\right)$ where the parameters
$y^{\alpha}, \alpha=0,1,2$ label the geodesic and $\lambda$ is an affine
parameter\index{affine parameter} on the geodesic, i.e. 
\be
t= \frac{d}{d\lambda}=\pd{x^{\mu}}{\lambda}\partial_{\mu}
\ee
is the tangent to the geodesics such that $t\cdot Dt^{\mu}=0$. Since the
parameter $\lambda$ is affine, $t^2\equiv -1$ for timelike geodesics (while
$t^2\equiv 0$ for null geodesics). The vectors
\be
\eta_{\alpha} = \frac{d}{dy^{\alpha}}=\pd{x^{\mu}}{y^{\alpha}}\partial_{\mu}
\ee
may be considered as a basis of `displacement' vectors across the congruence:
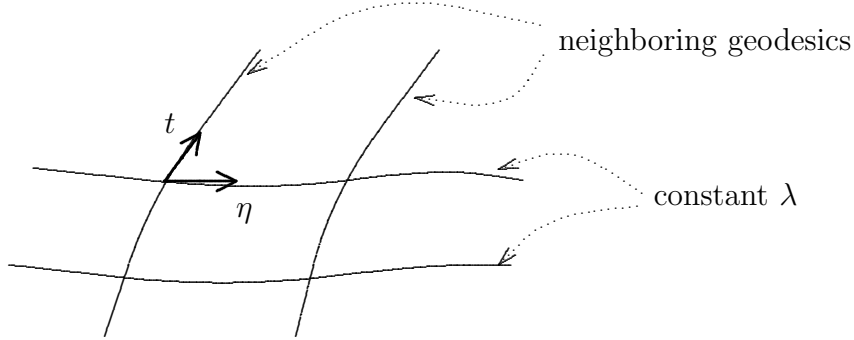
\begin{center}\input{p107-1.pictex}\end{center}
Note that $t$ and $\eta_{\alpha}$ commute (since we could choose 
coordinates $x^{\mu}$ s.t. $t=\fltpd{}{\lambda}$ and
$\eta_{\alpha}=\fltpd{}{y^{\alpha}}$), so
\bea
0 & = & t^{\nu}\partial_{\nu}\eta_{\alpha}^{\mu}-
\eta^{\nu}_{\alpha}\partial_{\nu}t^{\mu} \\
 & = & t^{\nu}\left(\partial_{\nu}\eta^{\mu}_{\alpha}+
\Gamma^{\mu}_{\I\sigma\nu}\eta^{\sigma}_{\alpha}\right)-
\eta^{\nu}_{\alpha}\left(\partial_{\nu}t^{\mu}+
\Gamma^{\mu}_{\I\sigma\nu}t^{\sigma}\right)
\\
 & = & t^{\nu}D_{\nu}\eta^{\mu}_{\alpha}-
\eta^{\nu}_{\alpha}D_{\nu}t^{\mu} \quad \mbox{(by symmetry of connection)}
\eea
or
\bebox{
t^{\nu}D_{\nu}\eta^{\mu}_{\alpha}=B^{\mu}_{\I \nu}\eta^{\nu}_{\alpha}
}
where
\be
B^{\mu}_{\I \nu}=D_{\nu}t^{\mu}
\ee
measures the failure of the displacement vectors $\eta_{\alpha}$ to be 
paralelly-transported along the geodesics, i.e. it measures \emph{geodesic
deviation}\index{geodesic!deviation}. \\

A geodesic nearby some fiducial geodesic may now be specified by a displacement
vector $\eta$, but this specification is not \emph{unique} because
$\eta'=\eta+at$ ($a=$ constant) is a displacement vector to the \emph{same}
geodesic.
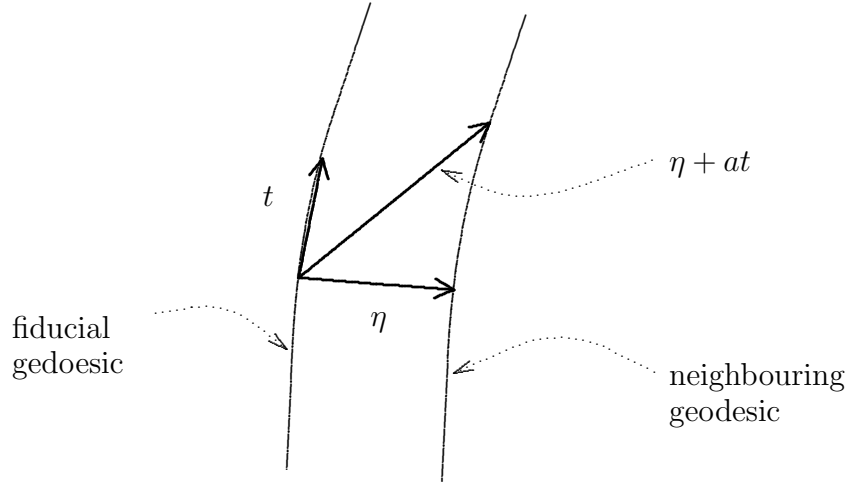
\begin{center}\input{p108-1.pictex}\end{center}
For timelike geodesics we can remove this ambiguity by requiring 
$\eta$ to be orthogonal to $t$, i.e.
\bebox{
\eta\cdot t=0 }
Strictly, speaking we can only make such a choice at a given value of 
$\lambda$, by choosing the origin of $\lambda$ across the congruence. However
\bea
\frac{d}{d\lambda}(\eta\cdot t) & = & 
\left(t\cdot D\eta^{\mu}\right)t_{\mu} 
\quad \mbox{(since $t\cdot Dt_{\mu}=0$)} \\
 & = & B^{\mu}_{\I\nu} \eta^{\nu} t_{\mu} = 
\left(\eta^{\nu}D_{\nu}t^{\mu}\right)t_{\mu} \\
 & = & \half \eta\cdot \partial t^2 =0 \, ,
\eea
since $t^2\equiv -1$ for timelike congruences, so if $\eta\cdot t$ is chosen to
vanish at one value of $\lambda$ it will do so for all $\lambda$. \\

For null congruences\index{congruence!null} the condition $\eta\cdot t=0$ is
not sufficient to eliminate the ambiguity in the choice of $\eta$ because 
\bea
\eta'\cdot t & = & (\eta+at)\cdot t = \eta\cdot t + a t\cdot t \\
 & = & \eta\cdot t 
\eea
when $t^2=0$, which means that $\eta'\cdot t=0$ whenever $\eta\cdot t=0$. 
The problem is that \emph{the 3-dim space of vectors orthogonal to $t$ now
includes $t$ itself}, so the displacement vectors $\eta$ orthogonal to $t$ 
specify only a \emph{\emph{two}-parameter family of geodesics}.  Displacement
vectors to the other null geodesics in the congruence have a component in the
direction of a vector $n$ that is \ul{not} orthogonal to $t$. The choice of $n$
is otherwise arbitrary (it is analogous to the choice of gauge in
electrodynamics), but it is \emph{convenient} to choose it such that
\bebox{n^2=0, \quad n\cdot t=-1} e.g. if $t$ is tangent to an outgoing radial
null geodesic, then $n$ is  tangent to an ingoing one.
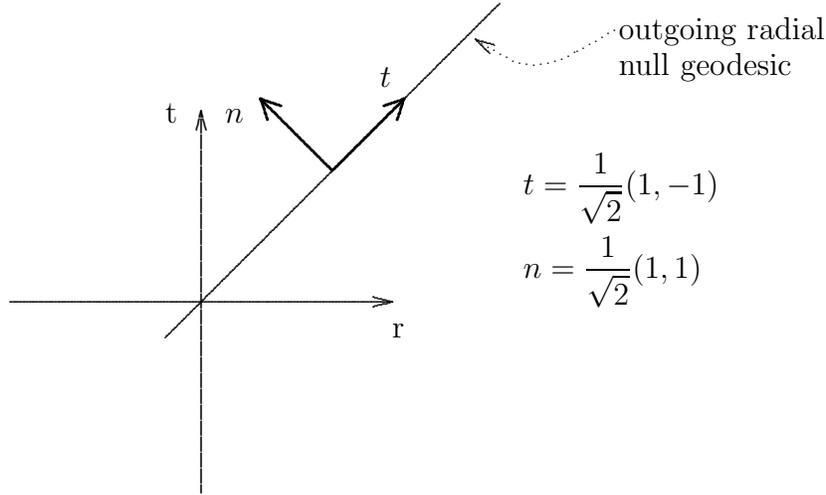
\begin{center}\input{p109-1.pictex}\end{center}
Consistency of the choice of $n$ requires that $n^2$ and $n\cdot t$ be 
independent of $\lambda$, which is satisfied if 
\bebox{t\cdot Dn^{\mu}=0}
i.e. we choose $n$ to be parallely-transported along the geodesics. \\

Having made a choice of the vector $n$, we may now uniquely specify a
two-parameter subset of geodesics of a null geodesic congruence by 
displacement vectors $\eta$ orthogonal to $t$ by requiring them to also
satisfy 
\bebox{
\eta\cdot n = 0
}
The vectors $\eta$ now span a two-dimensional subspace, $T_{\perp}$, of the
tangent space, that is orthogonal to both $t$ and $n$, i.e. $P\eta=\eta$, where
\be
P^{\mu}_{\I\nu} = \delta^{\mu}_{\I\nu}+n^{\mu}t_{\nu}+t^{\mu}n_{\nu}
\ee
projects onto $T_{\perp}$.
\smallskip

\fbox{\parbox{6in}{
\paragraph{Proposition} $P\eta=\eta\Rightarrow t\cdot D\eta^{\mu}=
\hat{B}^{\mu}_{\I\nu} \eta^{\nu}$, where
\be
\hat{B}^{\mu}_{\I \nu} = P^{\mu}_{\I\lambda}B^{\lambda}_{\I\rho}
P^{\rho}_{\I\nu}
\ee
i.e. if $\eta\in T_{\perp}$ initially, it remains in this subspace. \\
}}
\paragraph{Proof}
\bea
t\cdot D\eta^{\mu} & = & t\cdot D\left(P^{\mu}_{\I\nu}\eta^{\nu}\right) 
\quad \mbox{(if $P\eta=\eta$)} \\
 & = & P^{\mu}_{\I\nu} t\cdot D\eta^{\nu} \quad 
\mbox{(since $t\cdot Dn=t\cdot Dt=0$)} \\
 & = & P^{\mu}_{\I\nu} B^{\nu}_{\I\rho}\eta^{\rho} 
\quad \mbox{(by definition)} \\
 & = & P^{\mu}_{\I\nu} B^{\nu}_{\I\rho} P^{\rho}_{\I\lambda} 
\eta^{\lambda} \quad \mbox{(since $P\eta=\eta$)} \\
 & = & \hat{B}^{\mu}_{\I\nu} \eta^{\nu} \quad \Box.
\eea

$\hat{B}$ is effectively a $2\times 2$ matrix.  We now decompose it into its 
algebraically irreducible parts
\be
\hat{B}^{\mu}_{\I\nu}=\half \theta P^{\mu}_{\I\nu}+
\hat{\sigma}^{\mu}_{\I\nu}+\hat{\omega}^{\mu}_{\I\nu} 
\ee
where

\begin{tabular}{rclcc}
$\theta$ & = & $\hat{B}^{\mu}_{\I\mu}$ & (trace) & 
\emph{expansion} \\
$\hat{\sigma}_{\mu\nu}$ & = & $\hat{B}_{(\mu\nu)}-\half P_{\mu\nu}
\hat{B}^{\rho}_{\I\rho}$ & (symmetric, traceless) & \emph{shear} \\
$\hat{\omega}_{\mu\nu}$  & = & $\hat{B}_{[\mu\nu]}$  & 
(anti-symmetric) & \emph{twist}
\end{tabular}
\smallskip

Notation:
\bean
\hat{B}_{(\mu\nu)} & = & \half\left(\hat{B}_{\mu\nu}+\hat{B}_{\nu\mu}\right) \\
\hat{B}_{[\mu\nu]} & = & \half\left(\hat{B}_{\mu\nu}-\hat{B}_{\nu\mu}\right)
\eean

\paragraph{Lemma} $t_{[\mu}\hat{B}_{\nu\rho]} = t_{[\mu}B_{\nu\rho]}$

\paragraph{Proof} Using $t\cdot Dt=0$ and $t^2 = 0$, we have
\be
\hat{B}^{\mu}_{\I\nu} = B^{\mu}_{\I\nu}+
t^{\mu}\left(n_{\lambda}B^{\lambda}_{\I\nu}+n_{\lambda}
B^{\lambda}_{\I\rho}n^{\rho}t_{\nu}\right)+
\left(B^{\mu}_{\I\rho}n^{\rho}\right)t_{\nu}
\ee
Hence result. ($\left[\quad\right]$ indicates total anti-symmetrization 
on enclosed indices).
\smallskip

\fbox{\parbox{6in}{
\paragraph{Proposition} The tangents $t$ are normal to a family of null 
hypersurfaces iff $\hat{\omega}=0$. 
}}

\paragraph{Proof}  If $\hat{\omega}=0$, then
\bea
0 & = & t_{[\mu}\hat{\omega}_{\nu\rho]} \equiv t_{[\mu}\hat{B}_{\nu\rho]} \\
 & = & t_{[\mu}B_{\nu\rho]} \quad \mbox{(by Lemma)} \\
 & = & t_{[\mu}D_{\rho}t_{\nu]} 
\eea
so $t$ is normal to a family of hypersurfaces by Frobenius' 
theorem\index{Frobenius' theorem}.  (In this case we can take $t=l$).

Conversely, if $t$ is normal to a family of null hypersurfaces, then 
Frobenius' theorem implies $t_{[\mu}D_{\nu}t_{\rho]}=0$.  Then, reversing the
previous steps we find that,
\be
0 = t_{[\mu}\hat{\omega}_{\nu\rho]} = 
\frac{1}{3}\left(t_{\mu}\hat{\omega}_{\nu\rho}+
t_{\rho}\hat{\omega}_{\mu\nu}+t_{\nu}\hat{\omega}_{\rho\mu}
\right) 
\ee
Contract with $n$.  Since $n\cdot t=-1$ and $n\hat{\omega}=
\hat{\omega}n=0$ (because $\hat{\omega}$ contains the projection operator $P$),
we deduce that $\hat{\omega}=0$.  \\

If $\hat{\omega}=0$ we have a family of null hypersurfaces.  The family 
is parameterized by the displacement along $n$
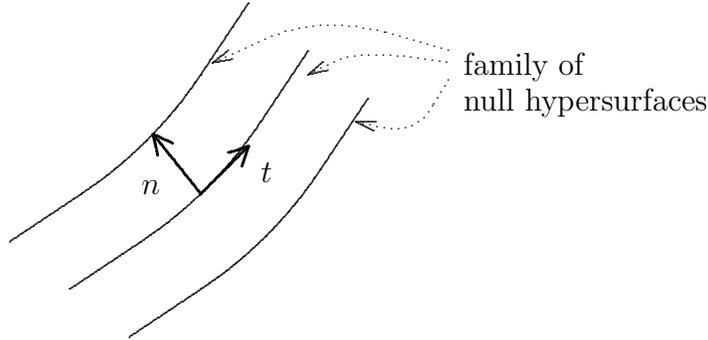
\begin{center}\input{p112-1.pictex}\end{center}

\subsection{Expansion and Shear}

Two linearly independent vectors $\eta^{(1)}$ and $\eta^{(2)}$ orthogonal 
to $n$ and $t$ determine an area element of $T_{\perp}$.  The shear
$\hat{\sigma}$ determines the change of \emph{shape} of this area element as
$\lambda$ increases.  The \emph{magnitude} of the area element defined by
$\eta^{(1)}$ and $\eta^{(2)}$ is
\be
a = \varepsilon^{\mu\nu\rho\sigma}t_{\mu}\eta_{\nu}
\eta^{(1)}_{\rho}\eta^{(2)}_{\sigma}
\ee
Since $t\cdot Dt=0$ and $t\cdot Dn=0$, we have
\bea
\frac{da}{d\lambda} & = & t\cdot \partial a = t\cdot 
Da= \varepsilon^{\mu\nu\rho\sigma}t_{\mu}n_{\nu}\left(t\cdot
D\eta^{(1)}_{\rho}\eta^{(2)}_{\sigma}+\eta^{(1)}_{\rho}t\cdot
D\eta^{(2)}_{\sigma}\right) \\
 & = &  \varepsilon^{\mu\nu\rho\sigma}t_{\mu}n_{\nu}
\left[\hat{B}_{\rho}^{\I\lambda}\eta^{(1)}_{\lambda}
\eta^{(2)}_{\sigma}+\eta^{(1)}_{\rho}
\hat{B}_{\sigma}^{\I\lambda}\eta^{(2)}_{\lambda}\right]
\\
 & = & 2 \varepsilon^{\mu\nu\rho\sigma}t_{\mu}n_{\nu} 
\hat{B}_{\rho}^{\I\lambda} \eta^{(1)}_{[\lambda}\eta^{(2)}_{\sigma]} \\
 & = & \theta a \quad \mbox{(see Question IV.2)}
\eea
i.e. $\theta$ measures the rate of increase of the magnitude of the 
area element.  If $\theta>0$ neighboring geodesics are \emph{diverging}, if
$\theta<0$ they are \emph{converging}.

\subsubsection{Raychaudhuri's equation for null geodesic congruences}

\bea
\frac{d\theta}{d\lambda} & = & t\cdot D
\left(B^{\mu}_{\I\nu}P^{\nu}_{\I\mu}\right) \\
 & = & P^{\nu}_{\I\mu} t\cdot DB^{\mu}_{\I\nu} 
\quad \mbox{(since $t\cdot Dt=0$ and $t\cdot Dn=0$)} \\
 & = & P^{\nu}_{\I\mu} t^{\rho}D_{\rho}D_{\nu}t^{\mu} \\
 & = & P^{\nu}_{\I\mu} t^{\rho}D_{\nu}D_{\rho}t^{\mu}+
P^{\nu}_{\I\mu}t^{\rho}\left[D_{\rho},D_{\nu}\right]t^{\mu} \\
 & = & P^{\nu}_{\I\mu}\left[\underbrace{D_{\nu}
\left(t\cdot Dt^{\mu}\right)}_0
-\left(D_{\nu}t^{\rho}\right)\left(D_{\rho}t^{\mu}
\right)\right]+P^{\nu}_{\I\mu}t^{\rho}R_{\rho\nu\I\sigma}^{\I\I\mu}
t^{\sigma} \\
 & = & -P^{\nu}_{\I\mu}B^{\mu}_{\I\rho}
B^{\rho}_{\I\nu}-t^{\rho}R_{\rho\sigma}
t^{\sigma}\quad \mbox{(using symmetries
of $R$)} \\
 & = & -P^{\nu}_{\I\mu}B^{\mu}_{\I\lambda}
P^{\lambda}_{\I\rho}B^{\rho}_{\I\nu}+
P^{\nu}_{\I\mu}B^{\mu}_{\I\lambda}t^{\lambda}n_{\rho}B^{\rho}_{\I\nu}
\nn
+P^{\nu}_{\I\mu}B^{\mu}_{\I\lambda}n^{\lambda}t_{\rho}
B^{\rho}_{\I\nu}-t^{\rho}t^{\sigma}R_{\rho\sigma} 
\\
 & = & -\hat{B}^{\mu}_{\I\rho} \hat{B}^{\rho}_{\I\nu}-
t^{\rho}t^{\sigma}R_{\rho\sigma} \quad \mbox{(using $t\cdot Dt\equiv 0$ and
$t^2\equiv 0$)}
\eea
or
\bebox{
\frac{d\theta}{d\lambda}=-\half\theta^2-\hat{\sigma}^{\mu\nu}
\hat{\sigma}_{\mu\nu}+\hat{\omega}^{\mu\nu}\hat{\omega}_{\mu\nu}-
R_{\mu\nu}t^{\mu}t^{\nu}
}
This is Raychaudhuri's equation for null geodesic 
congruences.\index{Raychaudhuri equation}

\subsubsection{Some consequences of Raychaudhuri's equation for 
null hypersurfaces}

\fbox{\parbox{6in}{
\paragraph{Proposition} The expansion $\theta$ of the null geodesic 
generator of a null hypersurface, $\mcN$, obeys the differential inequality 
\be
\frac{d\theta}{d\lambda}\le -\half \theta^2 
\ee
provided the spacetime metric solves Einstein's equations 
$G_{\mu\nu}=8\pi GT_{\mu\nu}$ and $T_{\mu\nu}$ satisfies the weak energy
condition. }}

\paragraph{Proof} $\hat{\sigma}^2 \ge 0$ because the metric in the 
orthogonal subspace $T_{\perp}$ (to $l$ and $n$) is positive definite.
$\hat{\omega}^2\ge 0$ also, but this comes in with wrong sign, however
$\hat{\omega}=0$ for a hypersurface.  Thus Raychaudhuri's equation implies
\bea
\frac{d\theta}{d\lambda} & \le & -\half \theta^2-R_{\mu\nu}l^{\mu}l^{\nu} \\
 & \le & -\half \theta^2-8\pi g T_{\mu\nu}l^{\mu}l^{\nu} \quad 
\mbox{(by Einstein's eq.)} \\
 & \le & -\half \theta^2 \quad \mbox{by weak energy condition}
\eea
\fbox{\parbox{6in}{
\paragraph{Corollary} If $\theta=\theta_0 < 0$ at some point $p$ on a null 
generator $\gamma$ of a null hypersurface, then $\theta\to -\infty$ along
$\gamma$ within an affine length $2/\left|\theta_0\right|$. }}

\paragraph{Proof} Let $\lambda$ be the affine parameter, with $\lambda=0$ at 
$p$.  Now
\be
\frac{d\theta}{d\lambda} \le -\half \theta^2 \quad \Leftrightarrow \quad 
\frac{d}{d\lambda}\left(\theta^{-1}\right)>\half \quad \Rightarrow \quad
\theta^{-1} \ge \half\lambda +\mbox{constant} 
\ee
where, since $\theta=\theta_0$ at $\lambda=0$, the constant cannot exceed 
$\theta_0^{-1}$.  Thus
\be
\theta^{-1}\ge \half\lambda +\theta_0^{-1} \quad \Rightarrow \quad 
\theta \le \frac{\theta_0}{1+\half\lambda \theta_0}
\ee
If $\theta_0<0$ the right-hand-side $\to -\infty$ when $\lambda=2/\left|
\theta_0\right|$, so $\theta\to -\infty$ within that affine length.

\paragraph{Interpretation}  When $\theta<0$ neighboring geodesics are 
converging.  The attractive nature of gravitation (weak energy condition) then
implies that they must continue to converge to a focus or a caustic. \\

\fbox{\parbox{6in}{
\paragraph{Proposition} If $\mcN$ is a Killing horizon then 
$\hat{B}_{\mu\nu}=0$ and 
\be
\frac{d\theta}{d\lambda}=0
\ee
}}

\paragraph{Proof}  Let $\xi$ be the Killing vector s.t. $\xi=fl$ 
$(l\cdot Dl=0)$ on $\mcN$ for some non-zero function $f$.  Then
\bea
\hat{B}_{\mu\nu} & = & \hat{B}_{(\mu\nu)} \quad \mbox{(since 
$\hat{\omega}=0$ for family of hypersurface)} \\
 & = & P_{\mu}^{\I\lambda}B_{(\lambda\rho)}P^{\rho}_{\I\nu}
\equiv P_{\mu}^{\I\lambda} D_{(\rho}l_{\lambda)}P^{\rho}_{\I\nu} \\
 & = & P_{\mu}^{\I\lambda}\left(\partial_{(\rho}f^{-1}
\right)\xi_{\lambda)}P^{\rho}_{\nu} \quad \mbox{(since
$D_{(\rho}\xi_{\lambda)}=0$)} \\
 & = & 0 \quad \mbox{(since $P\xi=\xi  P=0$)}
\eea
In particular $\theta=0$, \emph{everywhere on $\mcN$}, so 
$d\theta/d\lambda=0$.

\paragraph{Corollary}  For Killing horizon $\mcN$ of $\xi$
\bebox{
\left.R_{\mu\nu}\xi^{\mu}\xi^{\nu}\right|_{\mcN}=0
}

\paragraph{Proof}  Using $d\theta/d\lambda=0$ and $\hat{B}_{\mu\nu}=0$ in 
Raychaudhuri's equation.

\section{The Laws of Black Hole Mechanics}

Previously we showed that $\kappa^2$ is constant on a \emph{bifurcate} 
Killing horizon.  The proof fails if we have only part of a Killing horizon,
without the bifurcation 2-sphere, as happens in \emph{gravitational collapse}. 
In this case we need the:

\subsection{Zeroth law}  

\fbox{\parbox{6in}{If $T_{\mu\nu}$ obeys the dominant energy condition then 
the surface gravity $\kappa$ is constant on the future event horizon. }}

\paragraph{Proof}  Let $\xi$ be the Killing vector normal to $\mcH^+$ (here 
we use the theorem that $\mcH^+$ \emph{is} a Killing horizon).  Then since
$R_{\mu\nu}\xi^{\mu}\xi^{\nu}=0$ and $\xi^2=0$ on $\mcH^+$, Einstein's equations
imply
\be
0 = -\left. T_{\mu\nu}\xi^{\mu}\xi^{\nu}\right|_{\mcH^+} \equiv 
\left.J_{\mu}\xi^{\mu}\right|_{\mcH^+} 
\ee
i.e. $J=\left(-T^{\mu}_{\I\I\nu}\xi^{\nu}\right)\partial_{\mu}$ is 
tangent to $\mcH^+$.  It follows that $J$ can be expanded on a basis of tangent
vectors to $\mcH^+$
\be
J=a\xi+b_1\eta^{(1)}+b_2\eta^{(2)} \quad \mbox{on $\mcH^+$}
\ee
But since $\xi\cdot\eta^{(i)}=0$ this is spacelike or null 
(when $b_1=b_2=0$), whereas it must be \emph{timelike or null} by the
\emph{dominant energy condition}.  Thus, dominant energy $\Rightarrow J\propto
\xi$ and hence that
\bea
0 & = & \left.\xi_{[\sigma}J_{\rho]}\right|_{\mcH^+} = 
-\left.\xi_{[\sigma}T_{\rho]}^{\I\lambda}\xi_{\lambda}\right|_{\mcH^+} \\
 & = & \left. \xi_{[\sigma}R_{\rho]}^{\I\I\lambda}
\xi_{\lambda}\right|_{\mcH^+} \quad \mbox{(by Einstein's eq.)} \\
 & = & \left. \xi_{[\rho}\partial_{\sigma]}\kappa
\right|_{\mcH^+} \quad \mbox{(by result of Question IV.3)} \\
\eea
$\Rightarrow \partial_{\sigma}\kappa\propto \xi_{\sigma} 
\Rightarrow t\cdot \partial \kappa=0$ for any tangent vector $t$ to $\mcH^+$ 

$\Rightarrow$ \emph{$\kappa$ is constant on $\mcH^+$}.

\subsection{Smarr's Formula}

Let $\Sigma$ be a spacelike hypersurface in a stationary exterior black 
hole spacetime with an inner boundary, $H$, on the future event horizon and
another boundary at $i_0$.
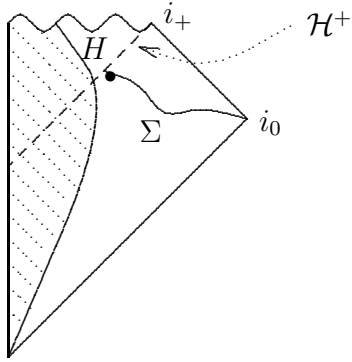
\begin{center}\input{p117-1.pictex}\end{center}
The surface $H$ is a 2-sphere that can be considered as the `boundary' of 
the black hole. \\

Applying Gauss' law to the Komar integral for $J$ we have
\bea
J & = & \frac{1}{8\pi G}\int_{\Sigma}dS_{\mu}D_{\nu}D^{\mu}m^{\nu}+
\frac{1}{16\pi G}\oint_H dS_{\mu\nu}D^{\mu}m^{\nu} \\
 & = & \frac{1}{8\pi G}\int_{\Sigma} dS_{\mu}R^{\mu}_{\I\nu} m^{\nu} +
J_H \quad \mbox{by Killing vector Lemma}
\eea
where $J_H$ is the integral over $H$.  Using Einstein's equation,
\be
J=\int_{\Sigma}dS_{\mu}\left(T^{\mu}_{\I\I\nu}m^{\nu}m^{\nu}-
\half Tm^{\mu}\right)+J_H 
\ee
In the absence of matter other than an electromagnetic field, we 
have $T_{\mu\nu}=T_{\mu\nu}(F)$, the stress tensor of the electromagnetic
field.  Since $g^{\mu\nu}T_{\mu\nu}(F)=T(F)=0$ we have
\bebox{
J=\int_{\Sigma}dS_{\mu}T^{\mu}_{\I\I\nu}(F)m^{\nu}+J_H 
\label{eq:Smarrstar}}
for an \emph{isolated} black hole (i.e. $T_{\mu\nu}=T_{\mu\nu}(F)$).

Now apply Gauss' law to the Komar integral for the total energy (= mass).
\bea
M & = & -\frac{1}{4\pi G}\int_{\Sigma}dS_{\mu}R^{\mu}_{\I\nu}k^{\nu}-
\frac{1}{8\pi G}\oint_H dS_{\mu\nu}D^{\mu}k^{\nu} \quad \mbox{(insert
$\xi=k+\Omega_H m$)} \nn \\
 & = & \int_{\Sigma}dS_{\mu}\left(-2T^{\mu}_{\I\I\nu}k^{\nu}+
Tk^{\mu}\right)-\frac{1}{8\pi G}\oint_H
dS_{\mu\nu}\left(D^{\mu}\xi^{\nu}-\Omega_H D^{\mu}m^{\nu}\right) \\
\eea
since $\Omega_H$ is constant on $H$.  For $T_{\mu\nu}=T_{\mu\nu}(F)$ 
$(T(F)=0)$we have
\be
M = -2\int_{\Sigma} dS_{\mu}T^{\mu}_{\I\I\nu}(F)k^{\nu}+2\Omega_HJ_H-
\frac{1}{8\pi G}\oint_H dS_{\mu\nu}D^{\mu}\xi^{\nu} 
\ee
for an isolated black hole.  Using (\ref{eq:Smarrstar}) we have
\be
M = -2\int_{\Sigma}dS_{\mu}T^{\mu}_{\I\I\nu}(F)\xi^{\nu}+2\Omega_H J -
\frac{1}{8\pi G}\oint_H dS_{\mu\nu}D^{\mu}\xi^{\nu}
\ee
For simplicity, we now suppose that $T_{\mu\nu}(F)=0$, i.e. the black hole 
has zero charge (see Questions III.7\&8 for general case).  Then
\be
M=2\Omega_H J-\frac{1}{8\pi G}\oint_H dS_{\mu\nu}D^{\mu}\xi^{\nu}
\ee

\paragraph{Lemma}
\be
dS_{\mu\nu}=\left(\xi_{\mu}n_{\nu}-\xi_{\nu}n_{\mu}\right)dA 
\quad \mbox{on $H$}
\ee
where $n$ is s.t. $n\cdot\xi=-1$.

\paragraph{Proof} $n$ and $\xi$ are normals to $H$, so we have to check 
coefficients.  In coordinates such that
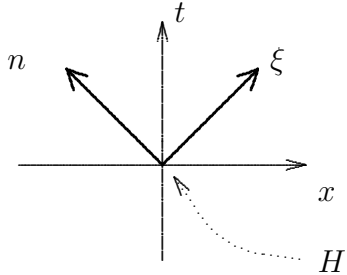
\begin{center}\input{p119-1.pictex}\end{center}
\bea
\xi_{\mu} & = & \frac{1}{\sqrt{2}}(1,1,0,0) \\
n_{\mu} & = & \frac{1}{\sqrt{2}}(1,-1,0,0)
\eea
we should have $\left|dS_{01}\right|=dA$.  We do if $dS_{\mu\nu}$ is as 
given.  [There is still a sign ambiguity.  Fix by requiring sensible results].

Thus
\bea
-\frac{1}{8\pi G}\oint_H dS_{\mu\nu}D^{\mu}\xi^{\nu} & = & 
-\frac{1}{4\pi G}\oint_H dA\underbrace{ (\xi\cdot D\xi)^{\nu}
}_{\kappa\xi^{\nu}} n_{\nu} \\
 & = & -\frac{\kappa}{4\pi G}\oint dA \underbrace{\xi\cdot n}_{-1} 
\qquad \mbox{($\kappa$ is constant by $0^{\subtext{th}}$ law)} \\
 & = & \frac{\kappa}{4\pi G} A 
\eea
where $A$ is the ``area of the horizon'' (i.e. $H$).

Hence
\bebox{
M=\frac{\kappa A}{4\pi} +2\Omega_H J
}
This is Smarr's formula\index{Smarr's formula} for the mass of a Kerr 
black hole. [Exercise:  Check, using previous results for $\kappa$, $\Omega_H$,
and $A$].  In the $Q\neq 0$ case, this formula generalizes to 
\be
M=\frac{\kappa A}{4\pi}+2\Omega_H J+\Phi_H Q
\ee
where $\Phi_H$ is the co-rotating electric potential\index{co-rotating 
electric potential} on the horizon (see Question III.6\&7).

\subsection{First Law}  

\fbox{\parbox{6in}{If a stationary black hole of mass $M$, charge $Q$ and 
angular momentum $J$, with future event horizon of surface gravity $\kappa$,
electric surface potential $\Phi_H$ and angular velocity $\Omega_H$, is
perturbed such that it settles down to another black hole with mass $M+\delta M$
charge $Q+\delta Q$ and angular momentum $J+\delta J$, then
\bebox{
dM=\frac{\kappa}{8\pi}dA+\Omega_H dJ+\Phi_H dQ 
}
}}
\newcounter{lawtwonotes}
\begin{list}{\arabic{lawtwonotes})}
{\usecounter{lawtwonotes}}
\item Definition of $\Phi_H$ and proof for $Q\neq 0$ in Q. III.6\&7.

\item This statement of the first law uses the fact that the event horizon 
of a stationary black hole must be a Killing horizon.
\end{list}

\paragraph{`Proof' for $Q= 0$ (Gibbons)}  Uniqueness theorems imply that
\be
M=M(A,J)
\ee
But $A$ and $J$ both have dimensions of $M^2$ $(G=c=1)$ so the function 
$M(A,J)$ must be \emph{homogeneous of degree $1/2$}.  By Euler's theorem for
homogeneous functions
\bea
A\pd{M}{A}+J\pd{M}{J} & = & \half M \\
 & = & \frac{\kappa}{8\pi}A+\Omega_H J \quad \mbox{by Smarr's formula}
\eea
Therefore 
\be
A\left(\pd{M}{A}-\frac{\kappa}{8\pi}\right)+J\left(\pd{M}{J}-\Omega_H\right)=0
\ee
But $A$ and $J$ are free parameters so 
\be
\pd{M}{A}=\frac{\kappa}{8\pi}, \quad \pd{M}{J}=\Omega_H
\ee

\subsection{The Second Law (Hawking's Area Theorem)}

\fbox{\parbox{6in}{If $T_{\mu\nu}$ satisfies the weak energy condition, and 
assuming that the cosmic censorship hypothesis is true then the area of the
future event horizon of an asymptotically flat spacetime is a non-decreasing
function of time. }} \\

Technically the cosmic censorship assumption is that the spacetime is `strongly
asymptotically predictable' which requires the existence of a globally
hyperbolic submanifold of spacetime containing both the exterior spacetime
\emph{and} the horizon.  A theorem of Geroch states that in this case there
exists a family of Cauchy hypersurfaces $\Sigma(\lambda)$ such that
$\Sigma(\lambda') \subset D^+\left(\Sigma(\lambda)\right)$ if 
$\lambda'>\lambda$.
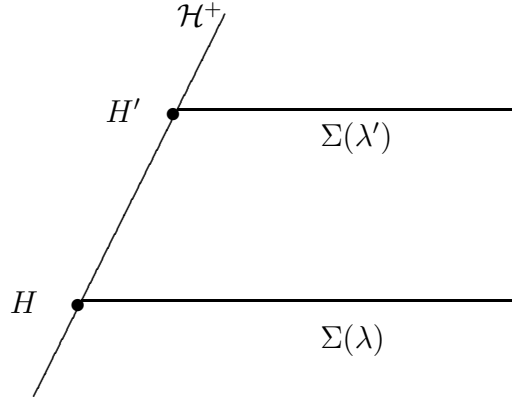
\begin{center}\input{p121-1.pictex}\end{center}
We can choose $\lambda$ to be the affine parameter on a null geodesic 
generator of $\mcH^+$.  The ``area of the horizon'' $A(\lambda)$ is the area of
the intersection of $\Sigma(\lambda)$ with $\mcH^+$.  The second law states that
$A(\lambda')\ge A(\lambda)$ if $\lambda'>\lambda$.

\paragraph{Idea of proof}  To show that $A(\lambda)$ cannot decrease with 
increasing $\lambda$ it is sufficient to show that each area element, $a$, of
$H$ has this property.  Recalling that
\be
\frac{da}{d\lambda}=\theta a
\ee
we see that the second law holds if $\theta\ge 0$ everywhere on $\mcH^+$.  
To see that this is true, recall that if $\theta<0$ the geodesics must converge
to a focus or caustic, i.e. nearby geodesics to a given one passing through a
point $p$ must intersect $\gamma$ at finite affine distance along it.  The first
point $q$ for which this happens is called the point conjugate to $p$ on
$\gamma$.
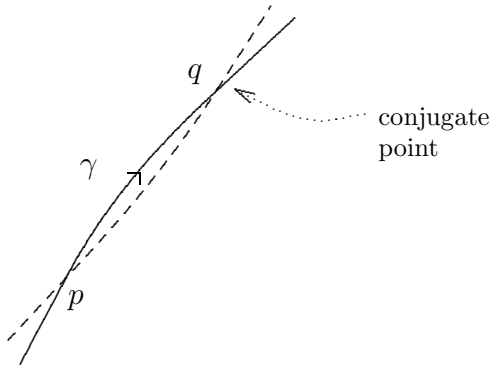
\begin{center}\input{p122-1.pictex}\end{center}
\emph{Points on $\gamma$ beyond $q$ are no longer null separated}.  They 
are \emph{timelike} separated from $p$.  An example illustrating this is light
rays in a flat 2-dim cylindrical spacetime.
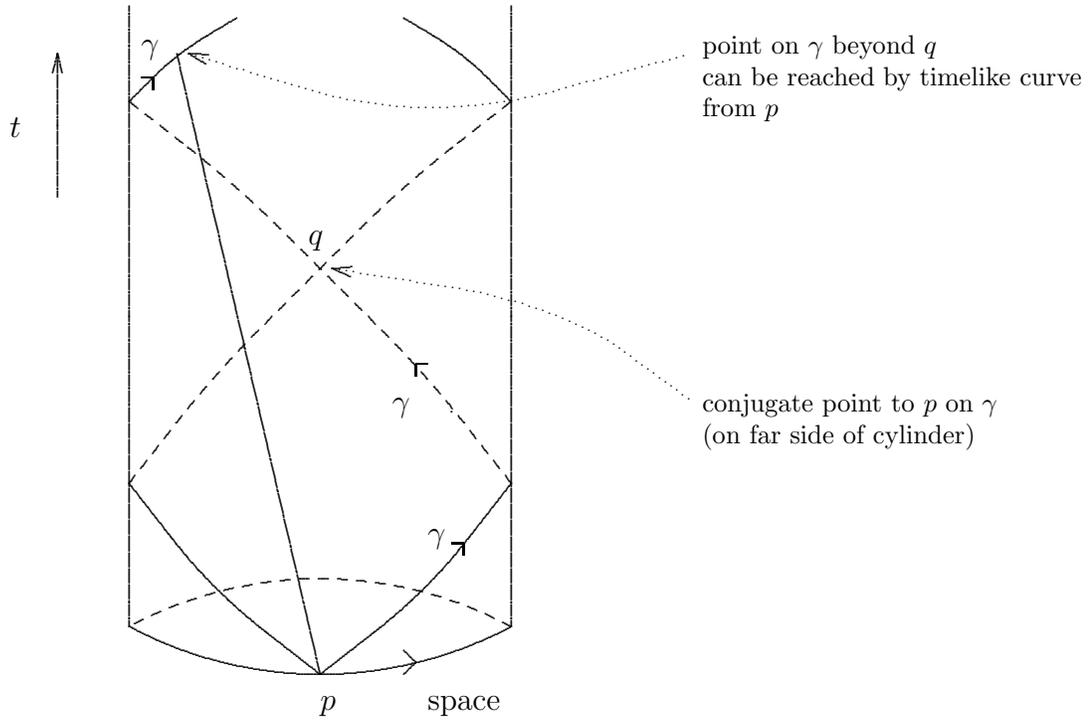
\begin{center}\input{p122-2.pictex}\end{center}
The existence of a conjugate point to the future of a null geodesic generator 
in $\mcH^+$ would mean that this generator of $\mcH^+$ has a finite endpoint,
in contradiction to Penrose's theorem, so the hypothetical conjugate point
cannot exist.  Thus it must be that $\theta\ge 0$ everywhere on $\mcH^+$ and
hence the second law.  

$\theta=0$ only for stationary spacetimes.

% clearpage inserted to account for big diagram which follows
\clearpage
\paragraph{Example}  Formation of black hole from pressure-free 
spherically-symmetric gravitational collapse.  Illustrate by a 
Finkelstein diagram
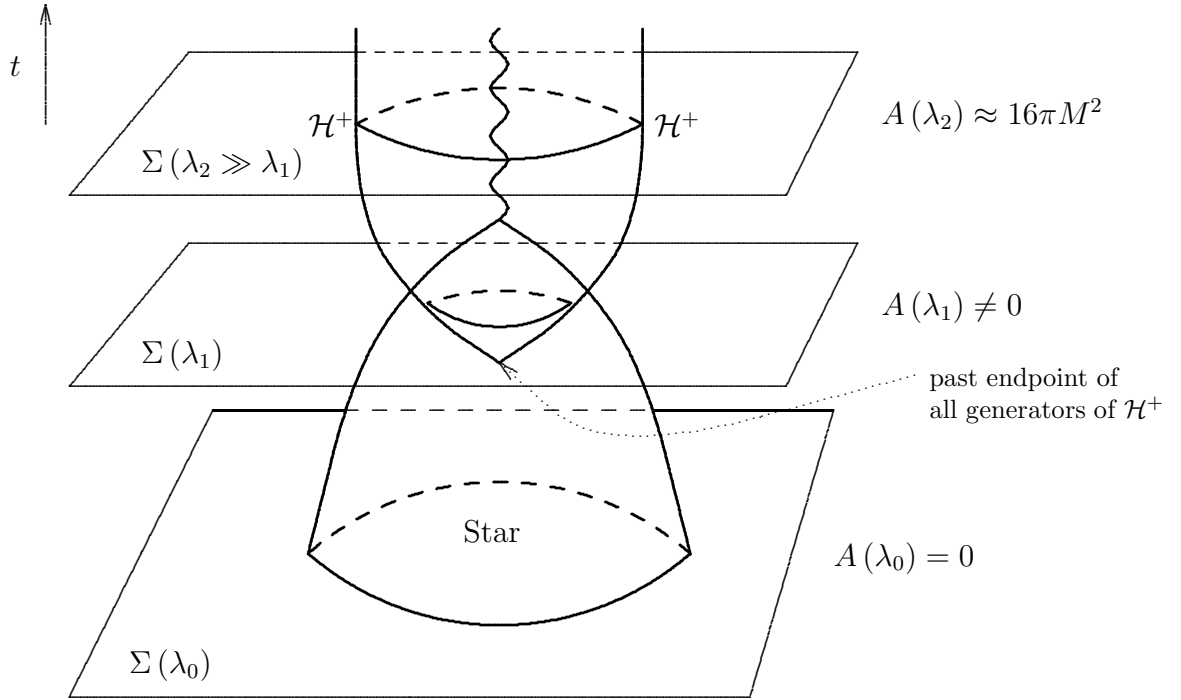
\begin{center}\input{p123-1.pictex}\end{center}
$A=0$ on $\Sigma\left(\lambda_0\right)$.  $A\neq 0$ on 
$\Sigma\left(\lambda_1\right)$ and it has increased to its final value of
$A=16\pi M^2$ for a stationary Schwarzschild black hole on
$\Sigma\left(\lambda_2\right)$. \\
% clearpage inserted to account for big diagram which follows
\clearpage
\subsubsection{Consequences of $2^{\subtext{nd}}$ Law}

\newcounter{conseq}
\begin{list}{(\arabic{conseq})}
{\usecounter{conseq}}

\item Limits to efficiency of mass/energy conversion in black hole 
collisions.  Consider Finkelstein diagram of two coalescing black holes.
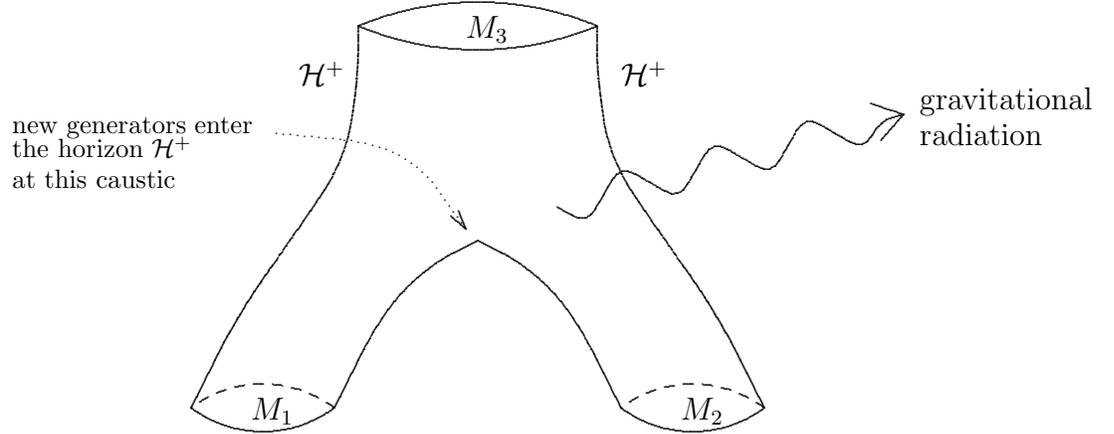
\begin{center}\input{p123-2.pictex}\end{center}
Then energy radiated is $M_1+M_2-M_3$, so the efficiency, $\eta$, of mass 
to energy conversion is
\be
\eta=\frac{M_1+M_2-M_3}{M_1+M_2} = 1-\frac{M_3}{M_1+M_2} 
\ee
Assuming that the two black holes are initially approximately stationary, 
so $A_1=16\pi M_1^2$ and $A_2=16\pi M_2^2$ the area theorem says that
\be
A_3 \ge 16\pi \left(M_1^2+M_2^2\right) 
\ee
But $16\pi M_3^2 \ge A_3$ (with equality at late times), so
\be
M_3\ge \sqrt{M_1^2+M_2^2}
\ee
Thus 
\be
\eta \le 1-\frac{\sqrt{M_1^2+M_2^2}}{M_1+M_2}\le 1-\frac{1}{\sqrt{2}}
\ee
The radiated energy could be used to do work, so the area theorem limits 
the useful energy that can be extracted from black holes in the same way that
the $2^{\subtext{nd}}$ law of thermodynamics limits the efficiency of heat
engines.

\item \emph{Black holes cannot bifurcate}.  Consider $M_3\to M_1+M_2$ 
(with $M_1>0$ and $M_2>0$).  The area theorem now says that
\be
M_3 \le \sqrt{M_1^2+M_2^2} \le M_1+M_2
\ee
but energy conservation requires $M_3 \ge M_1+M_2$ (with $M_3-M_1-M_2$ 
being radiated away).  We have a contradiction so the process cannot occur.

\end{list}

\chapter{Hawking Radiation}

\section{Quantization of the Free Scalar Field}

Let $\Phi(x)$ be a real scalar field satisfying the Klein-Gordon 
equation\index{Klein-Gordon equation}.
\be
\left(D^{\mu}\partial_{\mu}-m^2\right)\Phi(x)=0
\ee
Let $\left\{\phi_{\alpha}\right\}$ span the space ${\cal S}$ of solutions. We
shall assume that the spacetime is globally hyperbolic, i.e. that $\exists$ a
Cauchy surface $\Sigma$. A point in the space ${\cal S}$ then corresponds to a
choice of initial data on $\Sigma$. The space ${\cal S}$ has a natural
\emphin{symplectic structure}.
\be
\phi_{\alpha}\wedge \phi_{\beta}=\int_{\Sigma}dS_{\mu}
\phi_{\alpha}\stackrel{\leftrightarrow}{\partial}^{\mu}\phi_{\beta},\quad
\left(=-\phi_{\beta}\wedge \phi_{\alpha}\right)
\ee
where $\lrpd{}$ is defined by
\be
f \lrpd{} g = f \partial g -g \partial f
\ee
`Natural' means that $\wedge$ \emph{does not depend on the choice of $\Sigma$}.
\be
\left(\phi_{\alpha}\wedge\phi_{\beta}\right)_{\Sigma}-\left(\phi_{\alpha}
\wedge\phi_{\beta}\right)_{\Sigma'} = \int_S \dx{4}{x}
\sqrt{-g}D_{\mu}\left(\phi_{\alpha}\lrpd{\mu}\phi_{\beta}\right) 
\ee
\begin{center}\input{p125-1.pictex}\end{center}
But
\bea
D_{\mu}\left(\phi_{\alpha}\lrpd{\mu}\phi_{\beta}\right) & = & 
\phi_{\alpha}\left(D_{\mu}\partial^{\mu}\phi_{\beta}\right)-
\left(D_{\mu}\partial^{\mu}\phi_{\alpha}\right)\phi_{\beta}
\\ 
 & = & \phi_{\alpha}\left(m^2\phi_{\beta}\right)-
\left(m^2\phi_{\alpha}\right)\phi_{\beta} = 0 \ , 
\eea
using the Klein-Gordon equation in the last step.

The antisymmetric form $\phi_{\alpha}\wedge\phi_{\beta}$ can be brought 
to a canonical block diagonal form, with $2\times 2$ blocks
$\matrixtwo{0}{1}{-1}{0}$, by a change of basis (Darboux's theorem). Thus,
real solutions of the Klein-Gordon equation can be grouped in pairs
$(\phi,\phi')$ with $\phi\wedge\phi'=1$. It then follows that the complex
solution $\psi=\left(\phi-i\phi'\right)/\sqrt{2}$ has unit norm if we define its norm $||\psi||$ by
$||\psi||^2=\phi\wedge\phi'$ or, equivalently,
\be
||\psi||^2 = i\int_\Sigma dS_{\mu}\psi^*\lrpd{\mu}\psi\, .
\ee
More generally, we can introduce a complex basis $\{\psi_i\}$ of solutions
of the Klein-Gordon equation with hermitian inner product defined by
\be
(\psi_i, \psi_j) = i\int dS_{\mu}\, \psi_i^*\lrpd{\mu}\psi_j\ ,
\ee
and we can choose this basis such that $(\psi_i, \psi_j)=\delta_{ij}$. This
inner product is not positive definite, however, because 
$||\psi^*||^2 = -||\psi||^2$. In fact, we can choose the basis $\{\psi_i\}$ such
that
\be
\left( \begin{array}{rclcrcl}
\left(\psi_i,\psi_j\right) & = & \delta_{ij} & 
\quad & \left(\psi_i,\psi_j^*\right) & = & 0 \\ \\
\left(\psi_i^*,\psi_j\right) & = & 0 & \quad & 
\left(\psi_i^*,\psi_j^*\right) & = & -\delta_{ij} \end{array} \right)
\label{eq:inner_prod}
\ee

We could interpret the complex solution $\Psi=\sum_i a_i\psi_i$ as 
the wavefunction of a free particle since $(\;,\;)$ is positive-definite when
restricted to such solutions, but this cannot work when interactions are
present.  It is also inapplicable for \emph{real} scalar fields.  A real
solution $\Phi$ of the K-G equation can be written as
\be
\Phi(x)=\sum\left[ a_i\psi_i(x)+a_i^*\psi_i^*(x)\right]
\ee
To quantize we pass to the \emph{quantum} field
\be
\Phi(x)=\sum\left[a_i\psi_i(x)+a_i^{\dagger}\psi_i^*(x)\right]
\ee
where $\left\{a_i\right\}$ are now operators in a Hilbert space $\mcH$ with 
Hermitian conjugates $a_i^{\dagger}$ satisfying the commutation relations
\be
\left[a_i,a_j\right]=0, \quad \left[a_i,a_j^{\dagger}\right]=\delta_{ij} 
\quad (\hbar=1)
\ee
We choose the Hilbert space to be the Fock space built from a `vacuum' 
state $\ket{\mbox{vac}}$ satisfying
\bea
a_i\vac & = & 0 \quad \forall i \\
\left<\mbox{vac}|\mbox{vac}\right> & = & 1 
\eea
i.e. $\mcH$ has the basis
\bdm
\left\{\vac,a_i^{\dagger}\vac, a_i^{\dagger}a_j^{\dagger}\vac, \ldots \right\}
\edm
$\left<\;|\;\right>$ is a positive-definite inner product on this space.

This basis for $\mcH$ is determined by the choice of $\vac$, but this 
depends on the choice of complex basis $\left\{\psi_i\right\}$ of solutions of
the K-G equation satisfying (\ref{eq:inner_prod}).  There are \emph{many} such
bases. \\

Consider $\left\{\psi_i'\right\}$ where
\be
\psi_i'=\sum_j\left(A_{ij}\psi_j+B_{ij}\psi_j^*\right)
\label{eq:psi_i_prime}
\ee
This has the same inner product matrix (\ref{eq:inner_prod}) provided that
\bebox{
\begin{array}{rcl} AA^{\dagger}-BB^{\dagger} & = & \mathbf{1} \\ \\
\transpose{AB}-\transpose{BA} & = & 0  \end{array} \label{eq:ip_dagger}
}
Inversion of (\ref{eq:psi_i_prime}) leads to
\be
\psi_j = \sum_k A_{jk}'\psi_k'+B_{jk}'{\psi_k'}^*
\ee
where
\be
A'=A^{\dagger}, \quad B'=-\transpose{B}
\ee
\ul{Check}
\bea
\psi' & = & A\left(A'\psi'+B'{\psi'}^*\right)+B\left({A'}^*{\psi'}^*+{B'}^*\psi'\right) \\
 & = & \left(AA'+B{B'}^*\right)\psi'+\left(AB'+B{A'}^*\right){\psi'}^* \\
 & = & \left(AA^{\dagger}-BB^{\dagger}\right)\psi'-\left(A\transpose{B}-B\transpose{A}\right)\psi' \\
 & = & \psi'
\eea
But $A'$ and $B'$ must satisfy the same conditions as $A$ and $B$, i.e.
\bea
A'{A'}^{\dagger}-B'{B'}^{\dagger} & = & \mathbf{1} \\
A'\transpose{B'}-B'\transpose{A'} & = & 0 
\eea
Equivalently,
\bebox{
\begin{array}{rcl}
A^{\dagger}A-\transpose{B}B^* & = & \mathbf{1} \\
A^{\dagger}B-\transpose{B}A^* & = & 0 \end{array}
\label{eq:relations_star}
}
These conditions are not implied by (\ref{eq:ip_dagger}); the additional information contained in them is the invertibility of the change of basis.

In a general spacetime there is no `preferred' choice of basis satisfying 
(\ref{eq:inner_prod}) and so no preferred choice of vacuum.  In a stationary
spacetime, however, we can choose the basis $\left\{u_i\right\}$ of
\emph{positive frequency} eigenfunctions of $k$, i.e.
\be
k^{\mu}\partial_{\mu}u_i = -i\omega_i u_i, \quad \omega_i \ge 0 
\ee
Notes
\newcounter{pfenotes}
\begin{list}{(\arabic{pfenotes})}
{\usecounter{pfenotes}}
\item Since $k$ is Killing it maps solutions of the Klein-Gordon equation to 
solutions (\bold{Proof: Exercise}). 

\item $k$ is anti-hermitian, so it can be diagonalized with pure-imaginary 
eigenvalues.

\item Eigenfunctions with distinct eigenvalues are orthogonal so
\be
\left(u_i,u_j^*\right)=0
\ee
We can normalize $\left\{u_i\right\}$ s.t. $\left(u_i,u_j\right)=\delta_{ij}$, 
so the basis $\left\{u_i\right\}$ can be chosen s.t. (\ref{eq:inner_prod}) is
satisfied.  

\item We exclude functions with $\omega=0$.

\end{list}

For this choice of basis the vacuum state $\vac$ is actually the state of 
lowest energy.  The states $a_i^{\dagger}\vac$ are one-particle states,
$a_i^{\dagger}a_j^{\dagger}\vac$ two-particle states, etc., and
\be
N=\sum_i a_i^{\dagger}a_i
\ee
is the particle number operator\index{particle number operator}.

\section{Particle Production in Non-Stationary Spacetimes}

Consider a `sandwich' spacetime\index{sandwich spacetime} 
$M=M_- \cup M_0 \cup M_+$
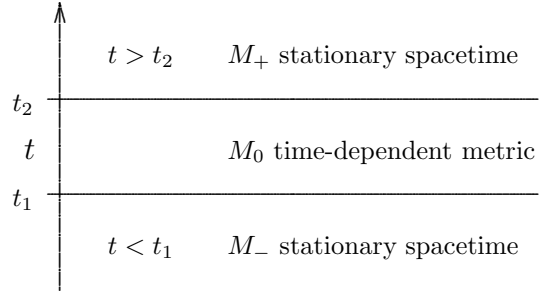
\begin{center}\input{p128-1.pictex}\end{center}
In $M_-$ we can choose to expand a scalar field solution of the 
Klein-Gordon equation as
\be
\Phi(x)=\sum_i \left[a_i u_i(x)+a_i^{\dagger}u_i^*(x)\right] \qquad 
\mbox{in $M_-$}
\ee
The functions $u_i(x)$ solve the KG equation in $M_-$ but \ul{not} in $M$, 
so its continuation through $M_0$ will lead to some new function $\psi_i(x)$ in
$M_+$, so
\be
\Phi(x)=\sum_i \left[a_i \psi_i(x)+a_i^{\dagger}\psi_i^*(x)\right] \qquad 
\mbox{in $M_+$}
\ee
Because the inner product $(\;,\;)$ was independent of the hypersurface 
$\Sigma$, the matrix of inner products will still be as before, i.e. as in
(\ref{eq:inner_prod}).  But, as we have seen this implies only that
\be
\psi_i = \sum_j \left(A_{ij}u_j+B_{ij}u_j^*\right)
\ee
for some matrices $A$ and $B$ satisfying (\ref{eq:ip_dagger}).  Thus, in $M_+$
\bea
\Phi(x) & = & \sum_i\left(a_i\psi_i+a_i^{\dagger}\psi_i^* \right) \\
 & = & \sum_i\left[a_i \sum_j\left(A_{ij}u_j+B_{ij}u_j^*\right)+
a_i^{\dagger}\sum_j\left(A_{ij}^*u_j^*+B_{ij}^*u_j\right)\right] \\
 & = & \sum_i \left[a_i' u_i(x)+{a_i'}^{\dagger}u_i^*(x)\right] 
\eea
where
\bebox{
a_j' = \sum_i\left(a_iA_{ij}+a_i^{\dagger}B_{ij}^*\right)
}
This is called a \emphin{Bogoliubov transformation}.  $A$ and $B$ are 
the \emph{Bogoliubov coefficients}.   \\

Note that (\bold{Exercise})
\be
\left.\begin{array}{rcl}
\left[a_i',a_j'\right] & = & 0 \\ \\ \left[a_i',{a_j'}^{\dagger}\right] 
& = & \delta_{ij} \end{array}\right\} \Leftrightarrow \mbox{ relations
(\ref{eq:relations_star}) satisfied by $A$ \& $B$}
\ee
If $B=0$ then (\ref{eq:ip_dagger}) and (\ref{eq:relations_star}) imply $A^{\dagger}A=AA^{\dagger}=1$, i.e. the 
change of basis from $\left\{u_i\right\}$ to $\left\{\psi_i\right\}$ is just a
unitary transformation which permutes the annihilation operators but does not
change the definition of the vacuum.

The particle number operator for the $i^{\subtext{th}}$ mode of $k$ is 
\be
\begin{array}{rclcr} N_i & = & a_i^{\dagger}a_i & \quad & \mbox{in $M_-$} \\ \\
N_i' & = & {a_i'}^{\dagger}a_i' & \quad & \mbox{in $M_+$} 
\end{array} 
\ee
The state with no particles in $M_-$ is $\vac$ s.t. $a_i\vac=0\; \forall i$.  
The expected number of particles in the $i^{\subtext{th}}$ mode in $M_+$ is then
\bea
\left<N_i'\right> & \equiv & \braket{\mbox{vac}}{N_i'}{\mbox{vac}} = 
\braket{\mbox{vac}}{{a_i'}^{\dagger}a_i'}{\mbox{vac}} \\
 & = & \sum_{j,k}\braket{\mbox{vac}}{\left(a_k B_{ki}\right)
\left(a_j^{\dagger}B_{ji}^*\right)}{\mbox{vac}} \\ 
 & = & \sum_{j,k}\underbrace{ \braket{\mbox{vac}}{a_ka_j^{\dagger}}
{\mbox{vac}}}_{\delta_{kj}} B_{ki}B_{ij}^{\dagger} \\
 & = & \left(B^{\dagger}B\right)_{ii}
\eea
The expected total number of particles is therefore 
$\tr\left(B^{\dagger}B\right)$.  Since $B^{\dagger}B$ is positive semi-definite,
this vanishes iff $B=0$. 

\section{Hawking Radiation}

The spacetime associated to gravitational collapse to a black hole cannot be 
everywhere stationary so we expect particle creation.  But the exterior
spacetime is stationary at late times, so we might expect particle creation to
be just a transient phenomenon determined by details of the collapse.

But the \emph{infinite time dilation} at the horizon of a black hole means 
that particles created in the collapse can take arbitrarily long to escape -
suggests a possible flux of particles at late times that is due to the existence
of the horizon and \emph{independent of the details of the collapse}.  There is
such a particle flux, and it turns out to be thermal - this is \emph{Hawking
radiation}\index{Hawking!radiation}

We shall consider only a massless scalar field $\Phi$ in a Schwarzschild black 
hole spacetime.  From Question IV.4 we learn that the positive frequency
outgoing modes of $\Phi$ have the behaviour
\be
\Phi_{\omega}\sim e^{-i\omega u}
\ee
near $\scri^+$.  Consider a geometric optics approximation in which a 
particle's worldline is a null ray, $\gamma$, of constant phase $u$, and trace
this ray backwards in time from $\scri^+$.  The later it reaches $\scri^+$ the
closer it must approach $\mcH^+$ in the exterior spacetime before entering the
star.
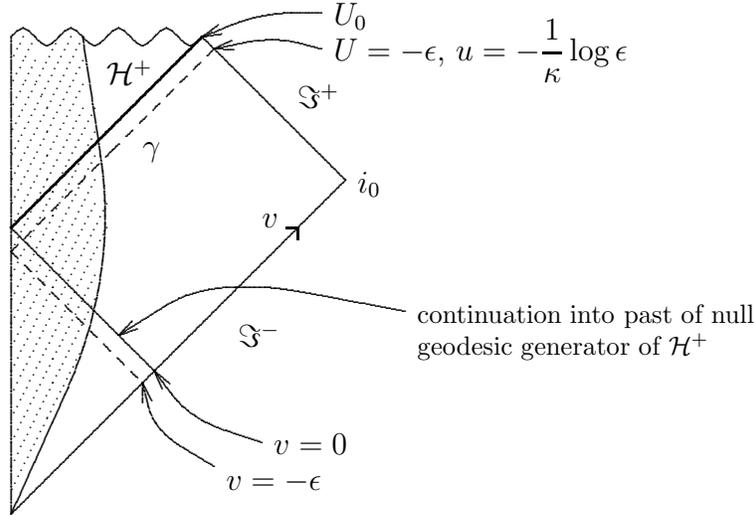
\begin{center}\input{p133-1.pictex}\end{center}
The ray $\gamma$ is one of a family of rays whose limit as $t\to\infty$ is a 
null geodesic generator, $\gamma_H$, of $\mcH^+$.  We can specify $\gamma$ by
giving its affine distance from $\gamma_H$ along an ingoing null geodesic
through $\mcH^+$
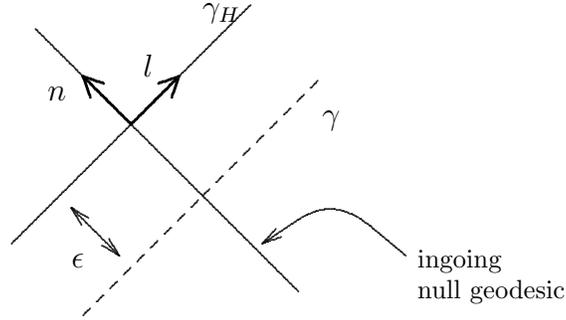
\begin{center}\input{p133-2.pictex}\end{center}
The affine parameter on this ingoing null geodesic is $U$, so $U=-\epsilon$.  
Equivalently
\be
u=-\frac{1}{\kappa}\log \epsilon \qquad \mbox{(on $\gamma$ near $\mcH^+$)}
\ee
so
\be
\Phi_{\omega}\sim \exp\left(\frac{i\omega}{\kappa}\log \epsilon\right) 
\qquad \mbox{near $\mcH^+$}
\ee
This oscillates increasingly rapidly as $\epsilon\to 0$, so 
\emph{the geometric optics approximation is justified at late times}. \\

We need to match $\Phi_{\omega}$ onto a solution of the K-G equation 
near $\scri^-$.  In the geometric optics approximation we just
parallely-transport $n$ and $l$ back to $\scri^-$ along the continuation of
$\gamma_H$.  Let this continuation meet $\scri^-$ at $v=0$.  The continuation of
the ray $\gamma$ back to $\scri^-$ will now meet $\scri^-$ at an affine distance
$\epsilon$ along an outgoing null geodesic on $\scri^-$
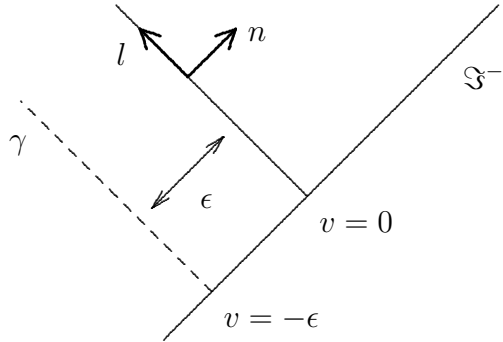
\begin{center}\input{p134-1.pictex}\end{center}
The affine parameter on outgoing null geodesics in $\scri^-$ is $v$ 
(since $ds^2=du\,dv+r^2d\Omega^2$ on $\scri^-$), so $v=-\epsilon$ on 
$\gamma$ so
\be
\Phi_{\omega}\sim \exp\left\{\frac{i\omega}{\kappa}\log(-v)\right\}
\ee
This is for $v<0$. For $v>0$ an ingoing null ray from $\scri^-$ passes 
through $\mcH^+$ and doesn't reach $\scri^+$, so
$\Phi_{\omega}=\Phi_{\omega}(v)$ on $\scri^-$, where
\be
\Phi_{\omega}(v) = \left\{ \begin{array}{ccc} 0 & \quad & v>0 \\
 \exp\left(\frac{i\omega}{\kappa}\log(-v)\right) & \quad & v<0 
\end{array}\right.
\ee

Take the Fourier transform,
\bea
\tilde{\Phi}_{\omega} & = & \int^{\infty}_{-\infty} 
e^{i\omega' v}\Phi_{\omega}(v)dv \\
 & = & \int^0_{-\infty}\exp\left\{i\omega'v +
\frac{i\omega}{\kappa}\log(-v)\right\}dv
\eea
\paragraph{Lemma} 
\bebox{
\tilde{\Phi}_{\omega}(-\omega')=-\exp\left(-\frac{\pi\omega}
{\kappa}\right)\tilde{\Phi}_{\omega}(\omega') \qquad \mbox{for $\omega'>0$}
}

\paragraph{Proof}  Choose branch cut in complex $v$-plane to lie along 
the real axis
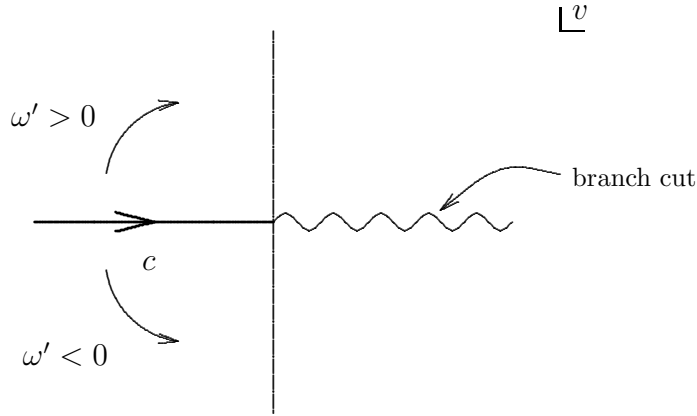
\begin{center}\input{p135-1.pictex}\end{center}
For $\omega'>0$ rotate contour to the positive imaginary axis and then 
set $v=ix$ to get
\bea
\tilde{\Phi}_{\omega}(\omega') & = & -i\int^{\infty}_0 \exp\left
\{-\omega'x+\frac{i\omega}{\kappa}\log\left(xe^{-i\pi/2}\right)\right\}dx \\
 & = & -\exp\left(\frac{\pi\omega}{2\kappa}\right) \int_0^{\infty} 
\exp\left\{-\omega'x+\frac{i\omega}{\kappa}\log(x)\right\}dx
\eea
Since $\omega'>0$ the integral converges.  When $\omega'<0$ we rotate 
the contour to the negative imaginary axis and then set $v=-ix$ to get
\bea
\tilde{\Phi}_{\omega}(\omega') & = & i\int_0^{\infty} \exp\left
\{ \omega'x+\frac{i\omega}{\kappa}\log\left(xe^{i\pi/2}\right)\right\}dx \\
 & = & \exp\left(-\frac{\pi\omega}{2\kappa}\right)\int^{\infty}_0 
\exp\left\{ \omega'x+\frac{i\omega}{\kappa}\log(x)\right\} dx 
\eea
Hence the result. 

\paragraph{Corollary}  A mode of \emph{positive} frequency $\omega$ on 
$\scri^+$, \emph{at late times}, matches onto \emph{mixed positive and negative}
modes on $\scri^-$.  We can identify (for positive $\omega'$)
\bea
A_{\omega\omega'} & = & \tilde{\Phi}_{\omega}(\omega') \\
B_{\omega\omega'} & = & \tilde{\Phi}_{\omega}(-\omega') = 
-e^{-\pi\omega/\kappa}\tilde{\Phi}_{\omega}(\omega')
\eea
as the Bogoliubov coefficients.  We see that
\bebox{
B_{ij}=-e^{-\pi \omega_i/\kappa}A_{ij}
}
But the matrices $A$ and $B$ must satisfy the Bogoliubov relations, e.g.
\bea
\delta_{ij} & = & \left(AA^{\dagger}-BB^{\dagger}\right)_{ij} \\
 & = & \sum_k A_{ik}A_{jk}^*-B_{ik}B^*_{jk} \\
 & = & \left[ e^{\pi\left(\omega_i+\omega_j\right)/\kappa}-1\right]\sum_k B_{ik}B_{jk}^* 
\eea
Take $i=j$ to get
\be
\left(BB^{\dagger}\right)_{ii} = \frac{1}{e^{2\pi\omega_i/\kappa}-1}
\ee
Now, what we actually need are the \emph{inverse} Bogoliubov coefficients corresponding to a positive frequency mode on $\scri^-$ matching onto mixed positive and negative frequency modes on $\scri^+$.  As we saw earlier, the inverse $B$ coefficient is
\be
B'=-\transpose{B}
\ee
The late time particle flux through $\scri^+$ given a vacuum on $\scri^-$ is
\be
\left<N_i\right>_{\scri^+} = \left(\left(B'\right)^{\dagger}B'\right)_{ii} = \left(B^*\transpose{B}\right)_{ii} = \left(B\transpose{B}\right)^*_{ii}
\ee
But $\left(B\transpose{B}\right)_{ii}$ is real so
\bebox{
\left<N_i\right>_{\scri^+} = \frac{1}{e^{2\pi\omega_i/\kappa}-1}
}
This is the Planck distribution\index{Planck distribution} for black body 
radiation\index{black body radiation} at the Hawking
temperature\index{Hawking!temperature}
\be
T_H=\frac{\hbar\kappa}{2\pi}
\ee

We conclude that at late times the black hole radiates away its energy at 
this temperature.  From Stephan's law\index{Stephan's law}
\be
\frac{dE}{dt}\simeq -\sigma AT_H^4, \qquad 
\left(\sigma=\frac{\pi^2k_B^4}{60\hbar^3c^2}\right)
\ee
where $A$ is the black hole area.  Since
\be
E=Mc^2,\quad A=\left(\frac{MG}{c^2}\right)^2, \quad k_BT_H \sim 
\frac{\hbar c^3}{GM}
\ee
we have
\be
\frac{dM}{dt} \sim \frac{\hbar c^4}{G^2M^2} 
\ee
which gives a lifetime 
\be
\tau \sim \left(\frac{G^2}{\hbar c^4}\right)M^3 
\ee
%which is larger than the age of the universe for $M\sim M_{\odot}$.

\paragraph{Note} The calculation of Hawking radiation assumed no 
backreaction, i.e. $M$ was taken to be constant.  This is a good approximation
when $dM/dt\ll M$, but fails in the final stages of evaporation.\section{Black
Holes and Thermodynamics}

Since $T=\frac{\hbar\kappa}{2\pi}$ is the black hole temperature, 
we can now rewrite the $1^{\subtext{st}}$ law of black hole mechanics as
\be
dM=TdS_{\subtext{BH}}+\Omega_H dJ +\Phi_H dQ, \qquad \mbox{($\Omega_H,\Phi_H$ 
intensive, $J,Q$ extensive)}
\ee
where
\bebox{
S_{\subtext{BH}}=\frac{A}{4\hbar}
}
is the black hole (or Beckenstein-Hawking) 
entropy\index{black hole!entropy}\index{Beckenstein-Hawking entropy}.  

Clearly, black hole evaporation via Hawking radiation will 
cause $S_{\subtext{BH}}$ to \emph{decrease} in violation of the
$2^{\subtext{nd}}$ law of black hole mechanics (derived on the assumption of
classical physics).  But the entropy is 
\be
S=S_{\subtext{BH}}+S_{\subtext{ext}}
\ee
where $S_{\subtext{ext}}$ is the entropy of matter in exterior spacetime.  
But because the Hawking radiation is \emph{thermal}, $S_{\subtext{ext}}$
increases with the result that $S$ is a non-decreasing function of time.  This
suggests:

\subsubsection{Generalized $2^{\subtext{nd}}$ Law of Thermodynamics}

$S=S_{\subtext{BH}}+S_{\subtext{ext}}$ is always a non-decreasing function 
of time (in any process). \\

This was first suggested by Beckenstein (without knowledge of the precise 
form of $S_{\subtext{BH}}$) on the grounds that the entropy in the exterior
spacetime could be decreased by throwing matter into a black hole.  This would
violate the $2^{\subtext{nd}}$ law of thermodynamics unless the black hole is
assigned an entropy.

\subsection{The Information Problem}

Taking Hawking radiation into account, a black hole that forms from 
gravitational collapse will eventually evaporate, after which the spacetime has
no event horizon.  This is depicted by the following CP diagram:
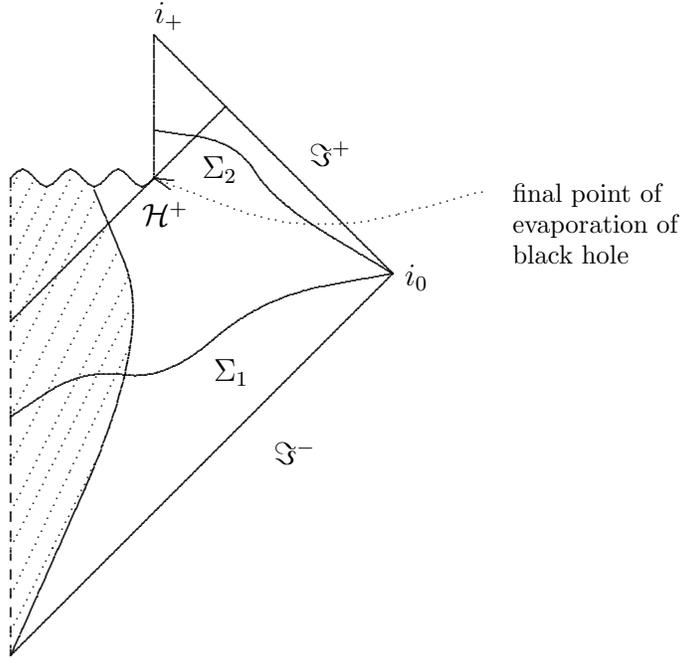
\begin{center}\input{p140-1.pictex}\end{center}
$\Sigma_1$ is a Cauchy surface for this spacetime, but $\Sigma_2$ is not 
because its past domain of dependence $D^-\left(\Sigma_2\right)$ does not
include the black hole region.  Information from $\Sigma_1$ can propagate into
the black hole region instead of to $\Sigma_2$.  Thus it appears that
information is `lost' into the black hole.  This would imply a
\emph{non-unitary} evolution from $\Sigma_1$ to $\Sigma_2$, and hence put QFT in
curved spacetime in conflict with a basic principle of Q.M.  However, from the
point of view of a static external observer, nothing actually passes through
$\mcH^+$, so maybe the information is not really lost.  A complete calculation
including all back-reaction effects might resolve the issue, but even this is
controversial since some authors claim that the resolution requires an
understanding of the Planck scale physics.  The point is that whereas QFT in
curved spacetime predicts $T_{\subtext{loc}}\to\infty$ on the horizon of a black
hole, this should not be believed when $kT$ reaches the Planck energy
$\left(\hbar c/G\right)^{1/2}c^2$ because i) Quantum Gravity\index{quantum
gravity} effects cannot then be ignored and ii) this temperature is then of the order
maximum (Hagedorn) temperature in string theory\index{string theory}. \\

\appendix
\chapter{Example Sheets}

%% TJWP removed plain tex commands
%\magnification\magstephalf
%\font\bb=msbm10 scaled 1200
%\font\small=cmr8
%\font\little=cmmi7
%%%%%%%%%%%%%%%%%%%%%%%

%% TJWP include page numbers
%\nopagenumbers
\def \pr{\partial}
\def\pmb#1{\setbox0=\hbox{#1}%
 \kern-.025em\copy0\kern-\wd0
 \kern.05em\copy0\kern-\wd0
 \kern-.025em\raise.0433em\box0 }
\def \bE{{\pmb {${\cal E}$}}}
\noindent

%\centerline{{\bf Example Sheet 1}}
%\vskip 10pt
\section{Example Sheet 1}

\noindent
{\bf 1.} Explain why 
\vskip 0.3cm
(i) GR effects are important for neutron stars but not for white
dwarfs
\vskip 0.3cm
(ii) inverse beta-decay becomes energetically favourable for
densities higher than those in white dwarfs.
\vskip 10 pt
\noindent
{\bf 2.} Use Newtonian theory to derive the Newtonian pressure
support equation
$$
P'(r) \equiv {dP\over dr} = -{Gm\rho\over r^2}\ ,
$$
where
$$
m=4\pi \int_0^r \tilde r^2 \rho(\tilde r) d\tilde r\ ,
$$
for a spherically-symmetric and static star with pressure $P(r)$ and 
density $\rho(r)$. Show that
%% TJWP no eqalign in latex  \eqalign{
\bean
\int_0^r P(\tilde r)\tilde r^3 d\tilde r & = & 
{P(r)r^4\over 4} -{1\over 4}\int_0^r P'(\tilde r) \tilde r^4 d\tilde 
r \\
 & = & {Gm^2(r)\over 32\pi} + {P(r)r^4\over 4}\ .
\eean
Assuming that $P'\le 0$, with $P=0$ at the star's surface, show that
$$
{d\over dr}\left[\left(\int_0^r P(\tilde r)\tilde r^3 d\tilde
r\right)^{3/4}\right] \le {3\sqrt{2}\over 4} P^{3/4} r^2\ .
$$
Assuming the bound
$$
P\ {\buildrel <\over\sim}\ (\hbar c) n_e^{4/3}\ ,
$$
where $n_e(r)$ is the electron number density, show that the total
mass, $M$, of the star satisfies
$$
M\ {\buildrel <\over\sim}\ \left({hc\over
G}\right)^{3/2}\left({\mu_e\over m_N}\right)^2 
$$
where $m_N$ is the nucleon mass and $\mu_e$ is the number of
electrons per nucleon. Why is it reasonable to bound the pressure as
you have done? Compare your bound with Chandresekhar's limit.

\vskip 10 pt
\noindent
{\bf 3.} A particle orbits a Schwarzschild black hole with
non-zero angular momentum per unit mass $h$. Given that $\sigma=0$ for
a massless particle and $\sigma=1$ for a massive particle, show that
the orbit satisfies  
$$ 
{d^2 u\over d\phi^2} + u = {M\sigma \over
h^2} + 3Mu^2 
$$
where $u=1/r$ and $\phi$ is the azimuthal angle. Verify that this
equation is solved by
$$
u= {1\over 6M} + {2\omega^2\over 3M} -{2\omega^2\over M\cosh ^2
(\omega\phi)}\ ,
$$
where $\omega$ is given by
$$
4\omega^2 = \pm\sqrt{\left(1 - {12M^2\sigma\over h^2}\right)}\ .
$$
where $\sigma=1$ for a massive particle and $\sigma=0$ for a massless
particle. Interpret these orbits in terms of the effective potential.
Comment on the cases $\omega^2=1/4$, $\omega^2=1/8$ and $\omega^2=0$.

\vskip 10 pt
\noindent
{\bf 4.} A photon is emitted outward from a point P outside
a Schwarzschild black hole with radial coordinate r in the range
$2M<r<3M$. Show that if the photon is to reach infinity the angle its
initial direction makes with the radial direction (as determined by 
a stationary observer at P) cannot exceed
$$
{\rm arcsin} \sqrt{{27M^2\over r^2}\left( 1-{2M\over r}\right)}\ .
$$

\vskip 10 pt
\noindent
{\bf 5.} Show that in region II of the Kruskal manifold one may
regard $r$ as a time coordinate and introduce a new spatial
coordinate $x$ such that
$$
ds^2 = -{dr^2\over \left({2M\over r} -1\right)} +\left({2M\over
r}-1\right)dx^2 + r^2d\Omega^2\ .
$$
Hence show that {\it every} timelike
curve in region II intersects the singularity at $r=0$ within a proper
time no greater than $\pi M$. For what curves is this bound attained?
Compare your result with the time taken for the collapse of a ball of
pressure free matter of the same gravitational mass $M$. Calculate
the binding energy of such a ball of dust as a fraction of its
(conserved) rest mass.
\vskip 10 pt
\noindent
{\bf 6.} Using the map
$$
(t,x,y,z) \mapsto X= \pmatrix{t+z & x+iy\cr x-iy & t-z}\ ,
$$
show that Minkowski spacetime may be identified with the space of
Hermitian $2\times 2$ matrices $X$ with metric
$$
ds^2 = -\det (dX)\ .
$$
Using the Cayley map $X\mapsto U={1+iX\over1-iX}$, show further that
Minkowski spacetime may be identified with the space of unitary
$2\times 2$ matrices $U$ for which $\det (1+U)\ne0$. Now show that any
$2\times 2$ unitary matrix $U$ may be expressed uniquely in terms of
a real number $\tau$ and two complex numbers $\alpha$, $\beta$, as 
$$
U=e^{i\tau}\pmatrix{\alpha & \beta\cr -\bar\beta & \bar\alpha}
$$
where the parameters $(\tau,\alpha,\beta)$ satisfy $|\alpha|^2
+|\beta|^2 =1$, and are subject to the identification 
$$
(\tau,\alpha,\beta) \sim (\tau +\pi, -\alpha, -\beta)\ .
$$
Using the relation
$$
(1+U)dX = -2i dU(1+U)^{-1}\ ,
$$
deduce that 
$$
ds^2 = {1\over (\cos \tau + {\cal R}e\,\alpha)^2}\left(-d\tau^2
+|d\alpha|^2 + |d\beta|^2\right) 
$$
is the metric on Minkowski spacetime and hence conclude that the
conformal compactification of Minkowski spacetime may be identified
with the space of unitary $2\times 2$ matrices, i.e the group
$U(2)$. Explain how $U(2)$ may be identified with a portion of the
Einstein static universe $S^3\times {\bb R}$.

\vfill\eject

%\centerline {{\bf Example Sheet 2}}
%\vskip 10 pt
\section{Example Sheet 2}

\noindent
{\bf 1.} Let $\zeta$ be a Killing vector field. Prove that
$$
D_\sigma D_\mu \zeta_\nu = R_{\nu\mu\sigma}{}^\lambda \zeta_\lambda\ ,
$$
where $R_{\nu\mu\sigma\lambda}$ is the Riemann tensor, defined by
$[D_\mu,D_\nu]\, v_\rho = R_{\mu\nu\rho}{}^\sigma v_\sigma$ for
arbitrary vector field $v$.

\vskip 10 pt
\noindent
{\bf 2.}  A conformal Killing vector is one for which  
$$
({\cal L}_\xi g)_{\mu\nu} = \Omega^2 g_{\mu\nu}\ .
$$
for some non-zero function $\Omega$.
Given that $\xi$ is a Killing vector of $ds^2$, show that it is a
conformal Killing vector of the conformally-equivalent
metric $\Lambda^2 ds^2$ for arbitrary (non-vanishing) conformal factor
$\Lambda$.
 
Show that the action for a {\it massless} particle,
$$ S[x,e]={1\over2}\int d\lambda\, e^{-1}\dot x^\mu\dot x^\nu
g_{\mu\nu}(x)\ , $$ is invariant, to first order in the constant
$\alpha$, under the transformation 
$$
x^\mu \rightarrow x^\mu + \alpha \xi^\mu(x) \qquad \qquad
e \rightarrow e + {1\over4}\alpha\, e g^{\mu\nu}({\cal L}_\xi
g)_{\mu\nu}
$$
if $\xi =\xi^\mu \partial_\mu$ is a conformal Killing vector. Show
that $\xi$ is the operator corresponding to the conserved charge
implied by Noether's theorem.

\vskip 10 pt
\noindent
{\bf 3.} Show that the extreme RN metric in isotropic coordinates is
$$
ds^2 = -\left(1+{M\over\rho}\right)^{-2}dt^2 + 
\left(1+{M\over\rho}\right)^{2}\left(d\rho^2 + \rho^2 d\Omega^2\right)
\qquad (\dagger)
$$
Verify that $\rho=0$ is at infinite proper distance from any finite
$\rho$ along any curve of constant $t$. Verify also that
$|t|\rightarrow \infty$ as $\rho\rightarrow 0$ along any timelike or
null curve but that a timelike or null ingoing radial geodesic reaches
$\rho=0$ for {\it finite} affine parameter. By introducing a null
coordinate to replace $\rho$ show that $\rho=0$ is merely a coordinate
singularity and hence that the metric ($\dagger$) is geodesically
incomplete. What happens to the particles that reach $\rho=0$?
Illustrate your answers using a Penrose diagram.

\vskip 10 pt
 \noindent
{\bf 4.} The action for a particle of mass $m$ and charge $q$ is
$$
S[x,e] =\int d\lambda\,\left[{1\over2} e^{-1}\dot x^\mu\dot x^\nu
g_{\mu\nu}(x) -{1\over2}m^2 e -q\, \dot x^\mu A_\mu(x)\right]\qquad
\qquad (*) 
$$
where $A_\mu$ is the electromagnetic 4-potential. Show
that if
$$
({\cal L}_\xi A)_\mu \equiv \xi^\nu\partial_\nu A_\mu +
(\partial_\mu \xi^\nu) A_\nu =0
$$
for Killing vector $\xi$, then $S$ is invariant, to first-order in
$\xi$, under the transformation $x^\mu\rightarrow x^\mu
+\alpha\xi^\mu(x)$. Verify that the corresponding Noether charge
$$
-\xi^\mu \left(m u_\mu -q A_\mu\right)\ ,
$$ 
where $u^\mu$ is the particle's 4-velocity, is a constant of the 
motion. Verify for the Reissner-Nordstrom solution of the vacuum
Einstein-Maxwell equations, with mass $M$ and charge $Q$, that ${\cal
L}_k A =0$ for $k={\partial\over \partial t}$ and hence deduce, for
$m\ne 0$, that  
$$
\left(1-{2M\over r} + {Q^2\over r^2}\right) {dt\over d\tau} =
\varepsilon -{q\over m}{Q\over r}\ ,
$$
where $\tau$ is the particle's proper time and $\varepsilon$ is the
energy per unit mass. Show that the trajectories $r(t)$ of
massive particles with zero angular momentum satisfy
$$
({dr\over d\tau})^2 = (\varepsilon^2 -1) + \left(1-\varepsilon
{qQ\over mM}\right){2M\over r} +\left(\left({q\over m}\right)^2
-1\right){Q^2\over r^2}\ .
$$
Give a physical interpretation of this result for the special case
for which $q^2=m^2$, $qQ=mM$, and $\varepsilon=1$.

\vskip 10 pt
\noindent
{\bf 5.} Show that the action
$$
S[p,x,e]=\int d\lambda \big\{ p_\mu \dot x^\mu
-{1\over2}e\, \big[g^{\mu\nu}(x)p_\mu p_\nu + m^2\big]\big\}
$$
for a point particle of mass $m$ is equivalent, for $q=0$, to the action
of Q.4. Show that $S$ is invariant to
first order in $\alpha$ under the transformation
$$
\delta x^\mu =\alpha K^{\mu\nu}p_\nu \qquad \delta p_\mu
=-{1\over2}\alpha\, p_\rho p_\sigma \partial_\mu K^{\rho\sigma}
$$
for any symmetric tensor $K_{\mu\nu}$ obeying the {\it Killing
tensor} condition
$$
D_{(\rho} K_{\mu\nu)}=0\ .
$$
Show that the corresponding Noether charge is proportional to
$K^{\mu\nu}p_\mu p_\nu$ and verify that it is a constant of the
motion. A trivial example is $K_{\mu\nu}=g_{\mu\nu}$; what is
the corresponding constant of the motion? Show that
$\xi_\mu\xi_\nu$ is a Killing tensor if $\xi$ is a Killing vector. [A
Killing tensor that cannot be constructed from the metric and Killing
vectors is said to be irreducible. In a general axisymmetric
metric there are no such tensors, and so only three constants of the
motion, but for geodesics of the Kerr-Newman metric there is a 
`fourth constant' of the motion corresponding to an
irreducible Killing tensor.]
%$$
%\eqalign{
%K_{\mu\nu}dx^\mu dx^\nu = - &{\Sigma a^2\cos^2\theta\over \Delta}dr^2
%+ {\Delta a^2 \cos^2\theta\over \Sigma}(dt -a\sin^2\theta\,
%d\phi)^2\cr & +{r^2\sin^2\theta\over\Sigma}[adt - (r^2 + a^2)d\phi]^2
%+ r^2\Sigma
% d\theta^2\ .}
%$$

\vskip 10 pt 
\noindent 
{\bf 6.} By replacing the time coordinate $t$
by one of the radial null coordinates
$$
u= t+ {M\over \lambda} \qquad v= t- {M\over \lambda}
$$
show that the singularity at $\lambda=0$ of the Robinson-Bertotti (RB)
metric
$$
ds^2 = -\lambda^2 dt^2 + M^2 \left({d\lambda\over \lambda}\right)^2 +
M^2 d\Omega^2 
$$
is merely a coordinate singularity.
Show also that $\lambda=0$ is a degenerate Killing Horizon with respect
to $\partial\over \partial t$. By
introducing the new coordinates $(U,V)$, defined by
$$
u= \tan \left({U\over 2}\right)\qquad v=-\cot \left({V\over2}\right)
$$
obtain the maximal analytic extension of the RB metric and deduce its
Penrose diagram.

\vfill\eject

%\centerline {{\bf Example Sheet 3}}
%\vskip 10pt
\section{Example Sheet 3}

\noindent 
{\bf 1.} Let $\varepsilon$ and $h$ be the energy and and angular
momentum per unit mass of a zero charge particle in free fall
within the equatorial plane, i.e on a timelike ($\sigma=1$) or null
($\sigma=0$) geodesic with $\theta=\pi/2$, of a Kerr-Newman black hole.
Show that the particle's Boyer-Lindquist radial coordinate $r$
satisfies   
$$ 
\left({dr\over d\lambda}\right)^2 =\varepsilon^2 - V_{eff}(r)\ ,
$$ 
where $\lambda$ is an affine parameter, and the effective potential
$V_{eff}$ is given by $$ 
V_{eff} =
\left(1-{2M\over r}+{e^2\over r^2}\right)\left(\sigma + {h^2\over
r^2}\right) + {2a\varepsilon h\over r^3}\left( 2M -{e^2\over r}\right)
+{a^2\over r^2}\left[\sigma-\varepsilon^2\left(1+{2M\over r}-{e^2\over
r^2}\right)\right] \ .
$$

\vskip 10 pt 
\noindent 
{\bf 2.} Show that the surface gravity of the event horizon of a Kerr
black hole of mass $M$ and angular momentum $J$ is given by
$$
\kappa = {\sqrt{M^4 -J^2}\over 2M( M^2 + \sqrt{M^4 -J^2})}\ .
$$

\vskip 10 pt 
\noindent 
{\bf 3.} A particle at fixed $r$ and $\theta$ in a stationary
spacetime, with metric $ds^2= g_{\mu\nu}(r,\theta)dx^\mu dx^\nu$, has
angular velocity $\Omega= {d\phi\over dt}$ with respect to infinity.
Show that $\Omega(r,\theta)$ must satisfy
$$
g_{tt} + 2g_{t\phi}\Omega + g_{\phi\phi}\Omega^2 \le 0
$$
and hence deduce that
$$
{\cal D}\equiv g_{t\phi}^2 - g_{tt}g_{\phi\phi} \ge 0
$$
Show that ${\cal D}=\Delta (r) \sin^2\theta$ for the Kerr-Newman 
metric in Boyer-Lindquist coordinates, where $\Delta= r^2-2Mr+a^2+e^2$.
What happens if $(r,\theta)$ are such that ${\cal D}<0$? For what
values of $(r,\theta)$ can $\Omega$ vanish? Given that $r_{\pm}$ are the 
roots of $\Delta$, show that when ${\cal D}=0$  
$$ 
\Omega = {a\over r_{\pm}^2 +a^2} \ .
$$
\vskip 10pt
\noindent
{\bf 4.} Show that the area of the event horizon of a Kerr-Newman
black hole is
$$
A= 8\pi\big[ M^2 - {e^2\over2} + \sqrt{ M^4 -e^2 M^2 -J^2 }\,\big]\ .
$$

\vskip 10 pt
\noindent
{\bf 5.} A perfect fluid has stress tensor
$$
T_{\mu\nu} = (\rho + P)u_\mu u_\nu + P g_{\mu\nu}\ ,
$$
where $\rho$ is the density and $P(\rho)$ the pressure. State the
dominant energy condition for $T_{\mu\nu}$ and show that
for a perfect fluid in Minkowski spacetime this condition
is equivalent to  
$$ 
\rho\ge |P|\ .
$$
Show that the same condition arises from the requirement of
causality, i.e. that the speed of sound, $\sqrt{|dP/d\rho|}$, not
exceed that of light, together with the fact that the pressure 
vanishes in the vacuum.
\vskip 10pt
\noindent
{\bf 6.} The vacuum Einstein-Maxwell equations are
$$
G_{\mu\nu}= 8\pi T_{\mu\nu}(F) \qquad D_\mu F^{\mu\nu}=0
$$
where $F_{\mu\nu}= \partial_{\mu}A_{\nu}- \partial_\nu A_\mu$, and
$$
T_{\mu\nu}(F)= {1\over 4\pi}\big(F_{\mu}{}^\lambda F_{\nu\lambda}
-{1\over4}g_{\mu\nu}F^{\alpha\beta}F_{\alpha\beta}\big)\ .
$$
Asymptotically-flat solutions are stationary and axisymmetric if
the metric admits Killing vectors $k$ and $m$ that can be taken to be
$k={\partial\over\partial t}$ and $m={\partial\over\partial \phi}$ near
infinity, and if (for some choice of electromagnetic gauge)
$$
{\cal L}_k A={\cal L}_m A=0\ ,
$$
where the Lie derivative of $A$ with respect to a vector $\xi$, 
${\cal L}_\xi A$, is as defined in Q.4 of Example Sheet 2. The event
horizon of such a solution is necessarily a Killing horizon of $\xi =
k+\Omega_H m$, for some constant $\Omega_H$. What is the physical
interpretation of $\Omega_H$? What is its value for the Kerr-Newman
solution? The co-rotating electric potential is defined by 
$$ 
\Phi = \xi^\mu A_\mu \ .
$$
Use the fact that $R_{\mu\nu}\xi^\mu\xi^\nu=0$ on a Killing horizon
to show that $\Phi$ is constant on the horizon. In particular, show
that for a choice of the electromagnetic gauge for which $\Phi=0$ at
infinity, 
$$
\Phi_H= {Qr_+\over r_+^2 +a^2}
$$
for a charged rotating black hole, where $r_+= M+\sqrt{M^2-Q^2-a^2}$.

\vskip 10pt
\noindent
{\bf 7.} Let $({\cal M},g,A)$ be an asymptotically flat, stationary,
axisymmetric, solution of the Einstein-Maxwell equations of Q.6 and
let $\Sigma$ be a spacelike hypersurface with one boundary at spatial
infinity and an internal boundary, $H$, at the event horizon of a black
hole of charge $Q$. Show that    
$$ 
-2\int_\Sigma dS_\mu T^\mu{}_\nu(F)\xi^\nu = \Phi_H Q  
$$
where $\Phi_H$ is the co-rotating electric potential on the horizon. 
Use this result to deduce that the mass $M$ of a charged rotating black
hole is given by 
$$
M= {\kappa A\over 4\pi} + 2\Omega_H J + \Phi_H Q\ .
$$
where $J$ is the total angular momentum.
Use this formula for $M$ to deduce the first law of black hole
mechanics for charged rotating black holes: 
$$
dM= {\kappa \over 8\pi}dA + \Omega_H dJ + \Phi_H dQ\ .
$$
[Hint: ${\cal L}_\xi (F^{\mu\nu}A_\nu)=0$ ]

\vfill\eject

%\centerline{{\bf Example Sheet 4}}
%\vskip 10pt
\section{Example Sheet 4}

\noindent
{\bf 1.} Use the Komar integral,
$$
J= {1\over 16\pi G}\oint_\infty dS_{\mu\nu}D^\mu m^\nu\ ,
$$
for the total angular momentum of an asymptotically-flat axisymmetric
spacetime (with Killing vector $m$) to verify that $J=Ma$ for the
Kerr-Newman solution with parameter $a$.

\vskip 10pt
\noindent
{\bf 2.} Let $l$ and $n$ be two linearly independent vectors and
$\hat B$ a second rank tensor such that
$$
\hat B_\mu{}^\nu l_\nu =\hat B_\mu{}^\nu n_\nu =0\ .
$$
Given that $\eta^{(i)}$ $(i=1,2)$ are two further linearly
independent vectors, show that
$$
\varepsilon^{\mu\nu\rho\sigma}l_\mu n_\nu \hat B_\rho{}^\lambda
\big(\eta^{(1)}_\lambda\eta^{(2)}_\sigma - 
\eta^{(1)}_\sigma\eta^{(2)}_\lambda\big) =  \theta\, 
\varepsilon^{\mu\nu\rho\sigma} l_\mu n_\nu
\eta_\rho^{(1)}\eta_\sigma^{(2)}\ .
$$
where $\theta= \hat B_\alpha{}^\alpha$.

\vskip 10pt
\noindent
{\bf 3.} Let ${\cal N}$ be a Killing horizon of a Killing vector field
$\xi$, with surface gravity $\kappa$. Explain why, for any third-rank
totally-antisymmetric tensor $A$, the scalar 
$\Psi = A^{\mu\nu\rho}(\xi_\mu D_\nu\xi_\rho)$ vanishes on ${\cal N}$.
Use this to show that
$$
(\xi_{[\rho}D_{\sigma]} \xi_\nu)(D^\nu\xi^\mu) =\kappa
\xi_{[\rho}D_{\sigma]}\xi^\mu \qquad ({\rm on}\ {\cal N})\ ,\qquad (*)
$$
where the square brackets indicate antisymmetrization on the enclosed
indices.

>From the fact that $\Psi$ vanishes on ${\cal N}$ it follows that its
derivative on ${\cal N}$ is normal to ${\cal N}$, and hence that
$\xi_{[\mu}\partial_{\nu]}\Psi=0$ on ${\cal N}$. Use this fact and the
Killing vector lemma of Q.II.1 to deduce that, on ${\cal N}$,
$$
(\xi_\nu R_{\sigma\rho[\beta}{}^\lambda\xi_{\alpha]}
+\xi_\rho R_{\nu\sigma[\beta}{}^\lambda\xi_{\alpha]}
+\xi_\sigma R_{\rho\nu[\beta}{}^\lambda\xi_{\alpha]})\xi_\lambda\ .
$$
Contract on $\rho$ and $\alpha$ and use the fact that $\xi^2=0$ on
${\cal N}$ to show that
$$
\xi^\nu\xi_{[\rho}R_{\sigma]\nu\mu}{}^\lambda\xi_\lambda =
-\xi_\mu\xi_{[\rho}R_{\sigma]}{}^\lambda\xi_\lambda
\qquad ({\rm on}\ {\cal N})\ ,\qquad (\dagger)
$$
where $R_{\mu\nu}$ is the Ricci tensor.

For any vector $v$ the scalar $\Phi=(\xi\cdot D\xi -\kappa\xi)\cdot v$
vanishes on ${\cal N}$. It follows that
$\xi_{[\mu}\partial_{\nu]}\Phi|_{\cal N} =0$. Show that this fact, the
result (*) derived above and the Killing vector lemma imply
that, on ${\cal N}$, 
%% $$ \eqalign{
\bean
\xi^\mu\xi_{[\rho}\partial_{\sigma]}\kappa & = & \xi^\nu
R_{\mu\nu[\sigma}{}^\lambda\xi_{\rho]}\xi_\lambda \\
& = &\xi^\nu\xi_{[\rho}R_{\sigma]\nu\mu}{}^\lambda\xi_\lambda\ ,
\eean
%% } $$
where the second line is a consequence of the cyclic identity
satisfied by the Riemann tensor. Now use $(\dagger)$ to show that, on
${\cal N}$,
%$$ \eqalign{
\bea
\xi^\mu\xi_{[\rho}\partial_{\sigma]}\kappa & = &
\xi_{[\sigma}R_{\rho]}{}^\lambda\xi_\lambda \\
&= & 8\pi G\, \xi_{[\sigma}T_{\rho]}{}^\lambda\xi_\lambda\ ,
\eea
%} $$
where the second line follows on using the Einstein equations. Hence
deduce the zeroth law of black hole mechanics: that, provided the
matter stress tensor satisfies the dominant energy condition, the
surface gravity of any Killing vector field $\xi$ is constant on
each connected component of its Killing horizon (in particular, on the
event horizon of a stationary spacetime). 
% TJWP removed \vfill\eject

\vskip 10pt
\noindent
{\bf 4.} A scalar field $\Phi$ in the Kruskal spacetime satisfies
the Klein-Gordon equation
$$
D^2\Phi -m^2\Phi =0\ .
$$
Given that, in static Schwarzshild coordinates, $\Phi$ takes the form
$$
\Phi = R_\ell(r) e^{-i\omega t} Y_{\ell}(\theta,\phi)
$$
where $Y_{\ell\, m}$ is a spherical harmonic, find the radial equation
satisfied by $R_\ell(r)$. Show that near the horizon at $r=2M$,
$\Phi\sim e^{\pm i\omega r^*}$, where $r^*$ is the Regge-Wheeler radial
coordinate. Verify that ingoing waves are analytic, in Kruskal
coordinates, on the future horizon, ${\cal H}^+$, but not, in general,
on the past horizon, ${\cal H}^-$, and conversely for outgoing waves.

Given that both $m$ and $\omega$ vanish, show that
$$
R_\ell = A_\ell P_\ell(z) + B_\ell Q_\ell(z)
$$
for constants $A_\ell,\, B_\ell$, where $z=(r-M)/M$, $P_\ell(z)$ is a
Legendre Polynomial and $Q_\ell(z)$ a linearly-independent solution.
Hence show that there are no {\it non-constant} solutions that are both
regular on the horizon, ${\cal H}= {\cal H}^+ \cup {\cal H}^-$, and
bounded at infinity.

\vskip 10 pt
\noindent
{\bf 5.} Use the fact that a Schwarzschild black hole radiates at the
Hawking temperature
$$
T_H ={1\over 8\pi M}
$$
(in units for which $\hbar$, $G$, $c$, and Bolzmann's constant all
equal $1$) to show that the thermal equilibrium of a black hole with an
infinite reservoir of radiation at temperature $T_H$ is unstable.

A finite reservoir of radiation of volume $V$ at temperature $T$ has
an energy, $E_{res}$ and entropy, $S_{res}$ given by
$$
E_{res} = \sigma VT^4 \qquad S_{res} ={4\over3}\sigma VT^3
$$
where $\sigma$ is a constant. A Schwarzschild black hole of mass $M$ is
placed in the reservoir. Assuming that the black hole has entropy
$$
S_{BH} =4\pi M^2\ ,
$$
show that the total entropy $S= S_{BH}+S_{res}$ is extremized 
for fixed total energy $E= M+E_{res}$, when $T=T_H$, Show that the
extremum is a maximum if and only if $V<V_c$, where the critical value
of $V$ is $$
V_c = {2^{20}\pi^4E^5\over 5^5\sigma }
$$
What happens as $V$ passes from $V<V_c$ to $V>V_c$, or
vice-versa?

\vskip 10pt
\noindent
{\bf 6.} The specific heat of a charged black hole of mass $M$, at
fixed charge $Q$, is 
$$
C\equiv T_H {\partial S_{BH}\over \partial
T_H}\bigg|_Q \ ,
$$
where $T_H$ is its Hawking temperature and $S_{BH}$ its
entropy. Assuming that the entropy of a black hole is given by $S_{BH}=
{1\over4}A$, where $A$ is the area of the event horizon, show that the
specific heat of a Reissner-Nordstrom black hole is
$$
C= {2S_{BH}\sqrt{M^2-Q^2}\over (M-2\sqrt{M^2-Q^2})}\ .
$$
Hence show that $C^{-1}$ changes sign when $M$ passes through
${2|Q|\over\sqrt{3}}$. 

Repeat Q.5 for a Reissner-Nordstrom black hole.
Specifically, show that the critical reservoir volume, $V_c$, is
infinite for $|Q|\le M \le {2|Q|\over\sqrt{3}}$. Why is this result
to be expected from your previous result for $C$?

%\end

%
\printindex
\end{document}